\documentclass[aps,pre,reprint,superscriptaddress,floatfix,showpacs]{revtex4-2}
\usepackage{amsmath,amssymb,amsfonts}
\usepackage{bm}
\usepackage{graphicx}
\usepackage{dcolumn}
\usepackage{hyperref}
\usepackage{xcolor}
\usepackage{physics}      
\usepackage{subcaption}
\usepackage{booktabs} 
\usepackage{siunitx}  
\newcommand{\Lop}{\mathbf{\hat{L}}}          
\newcommand{\pieq}{\bm{\pi_{\rm eq}}}   
\newcommand{\vrho}{\bm{\rho}}            
\newcommand{\lam}{\lambda}              
\newcommand{\tvd}{\mathrm{TVD}}
\newcommand{\bbone}{\mathbf{1}}         
\begin{document}

\title{Stochastic Counterdiabatic Driving via Biorthogonal Liouvillian Eigenmodes}

\author{Sandeep Suresh Cranganore}
\affiliation{Institute for Machine Learning (Ellis Unit), Johannes Kepler University, Linz}

\author{Sebastian Lehner}
\affiliation{Institute for Machine Learning (Ellis Unit), Johannes Kepler University, Linz}

\author{Johannes Brandstetter}
\affiliation{Institute for Machine Learning (Ellis Unit), Johannes Kepler University, Linz}
\affiliation{Mistral AI, Paris, France}

\author{Max Welling}
\affiliation{AMLab, Informatics Institute, University of Amsterdam, The Netherlands}
\affiliation{CuspAI, Cambridge, England}

\date{\today}

\begin{abstract}
Finite-time driving of stochastic systems generates excess
dissipation, causing the evolving probability distribution to
lag behind the instantaneous equilibrium, and consequently degrading the
convergence of nonequilibrium free energy estimators based on the
Jarzynski equality. Escorted free energy simulations address the non-adiabatic lag by engineering
control fields $\mathbf{u}$ that eliminate the lag, enforcing the
trajectory-wise equality
$\mathcal{W}_{\mathbf{u}} = \Delta\mathcal{F}$, and yielding zero-variance
estimators. However, constructing the escorting field in closed
form remains a challenge, approached variously through flow-field
methods, targeted free energy perturbation, or learned
diffeomorphisms.
In this work, we construct a complementary numerical framework
based on gauge-type transforms instead of generalized coordinate
transforms for perfect escorting based on the exact spectral
decomposition of the time-dependent Fokker--Planck generator.
The \emph{biorthogonal decomposition} of the Liouville operator
directly yields a counterdiabatic correction whose
action on the instantaneous equilibrium distribution exactly
cancels the non-adiabatic lag at arbitrary driving speed in
formal analogy with shortcuts-to-adiabaticity techniques such as
Berry's transitionless driving for quantum systems.
Numerical verification for simulations of an overdamped particle
in a time-varying double-well potential and harmonic traps
confirms that the counterdiabatic condition is satisfied to machine precision, with the non-adiabatic lag suppressed by roughly twelve orders of magnitude in total variation distance and sixteen orders in KL divergence relative to the unescorted dynamics. As a diagnostic, we demonstrate vanishing dissipated work $\mathcal{W}_{\rm diss}(t) \approx 0$ for the deterministically propagated Fokker–Planck density across all protocol speeds.
\end{abstract}
\maketitle

\section{Introduction}
\label{sec:intro}

The adiabatic theorem underpins a wide range of phenomena in
physics, from quantum state preparation~\cite{Albash2018} to
classical non-equilibrium thermodynamics~\cite{Jarzynski2011}.
In both settings, adiabatic following fails whenever there is a finite-time switching, i.e.,
when protocol speed is finite: a quantum system leaks into excited
states, while a classical diffusing particle lags behind the
instantaneous equilibrium distribution. To systematically quantify this lag and relate microscopic fluctuations to macroscopic laws, \emph{stochastic thermodynamics} has emerged as a foundational framework.This microscopic approach has since found widespread applications across multiple disciplines, ranging from molecular dynamics and soft-matter design to nonperturbative quantum chromodynamics~\citep{RevModPhys.70.323}. Crucially, it provided the arena for the groundbreaking work of \citet{PhysRevLett.78.2690}, which marked a major breakthrough in non-equilibrium physics. Later identified as an instance of annealed importance sampling (AIS)~\citep{neal1998annealedimportancesampling}, this framework allows for the estimation of equilibrium free energy differences from non-equilibrium work via the celebrated \emph{Jarzynski Equality} (JE):
\begin{equation}\label{eq:JE}
\langle e^{-\beta \mathcal{W}}\rangle = e^{-\beta \Delta \mathcal{F}},
\end{equation}
where $\beta = 1/(k_B T)$ is the inverse temperature. By Jensen's inequality applied to the convex exponential,
$\langle e^{-\beta\mathcal{W}}\rangle \geq e^{-\beta\langle\mathcal{W}\rangle}$,
Eq.~\eqref{eq:JE} immediately yields
$\langle\mathcal{W}\rangle \geq \Delta\mathcal{F}$,
recovering the second law for isothermal processes: the mean
dissipated work $\langle\mathcal{W}_{\rm diss}\rangle =
\langle\mathcal{W}\rangle - \Delta\mathcal{F} \geq 0$
is non-negative, with equality achieved only in the
quasi-static limit $\tau\to\infty$. The JE is an instance of a broader class of
exact nonequilibrium relations known as fluctuation
theorems~\citep{PhysRevE.60.2721}.
In particular, the Crooks fluctuation theorem~\citep{PhysRevE.60.2721}
relates the probability distributions of work in the forward
and reverse protocols,
$P_F(\mathcal{W})/P_R(-\mathcal{W}) = e^{\beta(\mathcal{W}-\Delta\mathcal{F})}$,
from which the Jarzynski equality follows by integration over
the forward work distribution.
These relations hold arbitrarily far from equilibrium and for
any protocol duration $\tau$.

However, the practical utility of Eq.~\eqref{eq:JE} as a free energy estimator is severely bottlenecked by its convergence properties under finite-time protocols. Driven by a time-varying external potential $V(\mathbf{x}, \zeta_t)$ with a control parameter $\zeta_t \equiv \zeta(t)$, the system develops a structural "lag" relative to the instantaneous Boltzmann distribution $\pi_{\text{eq}}(\mathbf{x}, \zeta_t) = \exp[-\beta (V(\mathbf{x}, \zeta_t) - \mathcal{F}(\zeta_t))]$ due to the non-adiabatic nature of the driving. These accumulated diabatic errors scale with the protocol speed, driving the system far from equilibrium and manifesting macroscopically as an excess dissipated work $\mathcal{W}_{\rm diss}(\mathbf{x}, \zeta_t)$. Consequently, the exponential average in the JE becomes heavily dominated by rare, microscopic trajectories that manage to counteract this dissipation, leading to severe sample variance and poor statistical convergence. To mitigate this limitation, a variety of seminal strategies have been developed to enhance convergence and optimize free energy estimation.

The idea of eliminating dissipation via an escorting control field
was first derived by \citet{PhysRevLett.100.190601}, who showed that
a flow field $\mathbf{u}(\mathbf{x},t)$ can be engineered that can reduce or even purge this lag entirely, yielding the escorted version of Eq.~\eqref{eq:JE}, which is often called the \emph{escorted Jarzynski
equality} (EJE): 
\begin{equation}\label{eq:eje}
\langle e^{-\beta\mathcal{W}_{\mathbf{u}}}\rangle_{\mathbf{u}} =
e^{-\beta\Delta\mathcal{F}},
\end{equation}
where, the conventional work is modified by the contributions from the escorting term:
\begin{equation}
  \mathcal{W}_{\mathbf{u}}(x, \zeta_t) := \mathcal{W}(x, \zeta_t) + \mathbf{u}(x,  t)\cdot\nabla V(x,  \zeta_t)
               - \beta^{-1}\nabla\cdot\mathbf{u} (x,  t) \ ,
\label{eq:modified_work}
\end{equation}
and the non-equilibrium work is defined as $\mathcal{W}(x, \zeta_t) = \int_0^t \mathrm{d}t' \ \dot{\zeta}_{t'}\, \left\langle
\frac{\partial V(x, \zeta_t')}{\partial \zeta}
\right\rangle_{\!\boldsymbol{\rho}(t')}$. Under perfect escorting $\mathbf{u}(\mathbf{x}, t)$, the additional terms in Eq.~\eqref{eq:modified_work} ensure that the 
system always remains in equilibrium, with respect to $V(\mathbf{x}, \zeta_t)$. Thus, for every trajectory, it holds that $\mathcal{W}_{\mathbf{u}} = \Delta \mathcal{F}$, resulting in a zero variance estimate of the free energy difference~\citep{PhysRevLett.100.190601, lee2026estimatingfreeenergydifferences}. Thus, one can arbitrarily accelerate the convergence relative to Eq.~\eqref{eq:JE}, by
designing the escorting fields such that the system remains near equilibrium throughout the finite-time process. 
\medskip 

The strategy of engineering auxiliary escorting fields to enforce tracking of a target distribution -- as if the system were evolving adiabatically -- falls under the overarching framework of shortcuts to adiabaticity (STA)~\cite{GueryOdelin2019, Kolodrubetz2017}. Historically, the concept of a shortcut was pioneered in the context of steering quantum states, introduced through the seminal frameworks of transitionless quantum driving by \citet{Berry_2009} and assisted adiabatic passage by \citet{Demirplak2003}. When these engineered fields are explicitly designed to cancel out internal non-adiabatic contributions, the protocol is designated as \emph{counterdiabatic driving} (CD)~\citep{PhysRevLett.111.100502}. Initially formalized for closed quantum systems, the Berry-type formulation and quantum adiabatic theorem has already been adapted in the context of stochastic dynamics~\citep{Iram2021}, while counterdiabatic principles in general have since been rigorously extended to classical stochastic thermodynamics architectures, providing a powerful toolkit for minimizing dissipation in fluctuating, finite-time regimes~\citep{PhysRevLett.111.100502, 10.1063/1.3544679, Jarzynski2013, Patra2017, PhysRevE.96.012144, Iram2021}. 
\medskip 

A common thread in prior work on classical stochastic STA is
that the counterdiabatic field is constructed from either a
series of generalized coordinate
transforms (diffeomorphisms)~\citep{PhysRevE.65.046122, 10.1063/1.3544679, 10.1063/5.0018903}
(see Appendix~\ref{app:zero_variance_mappings}), a
probability-current continuity
equation~\citep{Patra2017}, or MBAR-based
estimators~\citep{10.1063/1.2978177}.
Owing to the computational cost of these approaches,
machine-learning methods have become increasingly prominent,
including flow-matching~\citep{lipman2023flowmatchinggenerativemodeling,
rezende2016variational, nielsen2020survaeflowssurjectionsbridge}, (stochastic) normalizing
flows (NFs) ~\citep{DBLP:conf/nips/0035KN20}, learned switching
protocols~\citep{holdijk2026learning} and virtual escorted trajectories~\citep{lee2026estimatingfreeenergydifferences} all targeting efficient
estimation of free energy differences.
Controlled fluctuation theorems have similarly motivated neural
samplers that derive variational objectives directly from the
Jarzynski and Crooks identities and effectively optimise escorting fields via
Radon--Nikodym derivative-based objectives. These include Controlled Monte Carlo
Diffusion~\citep{vargas2024transport}, the Non-Equilibrium
Transport Sampler~\citep{albergo2025netsnonequilibriumtransportsampler},
and discrete extensions~\citep{Holderrieth2025LEAPS}.
\medskip 

In this work, we implement a spectral framework for the counterdiabatic driving (CD)~\citep{PhysRevLett.111.100502}, which we call \emph{Liouvillian counterdiabatic driving} (LCD) for classical stochastic systems governed by the Fokker-Planck equation. The underlying Fokker-Planck generator (Liouville operator) $\Lop(t)$ when discretized over the spatial coordinates is intrinsically a non-symmetric rate matrix that admits an exact biorthogonal decomposition~\citep{10.1307/mmj/1028989861,Brody_2014} whose structure directly encodes all the information of the non-adiabatic lag accumulated during finite-time driving. Exploiting this intrinsic feature, we derive a closed-form spectral formula to construct the counterdiabatic correction $\Lop_{\rm CD}(t)$ using the Liouville operator itself. We further show that this spectral correction collapses, under detailed balance,
to a rank-one operator determined in closed form by the equilibrium distribution; the biorthogonal machinery becomes equivalent to the known closed-form escorting for equilibrium problems, and its value is primarily structural---establishing the transitionless-driving analogy, exposing the spectral-gap origin of the lag. Our LCD systematically eliminates this lag, escorting the probability distribution along the instantaneous Boltzmann equilibrium $\pieq(t)$ at arbitrary protocol speeds. Consequently, our deterministic
verification shows that LCD enforces instantaneous equilibrium tracking ensuring zero dissipated work ($\mathcal{W}_{\rm diss} = 0$) for arbitrary protocol speeds (not to be confused with free-energy estimation over sampled trajectories, done via the (E)JE [Eqs.~(\ref{eq:JE}, \ref{eq:eje})] which is not performed
in this work) . Previous methods predominantly operate without any explicit reference to the spectral features of $\Lop(t)$, our approach fills this gap by working directly within its time-dependent biorthogonal eigenbasis, based on a similar approach developed in the context of quantum mechanics~\citep{Berry_2009}. This yields an exact, optimization-free construction on the discrete generator that requires no coordinate transformations or learned representations. 
\medskip 

To this end, we introduce our framework that has the following features:
\begin{itemize}
    \item \emph{Alternative to coordinate-based methods.}
        Prior constructions of the escorting field rely on
        finding a series of bijective coordinate
        reparametrisations (diffeomorphisms) of phase space~\citep{PhysRevE.65.046122,
        10.1063/1.3544679}. Our counterdiabatic correction is obtained directly from the spectral decomposition of the Fokker--Planck generator itself, via a time-dependent gauge-like transformation of
        the probability density, without any
        reference to Jacobian determinants.

    \item \emph{Formal correspondence to Berry/Demirplak--Rice methodology.}
      The spectral correction $\Lop_{\rm CD}(t)$ mirrors the structural form of the quantum counterdiabatic Hamiltonian, adapted to the non-symmetric Fokker--Planck generator via  biorthogonal decomposition. This enables the reduction (elimination) of non-adiabatic lag and as a result excess dissipated work. 
      
   \item \emph{Systematic spectral truncation and numerical scalability.}
      The counterdiabatic correction $\Lop_{\rm CD}(t)$ naturally admits a controlled low-rank approximation by restricting the biorthogonal expansion to the $M < N$ relaxation modes of the generator. Although it does not correspond to the perfect escorting, this offers a highly predictable numerical trade-off, where the number of spectral modes required to reach a target accuracy systematically scales with the driving protocol speed and more importantly the inverse spectral gap. This feature enables computationally efficient implementations in large-scale systems where tracking the full spectrum is unfeasible.
\end{itemize}
The remainder of this paper is organized as follows. Section~\ref{sec:discretisation} introduces the matrix representation of the Fokker-Planck equation obtained via a probability-conserving finite-difference spatial discretization, establishing the resulting non-symmetric structure of the discrete Liouville operator. Section~\ref{sec:problem} reformulates the finite-time Fokker-Planck dynamics within an adiabatic (co-moving) frame via a time-dependent gauge-type transformation, explicitly decoupling the evolution into an instantaneous-equilibrium component and its non-adiabatic component. Section~\ref{sec:biorthogonal} develops the biorthogonal decomposition of $\Lop(t)$ and derives the fundamental structural properties of the instantaneous eigenbasis $\{ \bm{r}_n(t), \bm{\ell}^T_n(t)\}$, showcasing how the right zero mode tracks the Boltzmann distribution $\bm{r}_0(t) = \pieq(t)$ while the left zero mode satisfies $\bm{\ell}^T_0 = \mathbf{1}^T$ as a direct consequence of probability conservation. Our main result—the classical analogue to Berry's transitionless quantum driving~\cite{Berry2009} for stochastic systems governed by the Fokker--Planck equation is derived in Section~\ref{sec:AGP}, yielding an exact formulation for the counterdiabatic correction $\Lop_{\rm CD}(t)$ constructed directly from the biorthogonal eigenpairs of $\Lop(t)$. Section~\ref{sec:numerics_results} presents numerical demonstrations across two prototypical model systems: a time-varying double-well (DW) potential and a harmonically confined Brownian particle in the overdamped setting. We assess tracking fidelity across varying protocol durations using the total variation distance, the Kullback-Leibler divergence ($\mathcal{D}_{\rm KL}$), and the trajectory-wise vanishing of the dissipated work $\mathcal{W}_{\rm diss}$. In Section~\ref{sec:spectral_truncation}, we investigate the numerical scalability of a spectrally truncated approximation to $\Lop_{\rm CD}(t)$ using a restricted subset of $M < N$ relaxation modes, establishing a systematic trade-off for improving convergence. Lastly, in Section~\ref{sec:spectral_gap_limit}, we identify potentials, where our spectral CD methodology has clear disadvantages, mainly tied to vanishing spectral gaps.  
\section{Matrix Representation of the Fokker-Planck Equation}
\label{sec:discretisation}
Consider an overdamped Brownian particle in a
time-dependent potential $V(x,\zeta(t))$, where $\zeta(t)$
is an externally controlled parameter.
The probability density $\rho(x,t)$ evolves according to the
Fokker--Planck equation written in the Liouville form~\citep{PhysRevE.56.5018} (in 1D space):
\begin{equation}
  \dfrac{\partial \rho(x,t)}{\partial t} 
  = \underbrace{
      \dfrac{\partial }{\partial x}\big[(\dfrac{\partial V(x, t)}{\partial x} )\,\rho(x, t) \big]
      + \beta^{-1} \dfrac{\partial^2 \rho(x, t)}{\partial x^2}
    }_{\displaystyle\mathcal{L}(x,t)\,\rho(x, t)},
\label{eq:FP_scalar}
\end{equation}
where, $\mathcal{L}(x, t) :=  \dfrac{\partial }{\partial x}\big[(\dfrac{\partial V(x, t)}{\partial x} \cdot) + \beta^{-1} \dfrac{\partial}{\partial x} \big]$ is the Liouville operator. The gradient of the time-varying potential, corresponds to the drift velocity: $\frac{\partial V(x, \zeta_t)}{\partial x} = - v(x, t)$. Driven by a time-varying external potential $V(x, \zeta_t)$ with a control parameter $\zeta_t$, finite-time protocols inevitably induce a lag in the form of non-adiabatic (diabatic) excitations that prevent the time-evolved Fokker-Planck probability density from tracking a target distribution with high fidelity. This lag manifests thermodynamically as an excess dissipated work $\mathcal{W}_{\rm diss} = \mathcal{W} - \Delta \mathcal{F}$, whose magnitude scales with the protocol speed $\vert{}\dot{\zeta}_t\vert{}$. In the quasi-static limit ($\dot{\zeta}_t \rightarrow 0$), the system evolves reversibly along a sequence of instantaneous equilibria without diabatic transitions, yielding a vanishing dissipation $\mathcal{W}_{\rm diss} \rightarrow 0$. As considered in other works, we have the same goal as in~\citep{PhysRevLett.100.190601}, to track the instantaneous equilibrium distribution 
\begin{equation}
  \pi_{\rm eq}(x,t) = \dfrac{e^{-\beta V(x,\zeta(t))}}{\mathcal{Z}(t)} := e^{-\beta \big[V(x,\zeta_t) - \mathcal{F}(\zeta_t)\big]},
\label{eq:Boltzmann}
\end{equation}
at all times during the Fokker--Planck evolution for any arbitrary finite-time protocol speeds. The instantaneous equilibrium distribution defined in Eq.~\eqref{eq:Boltzmann}, is a stationary solution of the Liouville operator satisfies $\mathcal{L}(x, t)\,\pi_{\rm eq}(x, t) = 0$ at every time-instant~\citep{PhysRevLett.100.190601, doi:10.1073/pnas.071034098}.

\subsection{Discretisation: Turning the PDE into a Matrix-Valued ODE}\label{subsec:pde_to_ode}
We discretise $x$ on $N$ equally spaced grid points
$\{x_i\}_{i=0}^{N-1}$ with spacing $\Delta x$.
The continuous density $\rho(x,t)$ can be written in terms of a column vector, with the spatial points discretized: 
\begin{equation}
  \vrho(t) =
  \begin{pmatrix}
    \rho_0(t) \ \cdots \ \rho_{N-1}(t)
  \end{pmatrix}^T
  \in \mathbb{R}^N,
  \ 
  \rho_i(t) \approx \rho(x_i,t)\,\Delta x,
\label{eq:rho_vector}
\end{equation}
with normalisation $\sum_i \rho_i = 1$.
The differential Liouville operator $\mathcal{L}(x,t)$ described in Eq.~\eqref{eq:FP_scalar} becomes an
$N\times N$ \emph{rate matrix} $\mathbf{\hat{L}}(t)$, and the scalar
PDE~\eqref{eq:FP_scalar} becomes
\begin{equation}
\dfrac{\partial \vrho(t)}{\partial t} = \hat{\mathbf{L}}(t)\,\vrho(t).
\label{eq:matrix_FP}
\end{equation}
\textbf{Structure of the rate matrix.}
The entries of $\mathbf{\hat{L}}(t)$ are transition rates between
neighbouring grid points, obtained from the
\emph{Sasa--Tasaki} discretisation~\citep{Sasa2006-gc} that describes local transition rates between adjacent lattice sites $i$ and $i+1$ in the form:
\begin{subequations}\label{eq:rates}
\begin{align}
  L_{i+1,i}(t) &= \frac{1}{\beta \Delta x^2}\,
    e^{-\beta\Delta V_i(t)/2}, 
\end{align} 
\begin{align}
    L_{i,i+1}(t) = \frac{1}{\beta \Delta x^2}\,
    e^{+\beta\Delta V_i(t)/2},
\end{align}
\end{subequations}
where $\Delta V_i = V(x_{i+1},\zeta_t) - V(x_i,\zeta_t)$,
with reflecting boundary conditions. It is important to note that $\Lop(t)$ is not a symmetric matrix by construction, but can be made symmetric (see Section~\ref{sec:biorthogonal} for details). Its diagonal entries enforce probability conservation:
$L_{ii} = -\sum_{j\neq i} L_{ji}$,
so that every column of $\Lop$ sums to zero:
\begin{equation}
  \sum_{i=0}^{N-1} L_{ij}(t) = 0
  \quad \text{for all } j,t.
\label{eq:col_sum_zero}
\end{equation}
By the Perron--Frobenius theorem, $\Lop(t)$ has a unique
zero eigenvalue, all other eigenvalues are strictly negative,
and the zero-eigenvalue right eigenvector is exactly
$\pieq(t)$. The rate matrix $\Lop(t)$ is a highly sparse matrix $\sim \mathcal{O}(d\,N)$, in our case amounting to only $(2d+1)N$ non-zero elements. Thus, for despite much finer discretizations, the rate matrix exhibits a sparse representation. We cover this in detail in the Appendix~\ref{sec:scalability}. 
\section{Non-Adiabatic Lag Via Frame Transformations}
\label{sec:problem}
In order to quantify the non-adiabatic excitations induced by finite-time protocols, we transform the probability vector via a time-dependent gauge-like transformation,
\begin{equation}\label{eq:gauge_transform}
\tilde{\vrho}(t) = \bm{D}^{-1}(t)\,\vrho(t).   
\end{equation}
Operationally, $\bm{D}^{-1}(t)$ projects $\boldsymbol{\rho}(t) \in \mathbb{R}^{N}$ onto a spectral domain, changing the basis from a natural spatial grid basis, where components denote local site probabilities to the instantaneous (time-dependent) eigenbasis of the system's dynamics. Crucially, unlike external coordinate-parametrizations used in normalizing flows, the underlying coordinate space remains untransformed; what varies is strictly the representation of the state vector itself, allowing us to cleanly isolate non-adiabatic transitions during the time-varying external potential $V(x, \zeta_t)$. Thus, this transformation maps the system's evolution directly into the adiabatic frame of reference~\citep{Berry_2009, GueryOdelin2019}. The explicit role of $\bm{D}(t)$ as the diagonalizing operator emerges naturally when examining the dynamics in the adiabatic frame. Differentiating this transformation [Eq.~\eqref{eq:gauge_transform}] with respect to time and substituting the Liouvillian evolution equation from Eq.~\eqref{eq:matrix_FP} yields:
\begin{align}
  \dfrac{\partial \tilde{\vrho}}{\partial t}
  &= \bm{D}^{-1}\dfrac{\partial \vrho}{\partial t} + (\dfrac{\partial \bm{D}^{-1}}{\partial t})\vrho \notag\\
  &= \bm{D}^{-1}\mathbf{L}\bm{D}\,\tilde{\vrho}
     + (\dfrac{\partial \bm{D}^{-1}}{\partial t}) \bm{D}\,\tilde{\vrho} \notag\\
  &= \Lambda\,\tilde{\vrho}
     - \bm{D}^{-1} \dfrac{\partial \bm{D}}{\partial t} \,\tilde{\vrho}.
\label{eq:adiabatic_frame_deriv}
\end{align}
where the similarity transformation $\mathbf{D}^{-1}\mathbf{L}\mathbf{D} = \mathbf{\Lambda}$ maps the Liouville operator to its diagonal spectrum of relaxation rates (see Section~\ref{sec:exact_diag} for numerical corroboration). The last equality uses $\frac{\partial (\bm{D}^{-1}\bm{D})}{\partial t}=0
\Rightarrow (\frac{\partial \bm{D}^{-1}}{\partial t}) \bm{D} = -\bm{D}^{-1}(\frac{\partial \bm{D}}{\partial t})$.
The equation of motion in the adiabatic frame then reads
\begin{equation}
  \dfrac{\partial \tilde{\vrho}}{\partial t }
  = \underbrace{\Lambda(t)\,\tilde{\vrho}}_{\text{adiabatic}}
  - \underbrace{\Gamma(t)\,\tilde{\vrho}}_{\text{non-adiabatic}},
\label{eq:adiabatic_frame}
\end{equation}
where $\Lambda(t)$ is diagonal matrix in this representation, while the latter term in Eq.~\eqref{eq:adiabatic_frame} is identified as the \emph{adiabatic gauge connection} (AGC) pertaining to the Fokker-Planck equation, a term often used in the STA community~\citep{Kolodrubetz2017, GueryOdelin2019}. 
\begin{equation}
  \bm{\Gamma}(t) \equiv \bm{D}^{-1}(t)\,\dfrac{\partial \bm{D}(t)}{\partial t}.
\label{eq:Gamma}
\end{equation}

\textbf{Physical interpretation.} 
In the quasi-static limit where the parameter velocity $|\dot{\zeta}_t|$ vanishes ($\Gamma \to 0$), the spectral modes completely decouple, reducing the adiabatic frame dynamics to $\partial \tilde{\rho}_n/\partial t = \lambda_n \tilde{\rho}_n$. Here, the zero mode ($\lambda_0 = 0$) acts as an invariant manifold while all higher-order excited modes relax to zero. Under finite-time driving ($\bm{\Gamma} \neq 0$), however, the off-diagonal elements encapsulated within the AGC dynamically couple the zero mode to these transient relaxation modes. This spectral mixing constitutes the exact mathematical origin of the non-adiabatic lag: because the basis is time-dependent, the probability density cannot instantly relax to the moving equilibrium profile. Instead, the driving continuously scatters probability density into higher relaxation modes. This is formalized in Eq.~\eqref{eq:adiabatic_frame_deriv} where a clean orthogonal split of the stochastic evolution into an instantaneous equilibrium (ground-state) contribution and non-adiabatic transitions out of the instantaneous equilibrium
manifold occurs. This geometric deviation directly generates the irreversible dissipated work $\mathcal{W}_{\rm diss}$ along a trajectory and must be exactly canceled or reduced at each time instant by the engineered counterdiabatic field.

\section{Biorthogonal Decomposition of the Liouville Operator}\label{sec:biorthogonal}

The Liouville operator $\Lop(t)$, written in rate-matrix form
as in Eq.~\eqref{eq:rates}, is a real-valued matrix that is generally
\emph{non-symmetric}. The non-symmetry of $\Lop$ means that left and right
eigenvectors are distinct objects that must be treated separately.
We define the right and left eigenproblems as
\begin{align}
  \Lop(t)\,\bm{r}_n(t)
      &= \lambda_n(t)\,\bm{r}_n(t),
  \label{eq:right_eig}\\
  \bm{\ell}_n^T(t)\,\Lop(t)
      &= \lambda_n(t)\,\bm{\ell}_n^T(t),
  \label{eq:left_eig}
\end{align}
where $\mathbf{r}_n(t)\in\mathbb{R}^N$ is a column vector and
$\bm{\ell}_n^T(t)\in\mathbb{R}^{1\times N}$ is a row vector.
The eigenvalues $\{\lambda_n\}$ are real and non-positive,
\begin{equation}
  0 = \lambda_0 > \lambda_1 \geq \lambda_2
  \geq \cdots \geq \lambda_{N-1},
\end{equation}
a consequence of the Perron--Frobenius theorem applied to the
rate matrix $\Lop(t)$. As the columns of $\bm{D}(t)$ are the right eigenvectors $\mathbf{r}_n(t)$, so the transformation
$\tilde{\boldsymbol{\rho}}(t) = \bm{D}^{-1}(t)\boldsymbol{\rho}(t)$ re-expands $\vrho$ from the natural grid basis
$\{e_i\}$ (components = probability at site $i$) into the
instantaneous eigenbasis $\{\mathbf{r}_n(t)\}$ of
$\Lop(t)$ (components = amplitude of relaxation mode $n$).

\medskip
\textbf{Biorthogonality.}
Because $\Lop(t)$ is non-symmetric due to the spatial discretization, its left and right
eigenvectors form a \emph{biorthogonal} system satisfying
\begin{equation}
  \bm{\ell}_n^T \ \mathbf{r}_m
  \;=\;
  \sum_{i=1}^{N} (\ell_n)_i\,(r_m)_i
  \;=\; \delta_{nm}.
\label{eq:biorthog}
\end{equation}
Collecting the right eigenvectors as columns of the matrix
$\bm{D} = [\mathbf{r}_0\;\mathbf{r}_1\;\cdots\;\mathbf{r}_{N-1}]$,
and the left eigenvectors as rows of $\bm{D}^{-1}$, the
biorthogonality relation~\eqref{eq:biorthog} is equivalent to
$\bm{D}^{-1}\bm{D} = \mathbb{I}$.
The associated completeness relation (resolution of the identity)
reads
\begin{equation}
  \mathbb{I}
  \;=\;
  \sum_{n=0}^{N-1} \mathbf{r}_n\,\bm{\ell}_n^T
  \quad\Longleftrightarrow\quad
  \bm{D}\,\bm{D}^{-1} = \mathbb{I},
\label{eq:completeness}
\end{equation}
so that any probability vector $\boldsymbol{\rho}(t)\in\mathbb{R}^N$
admits the expansion
\begin{equation}
  \boldsymbol{\rho}(t)
  \;=\; \sum_{n=0}^{N-1} \tilde{\rho}_n(t)\,\bm{r}_n(t),
  \qquad
  \tilde{\rho}_n(t) = \bm{\ell}_n^T(t)\ \boldsymbol{\rho}(t),
\label{eq:expansion}
\end{equation}
where, this corresponds to a change of basis in the space of probability vectors, not a change of coordinates in physical space.
\medskip 

\textbf{The zero mode and its left partner.}
The zero eigenvalue $\lambda_0 = 0$ carries a distinguished physical
meaning.
Its right eigenvector is the instantaneous equilibrium distribution~\citep{PhysRevX.12.021048}:
\begin{equation}
  \mathbf{r}_0(t) = \boldsymbol{\pi}_{\rm eq}(t), 
  \qquad
  \Lop(t)\,\mathbf{r}_0(t) = \mathbf{0}.
\label{eq:zero_right}
\end{equation}
The corresponding left eigenvector satisfies
$\bm{\ell}_0^T\, \Lop = \mathbf{0}^T$. Since $\Lop$ is a rate matrix with zero column sums,
$\sum_i L_{ij} = 0$ for all $j$, the all-ones row vector
$\mathbf{1}^T = (1,1,\ldots,1)$ is annihilated from the left:
\begin{equation}
  \bm{\ell}_0^T \;\equiv\; \mathbf{1}^T,
  \qquad
  \mathbf{1}^T\,\Lop(t) = \mathbf{0}^T.
\label{eq:zero_left}
\end{equation}
Physically, $\bm{\ell}_0^T = \mathbf{1}^T$ expresses
\emph{probability conservation}: since $\mathbf{1}^T\boldsymbol{\rho} = 1$
for all normalised $\boldsymbol{\rho}$, and
$\mathbf{1}^T\,\partial \boldsymbol{\rho}/\partial t
= \mathbf{1}^T\Lop\boldsymbol{\rho} = \mathbf{0}^T\boldsymbol{\rho} = 0$,
the total probability is conserved exactly. Moreover, it is important to emphasize that the detailed-balance condition holds throughout this paper; the non-symmetric nature of the transition rate matrix $\Lop(t)$ stems entirely from the spatial discretization of the continuous system. Consequently, the dynamics can be cast into a symmetric gauge via the similarity transformation 
\begin{equation}\label{eq:symm_liouville}
    \Lop_{\rm symm}(t) = \pieq^{-1/2}(t) \ \Lop(t) \ \pieq^{1/2}(t).    
\end{equation}
This transformation reduces the complexity of the spectral problem from a biorthogonal framework to a standard orthogonal eigendecomposition, albeit evaluated with respect to a weighted inner product under the metric $\pieq^{-1}(t)$. This symmetrized version is fully equivalent to our non-symmetric version. Under such a similarity transformation, one can view it analogous to the Hamiltonian~\cite{Berry_2009,Demirplak2003} in the quantum setting, which is an Hermitian operator, unless dealing with non-Hermitian Lindbladians in open quantum systems~\citep{PhysRevA.104.062421, Vacanti_2014}. Because we choose to work directly with the raw, spatially discretized rate matrix rather than its symmetrized counterpart, we need to perform a  biorthogonal decomposition where the left and right zero modes do not coincide, i.e., $\bm{\ell}_0^T = \mathbf{1}^T \neq \boldsymbol{\pi}_{\rm eq}^T(t)$. This structural asymmetry highlights the probability-preserving nature of the unskewed classical representation. We later cover the computational advantages of working in a symmetric representation of the rate matrix [Eq.~\eqref{eq:symm_liouville}] in Section~\ref{sec:symm_rep_numerics}.
\medskip

\textbf{Probability conservation condition in the adiabatic frame.} Expanding the laboratory frame Fokker--Planck distribution $\vrho(t)$ in terms of the biorthogonal eigenbasis via the resolution of identity described in Eq.~\eqref{eq:completeness}:
\begin{equation}
  \vrho(t)
  = \sum_{n=0}^{N-1} \tilde{\rho}_n(t)\, \bm{r}_n(t),
  \qquad
  \tilde{\rho}_n(t) \equiv \bm{\ell}^T_n(t) \,\vrho(t).
\label{eq:rho_expansion}
\end{equation}
Applying the all-ones vector $\bbone^T = (1, \cdots , 1)$ to both sides and
using $\bbone^T \bm{r}_n = \delta_{n0}$
([Eq.~\eqref{eq:zero_left}] and biorthogonality):
\begin{equation}
  \bbone^T \vrho(t)
  = \sum_{n=0}^{N-1} \tilde{\rho}_n(t)\,\bbone^T \bm{r}_n
  = \sum_{n=0}^{N-1} \tilde{\rho}_n(t)\,\delta_{n0}
  = \tilde{\rho}_0(t).
\label{eq:norm_derivation}
\end{equation}
Since $\vrho(t)$ is a probability vector,
$\bbone^T \vrho(t) = \sum_i \rho_i(t) = 1$ at all times,
and therefore:
\begin{equation}
  \tilde{\rho}_0(t)
  = \bm{\ell}^T_0 \,\vrho(t)
  = \bbone^T \vrho(t)
  = 1,
  \quad \forall\, t \in [0, \tau].
\label{eq:norm_adiabatic}
\end{equation}
This is not a dynamical equation but an identity expressing probability conservation:
it holds at every instant under any dynamics, regardless of
whether $\vrho(t)$ tracks $\pieq(t)$. The zeroth component $\tilde{\rho}_0$ is pinned always to unity by probability conservation, while the components
$\tilde{\rho}_n(t)$  for $n \geq 1$ are are mode amplitudes
(not probabilities) that can take any real value and unconstrained in this frame.
In particular, 
\begin{equation}
  \sum_{n=0}^{N-1} \tilde{\rho}_n(t)
  = 1 + \sum_{n=1}^{N-1}\tilde{\rho}_n(t)
  \;\neq\; 1 .
\label{eq:sum_not_one}
\end{equation}
Thus, as derived in Eq.~\eqref{eq:sum_not_one},in the adiabatic frame, the condition $\sum_i \tilde{\rho}_i = 1$ need not hold in general. 
Infact, the continuous version of Eq.\eqref{eq:sum_not_one} and deviations from equilibrium distribution under time-dependent protocols using biorthonormal scheme was already shown in the context of the adiabatic theorem adapted to stochastic systems~\citep{Iram2021}.
\section{Engineering the Perfect CD term for the Instantaneous Equilibrium Distribution}
\label{sec:AGP}
The dynamics of $\tilde{\rho}_n(t)$ as derived in Eq.~\eqref{eq:adiabatic_frame} can be obtained by substituting the expansion~[Eq.~\eqref{eq:rho_expansion}] into
the adiabatic-frame equation [Eq.~\eqref{eq:adiabatic_frame}]
and projecting onto mode $n$ by applying $\bm{\ell}_n(t)$ from the
left:
\begin{equation}
  \dfrac{\partial \tilde{\rho}_n(t)}{\partial t}
  = \lambda_n(t)\,\tilde{\rho}_n(t)
    - \sum_{m=0}^{N-1} \Gamma_{nm}(t)\,\tilde{\rho}_m(t),
\label{eq:tilde_rho_dynamics}
\end{equation}
where $\Gamma_{nm} = [\bm{D}^{-1} (\partial \bm{D}/\partial t)]_{nm}
= \bm{\ell}^T_n(t) (\partial \bm{r}_m(t)/\partial t)$.
For $n = 0$, i.e. corresponding to the instantaneous equilibrium distribution: since the eigenvalue is $\lambda_0 = 0$ the corresponding matrix elements of the non-adiabatic contributions are $\Gamma_{0m} = \bm{\ell}^T_0(t) \partial \bm{r}_m(t)/\partial t
= \bbone^T \partial \bm{r}_m/\partial t = \partial (\bbone^T \bm{r}_m)/\partial t = 0$
for all $m$ (because $\bbone^T \bm{r}_n = \delta_{n0}$ is
time-independent), Eq.~\eqref{eq:tilde_rho_dynamics}
gives $\frac{\partial \tilde{\rho}_0}{\partial t} = 0$, consistent with
$\tilde{\rho}_0 = 1$ for all $t$.
For $n \geq 1$: the first term drives
$\tilde{\rho}_n \to 0$ at rate $|\lambda_n|$ (relaxation),
while $\Gamma_{nm}$ sources $\tilde{\rho}_n$ from
mode $m$ (non-adiabatic coupling). 
\subsection{CD driving of the instantaneous equilibrium distribution}
We seek an operator $\Lop_{\rm CD}(t)$ such that the modified dynamics
\begin{equation}
  \dfrac{\partial \bm{\rho}(t)}{\partial t} = \left[\Lop(t) + \Lop_{\rm CD}(t)\right]\bm{\rho}(t),
\label{eq:modified_FP}
\end{equation}
admits $\pieq(t)$ as an exact solution for all $t$.
Substituting $\bm{\rho} = \pieq(t)$ into Eq.~\eqref{eq:modified_FP}
and using the fact that the equilibrium distribution is always a stationary solution of the dynamics $\Lop\,\pieq = 0$, gives the \emph{counterdiabatic condition} to follow the instantaneous equilibrium distribution at all times:
\begin{equation}
   \dfrac{\partial \pieq(t)}{\partial t}  = \Lop_{\rm CD}(t)\,\pieq(t).
\label{eq:CD_condition}
\end{equation}
This is the analogue of the transitionless quantum driving approaches of~\citep{Berry2009, Demirplak2005} adapted to the Fokker--Planck evolved instantaneous equilibrium distribution. The Eq.~\eqref{eq:CD_condition} has also appeared in the context of flow-fields~\citep{Patra2017} and counterdiabatic driving currents on discrete graphs~\citep{PhysRevX.12.021048}. Moreover, one can cast Eq.~\eqref{eq:CD_condition} in the time-ordered exponential: 
\begin{equation}
    \pieq(t)=\mathcal{T} \exp\bigg[\int^{t}_{0} \Lop_{\rm CD}(s) \ ds\bigg]\pieq(0).    
\end{equation}
\medskip 

\textbf{Spectral solution via biorthogonal decomposition.}
To solve~\eqref{eq:CD_condition}, we differentiate the
stationarity condition $\Lop(t)\,\pieq(t) = 0$~\citep{doi:10.1073/pnas.071034098, PhysRevLett.100.190601} with respect to~$t$:
\begin{equation}
  \Lop(t)\,\dfrac{\partial \pieq(t)}{\partial t} = -\dfrac{\partial \Lop(t)}{\partial t}\,\pieq(t).
\label{eq:diff_stationary}
\end{equation}
Since $\sum_i \dfrac{\partial \pieq^i}{\partial t} = 0$ (normalisation is preserved),
$\dfrac{\partial \pieq}{\partial t}$ lies in the range of $\Lop$ restricted to the
non-zero modes.
Projecting~\eqref{eq:diff_stationary} onto the biorthogonal
basis via the resolution of identity~\eqref{eq:completeness} and
using $\bm{\ell}^T_n  \Lop = \lam_n \bm{\ell}^T_n$, we obtain
\begin{equation}
  \dfrac{\partial \pieq(t)}{\partial t} = -\sum_{n\neq 0}
    \frac{\bm{\ell}^T_n (t)\,(\partial \Lop /\partial t)\,  \bm{r}_0(t)}{\lambda_n(t)}\,\bm{r}_n(t).
\label{eq:dpi_spectral}
\end{equation}
The $n=0$ term vanishes because $\bbone^T\ (\partial \pieq/\partial t) = 0$. The Eq.~\eqref{eq:dpi_spectral} can also be obtained from the component form derived in Eq.~\eqref{eq:tilde_rho_dynamics}. The matrix element $(n,0)$ of $\Gamma$ gives the rate at which the zero mode (instantaneous equilibrium distribution) leaks into other modes $n$:
\begin{equation}
  \Gamma_{n0}(t)
  = [\bm{D}^{-1} \frac{\partial \bm{D}}{\partial t}]_{n0}
  = \bm{\ell}^T_n \bigg(\frac{\partial \bm{r}_0}{\partial t}\bigg)
  = -\frac{\bm{\ell}^T_n \,(\partial \Lop/\partial t)\ \bm{r}_0}{\lambda_n},
\label{eq:Gamma_n0}
\end{equation}
where the last step follows from differentiating
$\Lop \bm{r}_0 = \mathbf{0}$.
The non-adiabatic leakage is therefore proportional to the
matrix element $\bm{\ell}^T_n\, (\partial \Lop/\partial t)\, \bm{r}_0$ and inversely proportional to the spectral gap $|\lambda_n|$. The spectrum $\{\lambda_n(t)\}$ of the Fokker-Planck generator $\Lop(t)$ characterizes an ensemble of relaxation rates. These characterizing the fluctuations about the instantaneous equilibrium, $0 = \lambda_0 > \lambda_1 \ge \lambda_2 \ge \dots \ge \lambda_{N-1},$ carrying units of inverse time ($\text{s}^{-1}$). Accordingly, each instantaneous eigenmode $\bm{r}_n(t)$ decays exponentially as $e^{\lambda_n t}$, establishing $|\lambda_n|^{-1}$ as an intrinsic relaxation timescale on which the system clears fluctuations away from the instantaneous equilibrium distribution. While this relaxation spectrum becomes continuous in the thermodynamic continuum limit ($N \to \infty$)---structurally akin to the continuous Laplacian $-\partial_x^2$ on $\mathbb{R}$---it is discretized here by our finite spatial grid ($N_x$). Crucially, the slowest non-zero mode $\lambda_1(t)$ defines the spectral gap of the generator, which governs the dominant, macroscopically observable relaxation timescale: $\tau_{\text{relax}}(t) = |\lambda_1(t)|^{-1}.$

One can now construct the Liouvillian counterdiabatic operator $\Lop_{\rm CD}$ required for adiabatic tracking of Eq.~\eqref{eq:CD_condition} as a sum of rank-1 operators
that, when acting on $\pieq$, reproduces~\eqref{eq:dpi_spectral}.
Using $\bm{\ell}^T_n\, \pieq = \delta_{n0}$ (biorthogonality) and
$\bm{\ell}^T_0\, \pieq = \bbone^T \pieq = 1$, the 
\emph{minimal-rank} solution, in the resolvent form:
\begin{equation}
\Lop_{\rm CD}(t)
  = -\sum_{n\neq 0}
    \frac{\bm{\ell}^T_n\, (\partial \Lop/\partial t)\, \bm{r}_0}{\lambda_n(t)}\, \bm{r}_n \bm{\ell}^T_0.
\label{eq:ACD_spectral}
\end{equation}
Each term in~\eqref{eq:ACD_spectral} is a rank-1 matrix
$\bm{r}_n \bm{\ell}^T_0$ weighted by the matrix element
$\bm{\ell}^T_n (\partial \mathbf{L}/\partial t) \bm{r}_0 /\lam_n$ (see derivation in Section~\ref{sec:spectral_derivation}). This corrective field is a biorthogonal adiabatic gauge connection matrix of our non-symmetric Fokker--Planck generator, in exact formal analogy with the transitionless quantum driving methodologies of \citet{Berry2009, Demirplak2005}. This general form admits a substantial simplification in equilibrium, which we
derive next. As shown in Eq.~\eqref{eq:LCD-rank1}, $\mathbf{1}^T \Lop_{\rm CD} =[\mathbf{1}^T (\partial \pieq/\partial t)] \mathbf{1}^T =[\partial (\mathbf{1}^T \pieq)]\mathbf{1}^T=0.$ (or equivalently, it follows from $\bbone^T \bm{r}_n = 0, \ n\neq 0$, by biorthogonality
($\bm{\ell}^T_0 \bm{r}_n = \delta_{0n} = 0$)),
the corrected generator $\Lop_{\mathrm{eff}}= \Lop+ \Lop_{\mathrm{CD}}$ conserves probability exactly. We emphasize, however, that $\Lop_{\mathrm{eff}}$ is \emph{not} a stochastic rate matrix: $\Lop_{\mathrm{CD}}=(\partial \pieq/\partial t)\mathbf{1}^{T}$ is dense with sign-indefinite off-diagonal entries, so $ \Lop_{\mathrm{eff}}$ generally has
negative off-diagonals. It is a probability-conserving linear operator that renders $\pieq(t)$ an instantaneous solution, not the generator of a Markov jump process. 

While \citet{Iram2021} has derived the mathematical deviation from the instantaneous equilibrium distribution via the Fokker-Planck adiabatic driving, their lag removal concerns which physically accessible control realizes the counterdiabatic generator  (involving no diagonalization) in an evolutionary/biological systems setting, whereas we treat its numerical construction and stability on a discretized Fokker--Planck operator. What Eq.~\eqref{eq:ACD_spectral} adds is an
explicit \emph{construction} of that generator from the spectral data of $\Lop(t)$, rather than an implicit definition or coordinate transforms. This is the step that makes the correction computable without an analytic target, exposes the inverse spectral gaps $\lambda_n^{-1}$ that control its conditioning, and admits the systematic low-rank truncation for escorting the Fokker--Planck equation and removing dissipated work.

\subsection{The counterdiabatic operator is rank one and has a closed form}\label{sec:rank1}
The spectral correction Eq.~\eqref{eq:ACD_spectral} admits a striking simplification for the instantaneous equilibrium problem. Even though, it is written as a sum over all $N-1$ relaxation
modes, every term carries the same left factor $\ell_0^{T}=\mathbf{1}^{T}$.
Factoring it out and recognising the bracket as the spectral representation of
$\partial \pieq(t)/\partial t$ [Eq.~\eqref{eq:dpi_spectral}], the sum collapses to a single rank-one operator,
\begin{equation}
  \Lop_{\mathrm{CD}}(t)=\bigl(\partial \pieq/ \partial t\bigr)\,\mathbf{1}^{T},
  \label{eq:LCD-rank1}
\end{equation}
which satisfies the counterdiabatic condition [Eq.~\eqref{eq:CD_condition}] exactly,
$\Lop_{\mathrm{CD}}\pieq
=(\partial \pieq/\partial t)(\mathbf{1}^{T} \pieq)
=\partial \pieq/\partial t$, since $\mathbf{1}^{T}\pieq=1$.
Using $\pi_{\mathrm{eq}}\propto e^{-\beta V(x,\zeta_t)}$, the source term is
available analytically,
\begin{equation}
  \frac{\partial \pieq}{\partial t}
  =-\beta\dot\zeta_t\,\pieq \odot
   \bigg(\frac{\partial \mathbf{V}}{\partial \zeta}-\langle\frac{\partial \mathbf{V}}{\partial \zeta}\rangle_{\pieq}\bigg),
  \quad
  \langle\frac{\partial \mathbf{V}}{\partial \zeta}\rangle_{\pieq}=\frac{d  \mathcal{F}}{d \zeta},
  \label{eq:dpieq-explicit}
\end{equation}
where $\odot$ is the Hadamard product over grid points and $\mathbf{V}(\zeta_t) \in \mathbb{R}^N$ is the column vector representation of the external potential evaluated at each spatial grid point. For the equilibrium (detailed-balance) problem the exact counterdiabatic generator therefore can be computed without involving eigenvalues nor the eigenvectors of $\Lop(t)$: it is fixed at $\mathcal{O}(N)$ cost by the Boltzmann weight and its parametric derivative alone. Eq.~\eqref{eq:dpieq-explicit} is the
matrix form of the classical counterdiabatic term obtained in the continuum by~\citet{Guéry-Odelin_2023},
\begin{equation*}
    \frac{\partial \pi_{\rm eq}(x, \zeta_t)}{\partial t} =-\beta\bigg[\frac{\partial \mathcal{W}}{\partial t}-\langle \frac{\partial \mathcal{W}}{\partial t}\rangle\bigg]\pi_{\rm eq}(x, \zeta_t).
\end{equation*}
Eqs.~\eqref{eq:ACD_spectral} and~\eqref{eq:LCD-rank1} are the same operator,
not two competing constructions. Differentiating the stationarity condition
$\Lop\pieq=0$ yields the singular linear system
$\Lop\,(\partial\pieq/\partial t)=-(\partial\Lop/\partial t)\,\pieq$
[Eq.~\eqref{eq:diff_stationary}], solvable because
$\mathbf{1}^{T}(\partial\Lop/\partial t)\pieq=0$ and with a unique solution once
the zero-mass condition $\mathbf{1}^{T}(\partial\pieq/\partial t)=0$ is imposed.
The spectral expansion~\eqref{eq:dpi_spectral} solves it through the
pseudo-inverse $\Lop^{\dagger}=\sum_{n\neq0}\lambda_n^{-1}\,\bm{r}_n \bm{\ell}_n^{T}$, so the
inverse rates $\lambda_n^{-1}$ are the eigen-weights of that pseudo-inverse
rather than a property of $\Lop_{\mathrm{CD}}$. The spectral expansion and the closed form~\eqref{eq:LCD-rank1} coincide mathematically; numerically, however, they behave differently. The closed form follows from direct parametric differentiation of the Boltzmann distribution $\pieq \propto e^{-\beta V}$, using only elementary, well-scaled operations on $\pieq$—no eigendecomposition, no operator inversion, and no division by vanishing scalars—and is therefore stable at any gap. The spectral representation instead exposes the inverse relaxation rates $\lambda_n^{-1}$, the eigenvalues of the Liouvillian pseudo-inverse, which make its assembly severely ill-conditioned as the gap closes. Crucially, the closed form removes the inverse gaps only from the construction of $\Lop_{\mathrm{CD}}$; the physical spectral gap persists in the dynamics, where it continues to set the stiffness of the propagation and the relaxation rate of any numerical deviation from $\pieq$.

Thus, the spectral approach enables the following (i) makes the transitionless-driving analogy manifest (see also \citep{Iram2021} for quantum adiabatic theorem adapted to Fokker--Planck equation), (ii) exposes the spectral-gap origin of the non-adiabatic lag induced during finite-time protocols -- Eq.~\eqref{eq:ACD_spectral} (detailed in Section~\ref{sec:spectral_gap_limit}) (iii) allows systematic truncation for engineering imperfect escorting (see Section~\ref{sec:spectral_truncation_main}). 
\begin{figure*}[!tbhp]
    \centering
    \begin{subfigure}[b]{0.49\textwidth}
        \centering
        \includegraphics[width=\textwidth]{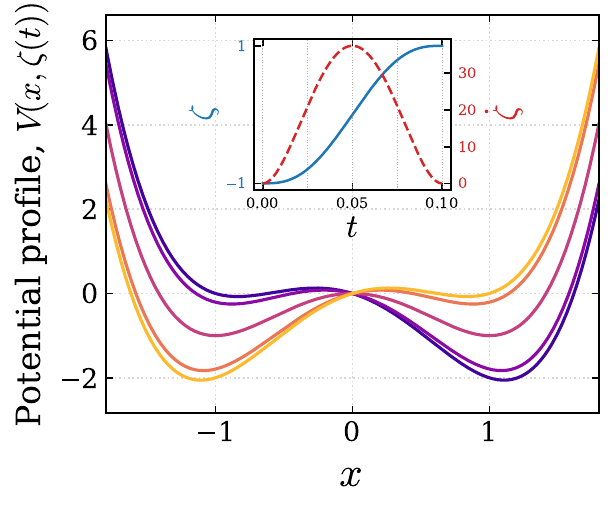}
        \caption{Time varying double-well potential.}
        \label{fig:potential_dw}
    \end{subfigure}
    \hfill
    \begin{subfigure}[b]{0.49\textwidth}
        \centering
        \includegraphics[width=\textwidth]{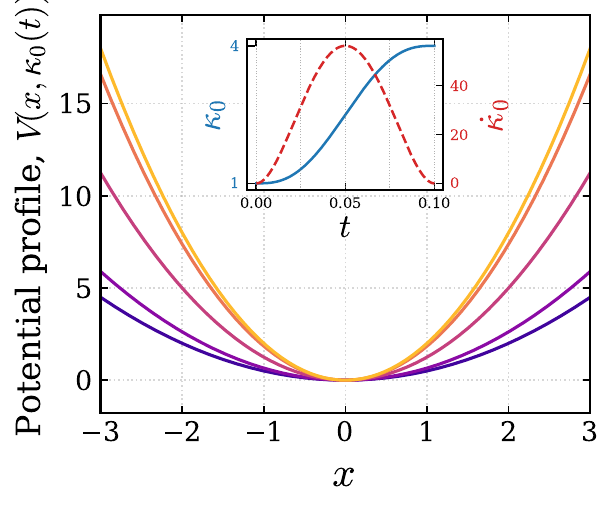}
        \caption{Time varying harmonic potential.}
        \label{fig:potential_harm}
    \end{subfigure}

    \caption{\textbf{Time-varying experimental potentials and control protocol profiles.}
  (a)~Double-well potential $V_{\rm dw}(x,\zeta_t) =
  x^4 - 2x^2 + \zeta_t x$ with tilt parameter $\zeta_t$
  driven from $\zeta_i = -1$ to $\zeta_f = +1$ over
  $t\in[0,0.1]$. The potentials start at $t=0.0$ (violet coded) and modulates to $t=0.1$ (orange coded) at the end of the protocol.
  Inset: protocol $\zeta(t)$ and driving rate
  $\dot{\zeta}(t)$.
  (b)~Harmonic trap $V_{\rm h}(x,\kappa_0(t)) =
  \tfrac{1}{2}\kappa_0(t)x^2$ with stiffness
  $\kappa_0(t) \equiv \zeta_t$ driven from $\kappa_i = 1$
  to $\kappa_f = 4$ over $t\in[0,0.1]$.
  Inset: protocol $\zeta(t)$ [$\kappa_0(t)$] and driving rate
  $\dot{\zeta}(t)$ [$\dot{\kappa}_0(t)$].
  Both protocols follow the smoothstep
  form [Eq.~\eqref{eq:protocol}], ensuring zero first and second
  derivatives at $t=0$ and $t=\tau$.}
    \label{fig:experimental_potentials}
\end{figure*}

\color{black}
\section{Numerical Setup}
\label{sec:numerical_setup}

The system~[Eq.~\eqref{eq:modified_FP}] is stiff, with fast
modes at rates proportional to $|\lambda_{N-1}|$ for a protocol
acting on a timescale $\tau$.
Explicit integrators (Euler, RK4) would require
$\Delta t < |\lambda_{N-1}|^{-1}$, making long simulations
prohibitively expensive.
We instead evolve $\boldsymbol{\rho}(t)$ via the matrix
exponential~\cite{PhysRevLett.133.057102},
\begin{equation}
  \boldsymbol{\rho}_{k+1}
  = \exp\!\bigl[\Omega(t_k,\,t_k+\Delta t)\bigr]\,
    \boldsymbol{\rho}_k,
\label{eq:expm}
\end{equation}
where $\Omega$ is the fourth-order Magnus exponent
(Appendix~\ref{sec:magnus_fourth}), evaluated at two
Gauss--Legendre nodes
$t_{1,2} = t_k + (1/2 \mp \sqrt{3}/6)\,\Delta t$.
This retains the unconditional stability of the matrix
exponential while achieving $\mathcal{O}(\Delta t^4)$
convergence. 

Both potentials are driven by the smooth protocol
\begin{equation}
  \zeta(t) = \zeta_i + (\zeta_f - \zeta_i)\,
  s^3(6s^2 - 15s + 10),
  \quad
  s = t/\tau,
\label{eq:protocol}
\end{equation}
whose first and second derivatives vanish at $t = 0$ and
$t = \tau$.
We consider two model systems:
a double-well (DW) potential as shown in Fig.~\ref{fig:potential_dw}, with  
$V_{\rm dw}(x,\zeta_t) = x^4 - 2x^2 + \zeta_t\, x$
with tilt parameter driven from $\zeta_i = -1$ to
$\zeta_f = +1$, discretised on $N = 80$ grid points over
$x \in [-2.5,\,2.5]$ 
($\Delta x \approx 0.063$); and a harmonic trap
$V_{\rm h}(x,\kappa_0) = \tfrac{1}{2}\kappa_0(t)\,x^2$
with stiffness $\kappa_0(t) \equiv \zeta_t$ driven from
$\kappa_i = 1$ to $\kappa_f = 4$, discretized on $N = 80$
grid points over $x \in [-4.0,\,4.0]$
($\Delta x \approx 0.101$).
Both systems are evolved at inverse temperature $\beta = 1$. 
\medskip 

As already described in Section~\ref{subsec:pde_to_ode}, we employ the Sasa--Tasaki
discretisation~\cite{Sasa2006-gc,
sasa2005steadystatethermodynamics} to construct the rate matrix (for both the bare and CD-evolved rate matrix). To minimize temporal discretization errors across the wide range of protocol durations $\tau \in [10^{-3}, 10]$, the integration time-step is dynamically scaled between $\Delta t = 10^{-7}$ and $\Delta t = 10^{-4}$. By choosing smaller time-steps for shorter protocol durations, we ensure high temporal resolution in the strongly driven regime, where the parameter velocity $\vert{}\dot{\zeta}_t\vert{}$ is large, thereby maintaining a uniformly high numerical tracking fidelity across all driving speeds. Additionally, in the numerics, the right zero eigenvector $\bm{r}_0(t)$ is normalized to unit sum, $\sum_i (r_0)_i = 1$, uniquely identifying it
with the instantaneous Boltzmann distribution
$\pi_{\rm eq}(x_i, t) \propto e^{-\beta V(x_i,\zeta_t)}$.
\begin{figure*}[!tbhp]
    \centering
    \begin{subfigure}{\linewidth}
        \centering
        \includegraphics[width=\linewidth]{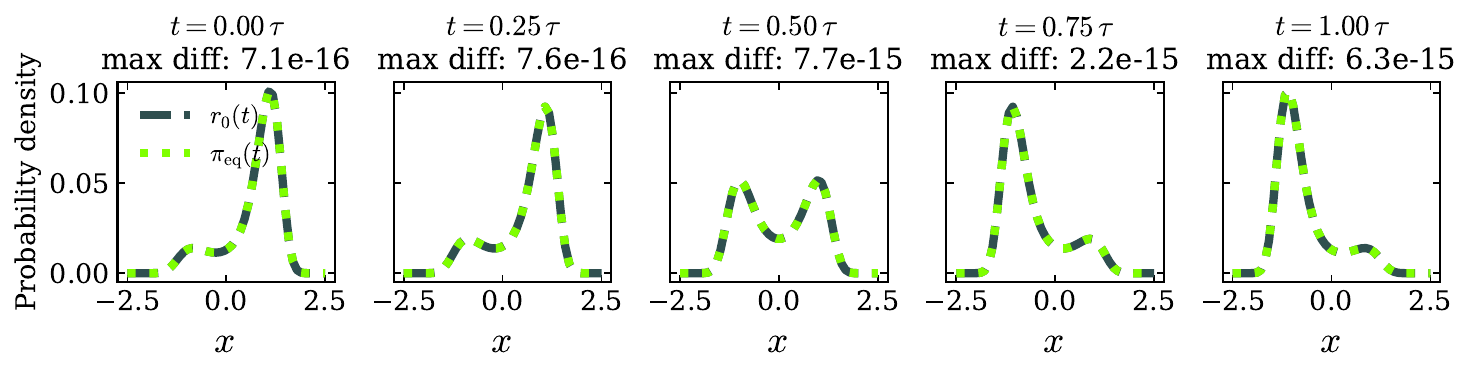}
        \caption{Double-well potential zero-mode verification.}
        \label{fig:r0_agreement_dw}
    \end{subfigure}
    
    \vspace{0.1cm} 
    
    \begin{subfigure}{\linewidth}
        \centering
        \includegraphics[width=\linewidth]{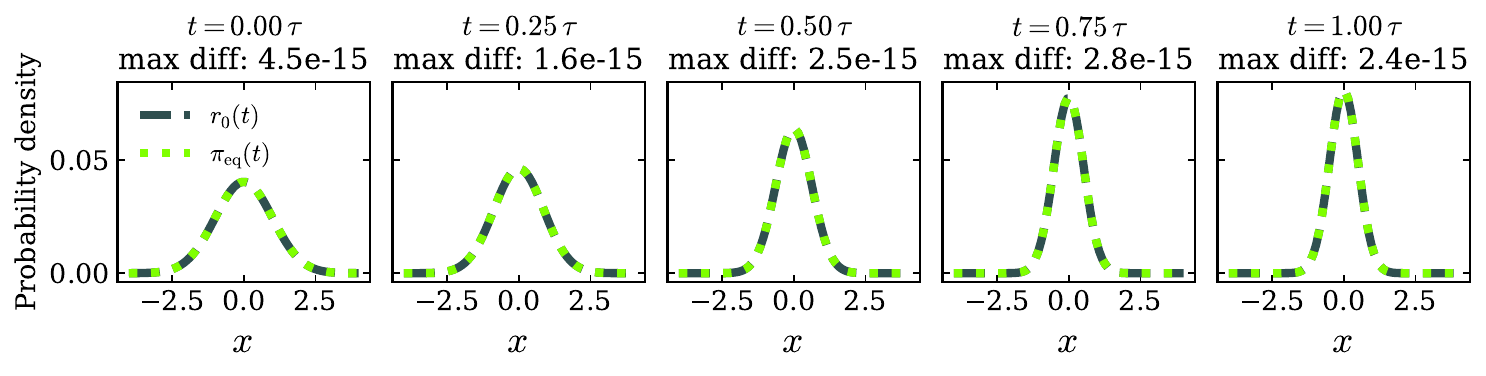}
        \caption{Harmonic trap potential zero-mode verification.}
        \label{fig:r0_agreement_harm}
    \end{subfigure}

    \caption{\textbf{Spectral verification of the discrete stationary zero-mode.} 
    The numerically diagonalized right eigenvector $\bm{r}_0(t)$ (dashed grey profile) corresponding to the zero eigenvalue is coincides upto machine precision with the analytical instantaneous Boltzmann distribution $\pi_{\mathrm{eq}}(t)$ (dotted green profile).}
    \label{fig:eq_r0_agreement_combined}
\end{figure*}
\subsection{Computation of the dissipated work during evolution}
\label{sec:wdiss}
The thermodynamic work performed on the system during a finite-time protocol $\zeta_t$ is defined in our spatially discretized setup as the time-integral of the instantaneous power:
\begin{align}\label{eq:work_def}
\mathcal{W}(t) &= \int_0^t \left[ \frac{\partial \mathbf{V}(\zeta_{s})}{\partial s} \right]^T \boldsymbol{\rho}(s)  ds, \nonumber \\ 
&= \int_0^t \sum_{i=1}^N \frac{\partial V(x_i, \zeta_{s})}{\partial s} \rho_i(s)  ds,
\end{align}

where, $\mathbf{V}_i(\zeta_t) = V(x_i, \zeta_t)$. The average is taken over the time-evolved distribution $\boldsymbol{\rho}(t')$ at each instant. Depending on the context, $\boldsymbol{\rho}(t')$ corresponds to either the bare distribution [Eq.~\eqref{eq:matrix_FP}] or the counterdiabatic quantity evolving under the escorted generator [Eq.~\eqref{eq:modified_FP}]. In the latter case, $\mathcal{W}(t)$ represents the total escorted work. The excess dissipated work is subsequently quantified as
\begin{equation}\label{eq:wdiss_def}
\mathcal{W}_{\rm diss}(t)
= \mathcal{W}(t) - \Delta \mathcal{F}(\zeta_t),
\end{equation}
where $\Delta \mathcal{F}(\zeta_t) = \mathcal{F}(\zeta_t) - \mathcal{F}(\zeta_0)$. For a perfect counterdiabatic driving, $\boldsymbol{\rho}(t') = \pieq(t')$, i.e., the excess dissipation vanishes identically throughout the protocol, since in
this case
$\langle \frac{\partial V}{\partial t}\rangle_{\boldsymbol{\rho}}
= \langle \frac{\partial V}{\partial t}\rangle_{\boldsymbol{\pi}_{\rm eq}}
= d F/d t$
at each instant. For numerically integrating Eq.~\eqref{eq:work_def}, we use the Simpson's method, which is accurate to $\mathcal{O}(\Delta t^4)$ as detailed in Section~\ref{sec:simpson} and consistent with the fourth-order Magnus-based evolution (see Section~\ref{sec:magnus_fourth}).

\section{Main Experimental Results}\label{sec:numerics_results}
To demonstrate the efficacy of the proposed Liouvillian counterdiabatic driving [Eq.\eqref{eq:modified_FP}] framework, we study an overdamped Brownian particle in one dimension subjected to the two aforementioned driving fields: a time-varying double-well potential and a modulated harmonic potential (see Section~\ref{sec:numerical_setup} and Fig.~\ref{fig:experimental_potentials}). For both landscapes, the control parameter is swept across a wide range of protocol durations $\tau \in \{0.001, 0.01, 0.1, 0.25, 0.5, 0.75, 1.0, 2.0, 4.0, 6.0, 10.0\}$, allowing us to evaluate the methodology from the highly non-equilibrium, fast-switching regime ($\tau = 10^{-3}$) to the near-adiabatic limit ($\tau = 10.0$). We systematically quantify the performance of our spectral framework against three rigorous operational criteria: (i) \emph{Zero-mode tracking fidelity}—verifying that the right zero-mode eigenvector $\bm{r}_0(t)$ coincides with the instantaneous equilibrium distribution $\pieq(t)$ up to machine precision at all times; (ii) \emph{Elimination of non-adiabatic lag}—evaluating the ability of the driven dynamics to precisely track $\pieq(t)$ at each time instant, irrespective of the protocol speed $\vert{}\dot{\zeta}_t\vert{}$; (iii) and \emph{Suppression of dissipation}—confirming the vanishing of the excess dissipated work at every instant during the entire Fokker--Planck evolution, $\mathcal{W}_{\rm diss}(\zeta_t) \to 0$ as achieved by the flow field in EJE (see [Eqs.~(\ref{eq:eje}, \ref{eq:modified_work})] bounded only by numerical discretization errors of the matrix-valued evolution.
\begin{figure*}[t]
    \centering
    \begin{subfigure}{0.49\linewidth}
        \centering
        \includegraphics[width=\linewidth]{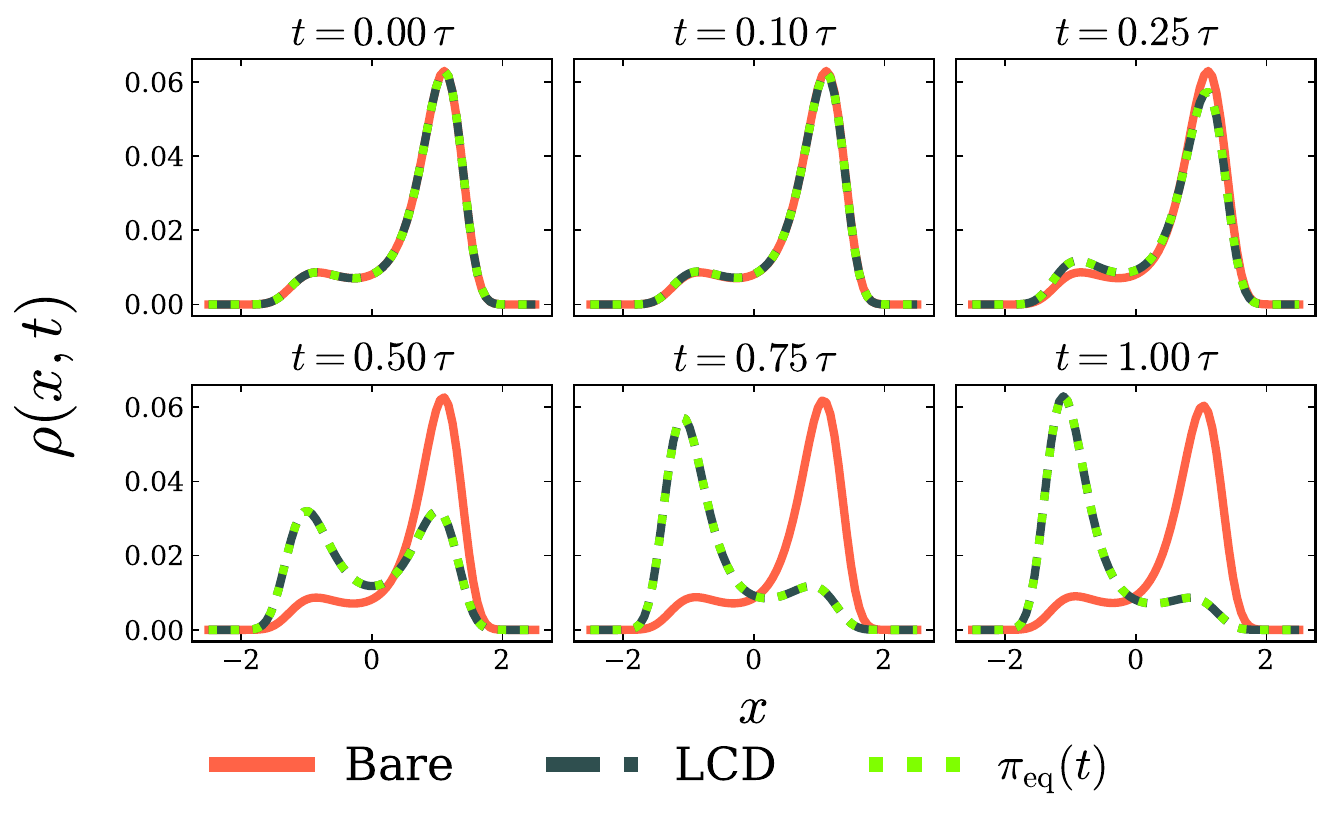}
        \caption{FP distribution evolution under DW potential.}
        \label{fig:snapshots_dw}
    \end{subfigure}
    \vspace{0.1cm} 
    \begin{subfigure}{0.49\linewidth}
        \centering
        \includegraphics[width=\linewidth]{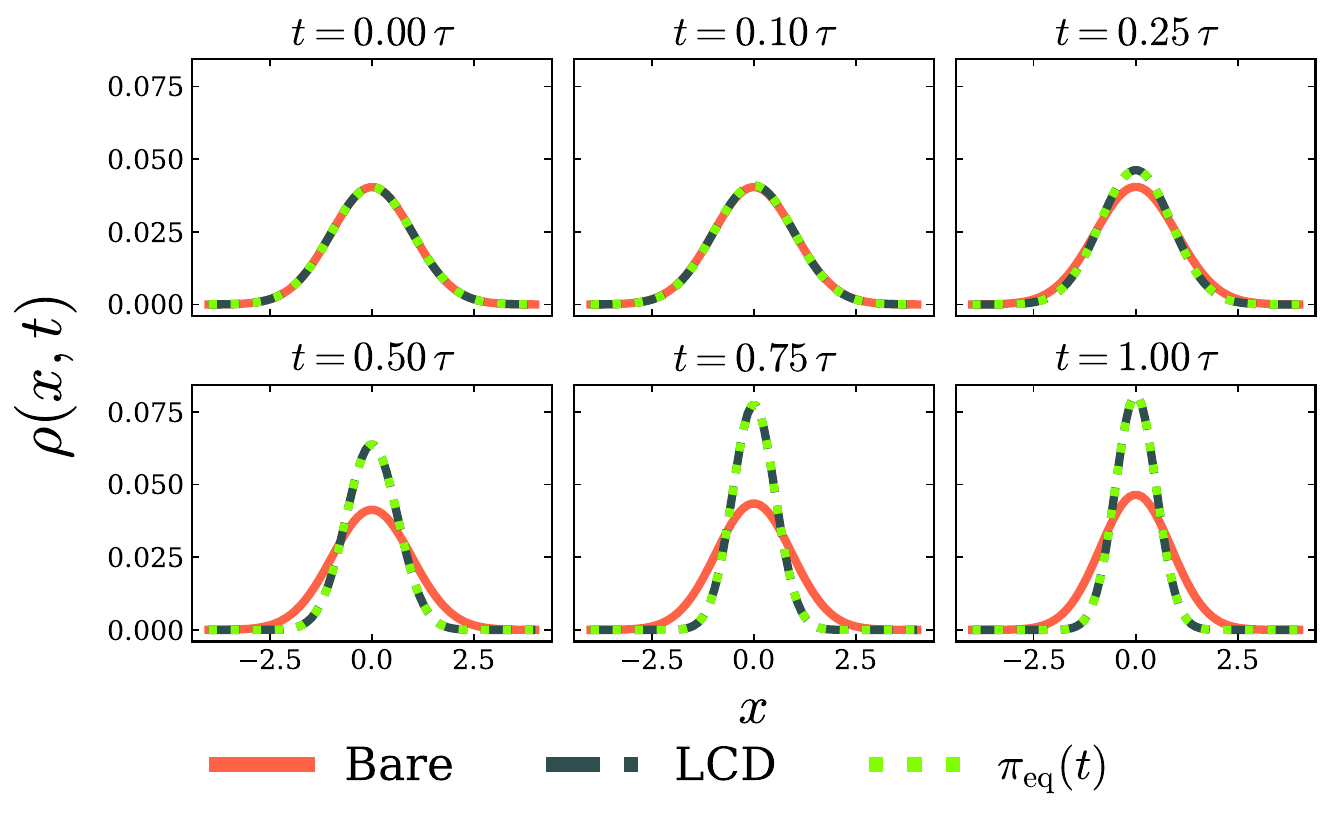}
       \caption{FP distribution evolution under harmonic potential.}
        \label{fig:snapshots_harm}
    \end{subfigure}

    \caption{\textbf{Evolutionary snapshots of the probability density distributions.} 
    Spatiotemporal density profiles $\rho(x,t)$ comparing the bare driving dynamics (solid orange line) and our spectral-based Liouvillian CD driving (LCD) (grey) against the instantaneous target Boltzmann distribution $\pi_{\mathrm{eq}}(x, t)$ (dashed green) across selected temporal fractions of the protocol for (a) the double-well landscape and (b) the harmonic confinement trap.}
    \label{fig:snapshots_combined}
\end{figure*}
Additionally, we systematically investigate a spectrally truncated approximation of the counterdiabatic correction in Eq.~\eqref{eq:ACD_spectral} and also identify specific class of potentials where our spectral methodology breaks down.

\subsection{Zero-mode alignment with instantaneous equilibrium}
As established in Section~\ref{sec:biorthogonal}, a fundamental requirement of the matrix-valued Liouvillian framework is that the right zero-mode eigenvector identically mirrors the instantaneous equilibrium distribution. This condition is robustly corroborated by our numerical experiments: $\bm{r}_0(t)$ coincides with the instantaneous equilibrium distribution $\bm{\pi}_{\mathrm{eq}}(t)$ at all times, irrespective of the driving rate or potential landscape. To verify the algebraic definition in Eq.~\eqref{eq:zero_right}, we evaluate this spectral baseline across uniform temporal snapshots, $t/\tau \in \{0, 0.25, 0.5, 0.75, 1.0\}$, for both the double-well -- Figure~\ref{fig:r0_agreement_dw} and harmonic -- Figure~\ref{fig:r0_agreement_harm} landscapes. In all cases, the states exhibit absolute structural convergence down to machine precision. The global coordinate-wise deviation remains strictly bounded by $\max_t \Vert \bm{r}_0(t) - \bm{\pi}_{\mathrm{eq}}(t) \Vert_\infty = \mathcal{O}(10^{-15})$, confirming that the discrete zero-mode manifold ($\lambda_0=0$) precisely maps to the instantaneous equilibrium target under arbitrary driving profiles.

\subsection{Metrics and Tracking Fidelity}\label{sec:metrics_and_fidelity}
Propagating the Fokker-Planck density under the effective counterdiabatic (CD) Liouvillian $\Lop_{\rm eff}(t) = \Lop(t) + \Lop_{\rm CD}(t)$ [Eq.~\eqref{eq:modified_FP}] is engineered to eliminate the non-adiabatic lag induced by finite-time driving. The spectral correction term $
\Lop_{\rm CD}(t) = -\dot{\zeta_t}\sum_{n\neq 0} \frac{\bm{\ell}^T_n (\partial \Lop(t)/\partial \zeta) \bm{r}_0}{\lambda_n(t)}\, \bm{r}_n \bm{\ell}^T_0$, 
perfectly escorts the state vector along the instantaneous equilibrium manifold. To quantify the fidelity of this tracking shortcut, we evaluate the total variation distance (TVD),
\begin{equation}
\tvd\big(\boldsymbol{\rho}(t), \boldsymbol{\pi}_{\rm eq}(t)\big) = \tfrac{1}{2}\|\boldsymbol{\rho}(t)-\boldsymbol{\pi}_{\rm eq}(t)\|_1,
\end{equation}
and the Kullback-Leibler (KL) divergence, $\mathcal{D}_{\rm KL}\big(\boldsymbol{\rho}(t)\|\pieq(t)\big)$, between the time-evolved density and the instantaneous target.
\medskip 

\begin{figure*}[!tbhp]
    \centering
    \begin{subfigure}[t]{0.49\textwidth}
        \centering
        \includegraphics[width=\textwidth]{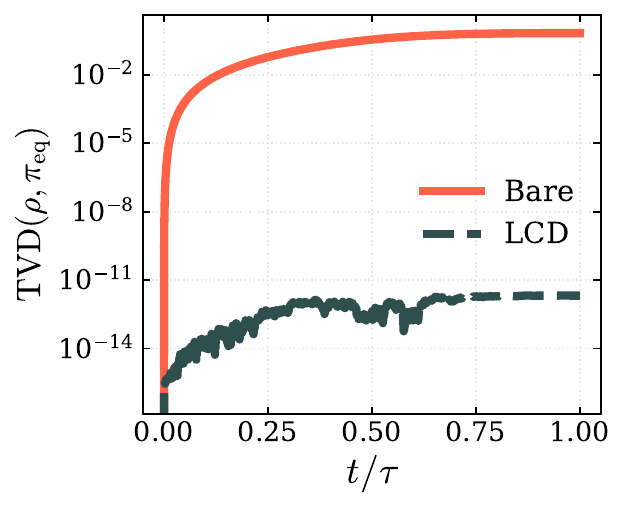}
        \caption{TVD against normalized time evolved under DW potential.}
        \label{fig:dw_tvd}
    \end{subfigure}
    \hfill
    \begin{subfigure}[t]{0.49\textwidth}
        \centering
        \includegraphics[width=\textwidth]{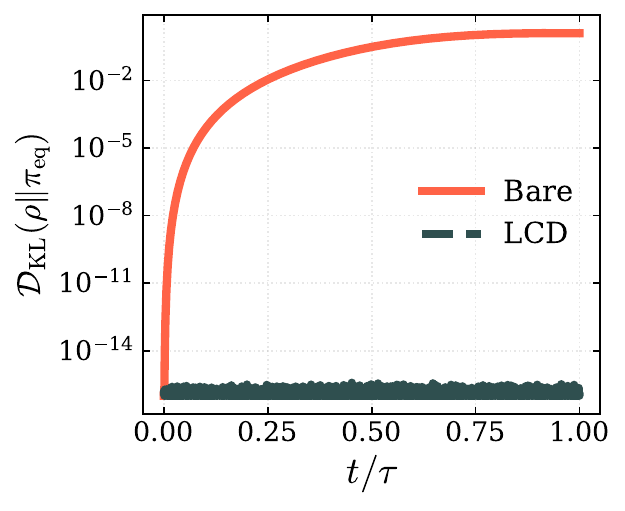}
        \caption{KL divergence against normalized time evolved under DW potential.}
        \label{fig:dw_kl}
    \end{subfigure}
    
    
    \begin{subfigure}[b]{0.49\textwidth}
        \centering
        \includegraphics[width=\textwidth]{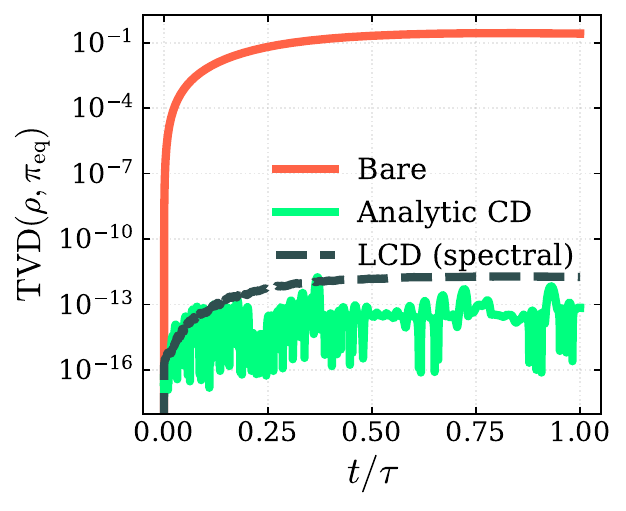}
        \caption{TVD against normalized time evolved under an harmonic trap potential.}
        \label{fig:harm_tvd}
    \end{subfigure}
    \hfill
    \begin{subfigure}[b]{0.49\textwidth}
        \centering
        \includegraphics[width=\textwidth]{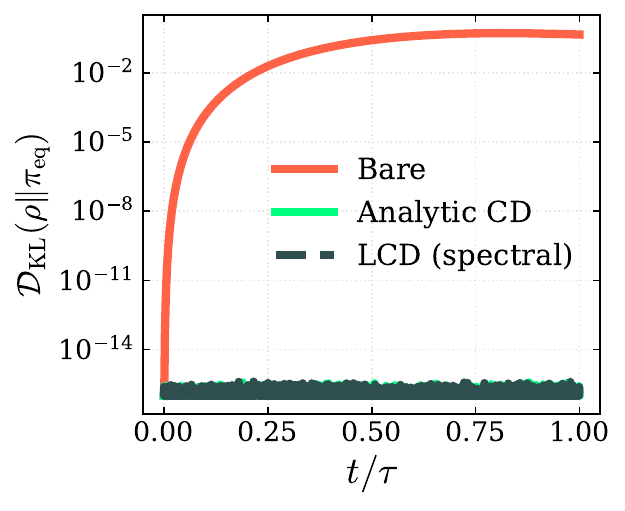}
        \caption{KL divergence against normalized time evolved under an harmonic trap potential.}
        \label{fig:harm_kl}
    \end{subfigure}
    \caption{\textbf{Informational metrics comparison across potentials.} Total variational distance (TVD) and KL divergence between the time-evolved FP density and the instantaneous equilibrium distribution.  
    Top row (a, b) shows results of the TVD and KL when evolved under a DW potential respectively, whereas the bottom row (c, d) details the same for the harmonic potential. Bare (uncontrolled) evolution (solid orange line), LCD (dashed grey line) and if available, Analytic CD (solid green line).}
    \label{fig:metric_profiles}
\end{figure*}

\emph{Double-well potential.}---The tracking metrics across different protocol durations are compiled in Table~\ref{tab:metrics}. Compared to the uncontrolled bare evolution [Eq.~\eqref{eq:matrix_FP}], the CD framework yields an overwhelming improvement: the maximum TVD remains strictly bounded within $\mathcal{O}(10^{-12})$--$\mathcal{O}(10^{-11})$, while the corresponding KL divergence sits at the machine-precision floor of $\mathcal{O}(10^{-16})$. The time-resolved profiles of these metrics are shown in Figs.~\ref{fig:dw_tvd} and \ref{fig:dw_kl} for a representative protocol ($\tau = 0.1$), demonstrating continuous, uniform lag suppression.

\begin{table*}[!tbhp]
\caption{
  Maximum KL divergence and maximum TVD for bare FP and LCD driven FP at several protocol
  durations $\tau$ for both the double well and harmonic trap potentials.
}
\label{tab:metrics}
\begin{ruledtabular}
\begin{tabular}{l cccc c cccc}
 & \multicolumn{4}{c}{\textbf{Double Well}} & & \multicolumn{4}{c}{\textbf{Harmonic Trap}} \\
\cline{2-5} \cline{7-10}
\rule{0pt}{3ex}
Protocol duration ($\tau$) & $\mathcal{D}_\text{KL}^{\rm bare}$ & $\mathcal{D}_\text{KL}^{\rm LCD}$ & $\tvd^{\mathrm{bare}}$ & $\tvd^{\mathrm{LCD}}$ & & $\mathcal{D}_\text{KL}^{\rm bare}$ & $\mathcal{D}_\text{KL}^{\rm LCD}$ & $\tvd^{\mathrm{bare}}$ & $\tvd^{\mathrm{LCD}}$ \\
\hline
\rule{0pt}{3ex}
 $10^{-3}$($\Delta t=10^{-7}$) &  1.40 & $4.68\times10^{-16}$ & 0.68 & $3.69\times10^{-12}$ & &  0.80 & $2.25\times10^{-16}$ & 0.32 & $2.3\times10^{-12}$ \\
 0.01 ($\Delta t=10^{-6}$) &  1.39 & $3.63\times10^{-16}$ & 0.67 & $1.71\times10^{-12}$ & &  0.76 & $4.93\times10^{-16}$ & 0.32 & $2.2\times10^{-12}$ \\
 0.10 ($\Delta t=10^{-5}$) &  1.27 & $3.88\times10^{-16}$ & 0.66 & $2.14\times10^{-12}$ & &  0.52 & $4.26\times10^{-16}$ & 0.27 & $1.95\times10^{-12}$ \\
 0.25 ($\Delta t=10^{-5}$)  &  1.13 & $3.98\times10^{-16}$ & 0.64 & $1.20\times10^{-12}$ & &  0.32 & $4.33\times10^{-16}$ & 0.23 & $1.61\times10^{-12}$ \\
 0.50 ($\Delta t=10^{-5}$)  &  0.97 & $4.13\times10^{-16}$ & 0.60 & $1.08\times10^{-12}$ & &  0.19 & $4.56\times10^{-16}$ & 0.18 & $1.30\times10^{-12}$ \\
 0.75 ($\Delta t=10^{-5}$) &  0.85 & $4.26\times10^{-16}$ & 0.57 & $9.74\times10^{-13}$ & &  0.12 & $4.28\times10^{-16}$ & 0.15 & $1.01\times10^{-12}$ \\
 1.00 ($\Delta t=10^{-5}$) &  0.75  & $4.25\times10^{-16}$ & 0.53 & $8.76\times10^{-13}$ & &  0.09  & $4.60\times10^{-16}$ & 0.13 & $9.34\times10^{-13}$ \\
 2.00 ($\Delta t=10^{-5}$) & 0.48 & $4.76\times10^{-16}$ & 0.44 & $1.01\times10^{-12}$ & & 0.03 & $4.76\times10^{-16}$ & 0.09 & $5.88\times10^{-13}$ \\
 4.00 ($\Delta t=10^{-4}$) & 0.25 & $4.19\times10^{-16}$ & 0.32 & $1.28\times10^{-12}$ & & 0.01 & $4.21\times10^{-16}$ & 0.05 & $4.05\times10^{-13}$ \\
 6.00 ($\Delta t=10^{-4}$) & 0.15 & $4.36\times10^{-16}$ & 0.25 & $7.14\times10^{-13}$ & & 0.006 & $4.70\times10^{-16}$ & 0.04 & $2.61\times 10^{-13}$ \\
 10.0 ($\Delta t=10^{-4}$) & 0.07 & $ 4.27\times10^{-16}$ & 0.18 & $5.97\times10^{-13}$ & & 0.002 & $ 4.85\times10^{-16}$ & 0.02 & $1.78\times10^{-13}$ \\
\end{tabular}
\end{ruledtabular}
\end{table*}

\emph{Harmonic potential.}---As shown in Table~\ref{tab:metrics}, the numerical CD framework exhibits equally robust performance under a driven harmonic potential. To establish a rigorous benchmark, we compare our numerical results against the exact analytical CD driving scheme derived in Section~\ref{sec:analytic_evol}. The numerical framework recovers the analytical target up to $10$--$12$ orders of magnitude over the bare dynamics. The minor discrepancy between the numerical and exact analytical tracking metrics represents minor time-stepping and discretization artifacts arising from propagating the full matrix-valued generator rather than scalar-valued evolution used for the analytic CD simulations. Crucially, the KL divergence remains pinned at $\mathcal{O}(10^{-16})$, matching the exact analytical solution to within numerical noise. This near-perfect overlap and the systematic elimination of the non-adiabatic lag are visually confirmed by the temporal evolution of the metrics for $\tau = 0.1$ in Figs.~\ref{fig:harm_tvd} and \ref{fig:harm_kl}, as well as the spatial snapshots provided in Fig.~\ref{fig:snapshots_harm}. 

We note that, KL is dominated by the far tails where $\pieq$ is exponentially small, so it saturates near machine precision, whereas TVD reflects the bulk error set by the quadrature/construction.

\begin{figure*}[!tbhp]
\centering
\begin{subfigure}[t]{0.49\textwidth}
        \centering
        \includegraphics[width=\textwidth]{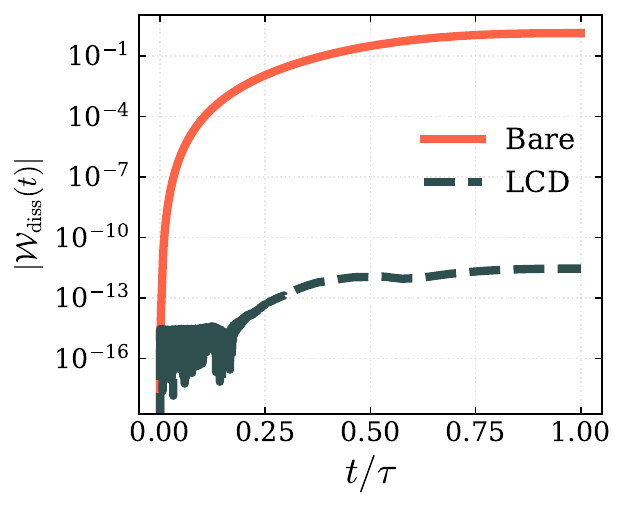}
        \caption{Dissipated work for overdamped FPE evolved under DW potential for $\tau=0.1$}
        \label{fig:dw_wdiss}
    \end{subfigure}
    \begin{subfigure}[t]{0.49\textwidth}
        \centering
        \includegraphics[width=\textwidth]{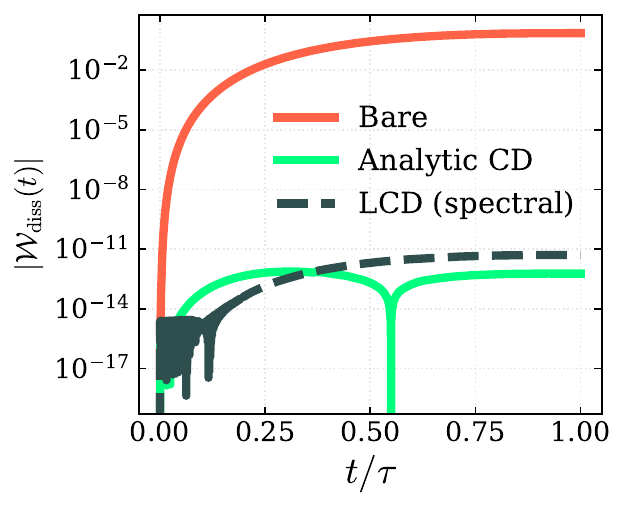}
        \caption{Dissipated work for overdamped FPE evolved under harmonic potential for $\tau=0.1$}
        \label{fig:harm_wdiss}
    \end{subfigure}
\caption{\textbf{Dissipation during Fokker--Planck time-evolution.} Dissipated work magnitude plotted against normalized time $t/\tau$. Bare (uncontrolled) evolution (solid orange line), LCD (dashed grey line) and if available, Analytic CD (solid green line).}
\end{figure*}

\subsection{Trajectory-Resolved Dissipated Work}\label{sec:traj_dissipated}

Beyond the terminal values, a complete assessment of the escort requires the
ensemble-averaged dissipated work resolved continuously along the driving path,
$\mathcal{W}_{\rm diss}(t) = \mathcal{W}(t) - \Delta\mathcal{F}(t)$. The
thermodynamic and distributional pictures are tied by a single identity: because
$\langle \partial V/\partial t\rangle_{\pi_{\rm eq}} = d\mathcal{F}/dt$ at every instant
[Section~\ref{sec:wdiss}], perfect tracking of the instantaneous equilibrium,
$\boldsymbol{\rho}(t) = \boldsymbol{\pi}_{\rm eq}(t)$, is \emph{equivalent} to
vanishing dissipation, $\mathcal{W}_{\rm diss}(t) = 0$. The trajectory-resolved
$\mathcal{W}_{\rm diss}(t)$ is therefore the thermodynamic corollary of the
distributional tracking established in Section~\ref{sec:wdiss}; reporting it additionally confirms that the escorting equality holds for the work functional.  While the LCD isolates the system from non-adiabatic excitations, ensuring that the dissipated work vanishes, this tracking is bought at the expense of a transient energetic cost. The counterdibatic  $\hat{\mathbf{L}}_{\text{CD}}(t)$ must perform work along the protocol path to suppress mode excitations, a thermodynamic cost of this shortcut that scales aggressively as the protocol duration $\tau$ approaches the limits of the system's spectral gap (see for instance \citep{PhysRevLett.118.100601}). Computing this excess dissipated work during the Fokker--Planck evolution is detailed in Section~\ref{sec:wdiss}.
\medskip

\emph{Double-well potential.}---The dissipation profile for a rapid switch ($\tau = 0.1$) is shown in Fig.~\ref{fig:dw_wdiss}. In the bare evolution the
accumulated dissipated work climbs steadily as the landscape is modulated; under
the LCD it is suppressed continuously. As reported in Table~\ref{tab:work}, the
maximal transient deviation $\max_t |\mathcal{W}_{\rm diss}(t)|$ remains at the
$\mathcal{O}(10^{-12})$ floor set by the fourth-order work quadrature and Magnus
evolution [Section~\ref{sec:wdiss}]---roughly four orders of magnitude above the
double-precision floor---confirming that the lag is canceled continuously rather
than only at the protocol endpoints.

\begin{figure*}[!t]
    \centering
    \begin{subfigure}{\linewidth}
        \centering
        \includegraphics[width=\linewidth]{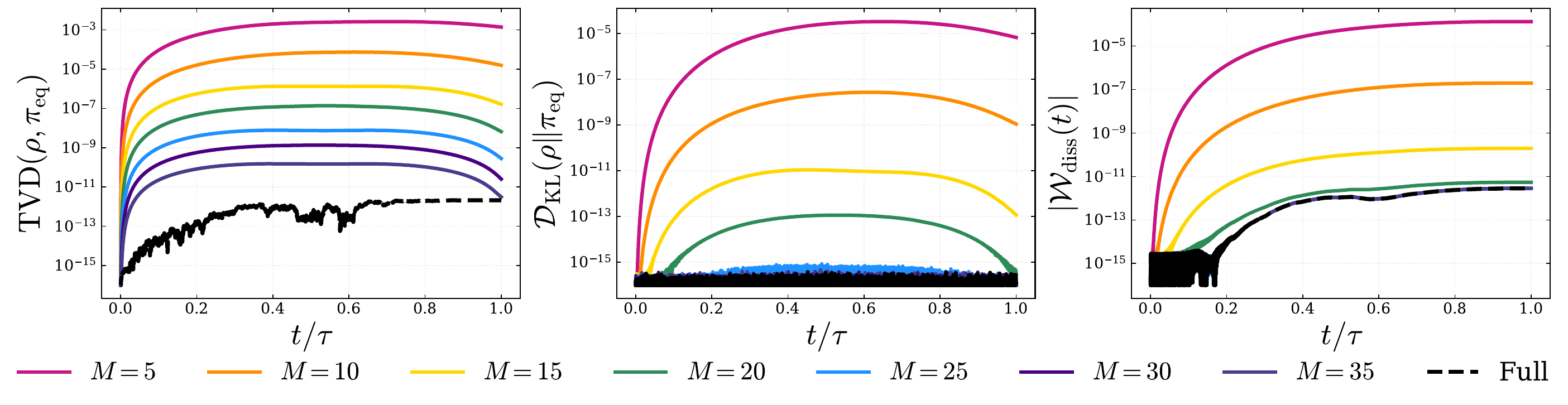}
        \caption{Truncated LCD performance for DW potential.}
        \label{fig:dw_trunc}
    \end{subfigure}
    
    \vspace{0.2cm} 
    
    \begin{subfigure}{\linewidth}
        \centering
        \includegraphics[width=\linewidth]{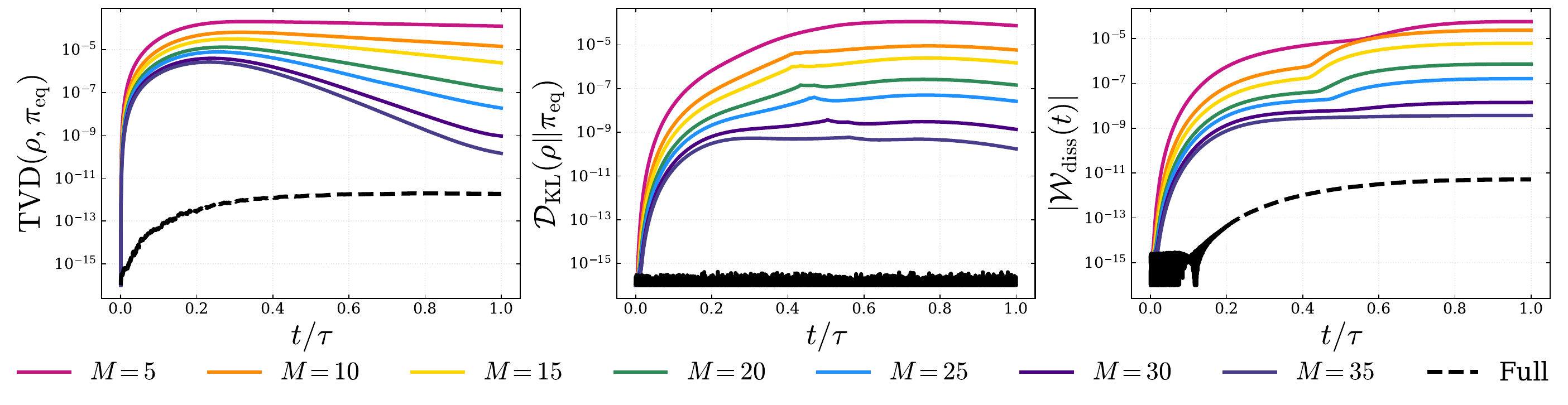}
        \caption{Truncated LCD performance for harmonic potential.}
        \label{fig:harm_trunc}
    \end{subfigure}
\caption{Convergence metrics of the spectrally truncated counterdiabatic expansion as a function of the active mode cutoff $M$. (a) Evolution under the double-well potential and (b) the driven harmonic trap. For both physical landscapes, the tracking fidelity—quantified via the total variation distance (TVD), Kullback-Leibler divergence ($\mathcal{D}_{\rm KL}$), and transient dissipated work $\mathcal{W}_{\rm diss}(t)$—exhibits rapid exponential convergence toward the machine-precision floor as $M$ approaches the full basis limit ($N=80$). This uniform decay demonstrates that high-fidelity non-adiabatic lag suppression is preserved even under substantial modal reduction, rendering large-scale system tracking computationally viable.}
\label{fig:truncation_convergence}
\end{figure*}
\begin{figure}[!tbhp]
\centering
\includegraphics[width=0.49\textwidth]{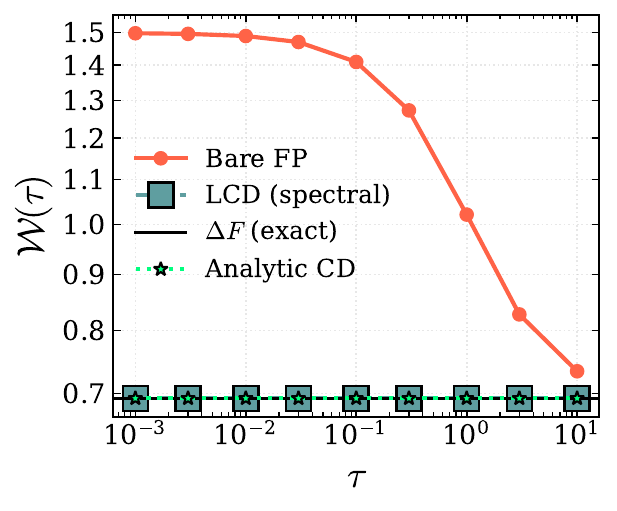}
\caption{ \textbf{Work versus protocol duration for the harmonic trap.}
  Total work $\mathcal{W}(\tau)$ as a function of protocol duration
  $\tau$ for the bare (red, solid) and LCD (blue, square)
  evolutions, compared with the analytic CD (green, star).
  The horizontal line marks the exact free energy difference
  $\Delta\mathcal{F} = \tfrac{1}{2}\ln 4 \approx 0.6931$.
  Under bare dynamics, $\mathcal{W}(\tau) > \Delta\mathcal{F}$ for all
  finite $\tau$, approaching $\Delta\mathcal{F}$ only in the
  quasi-static limit.
  Both the spectral LCD and the analytic CD yield
  $\mathcal{W}^{\rm LCD}(\tau) = \Delta\mathcal{F}$ to within
  $\sim\!\mathcal{O}(10^{-12})$ across all protocol durations
  tested, confirming that the escorting condition
  $\mathcal{W}_{\rm diss}(\tau) \approx 0$ holds irrespective of
  driving speed.}
\label{fig:delta_f_sweep}
\end{figure}
\medskip 

\emph{Harmonic potential.}---The cancellation is further benchmarked against the
exact analytic CD solution for the driven harmonic trap
(Section~\ref{sec:analytic_evol}). Figure~\ref{fig:harm_wdiss} shows
$|\mathcal{W}_{\rm diss}(t)|$ at $\tau = 0.1$ for the bare, analytic-CD, and
spectral-LCD protocols; broader metrics are collected in Table~\ref{tab:work}.
The bare dynamics accumulates substantial dissipation, whereas both the analytic
and spectral CD drives maintain $|\mathcal{W}_{\rm diss}(t)| \lesssim 10^{-12}$ at
every instant, so the spectral construction reproduces the continuum shortcut to
within the time-discretisation accuracy of the Simpson integral. The equilibrium
free-energy difference for this system is
$\Delta\mathcal{F} = \tfrac{1}{2}\ln(\kappa_f/\kappa_i) = \ln 2 \approx 0.6931$,
independent of the protocol duration $\tau$ (derivation in
Section~\ref{sec:free_energy_estimates}).

Figure~\ref{fig:delta_f_sweep} shows the total work $\mathcal{W}(\tau)$ across
protocol durations. Under LCD, $\mathcal{W}^{\rm LCD}(\tau) = \Delta\mathcal{F}$ to
within $\sim 10^{-12}$ for all $\tau \in [10^{-3},\,10]$
(Table~\ref{tab:work_metrics_harmonic}), whereas the uncontrolled work ranges from
$\mathcal{W} \approx 1.50$ at $\tau = 10^{-3}$ to $\mathcal{W} \approx 0.73$ at
$\tau = 10$, approaching $\Delta\mathcal{F}$ only in the quasi-static limit. This
equality is the expected consequence of exact tracking: with
$\boldsymbol{\rho}(t) = \boldsymbol{\pi}_{\rm eq}(t)$, the work reduces to
$\mathcal{W}(\tau) = \int_0^\tau \langle\partial  V /\partial t\rangle_{\pi_{\rm eq}}\,dt
= \Delta\mathcal{F}$. We stress a methodological distinction from
trajectory-sampling approaches: rather than estimating $\Delta\mathcal{F}$ from a
Jarzynski average $\langle e^{-\beta\mathcal{W}}\rangle$ over stochastic
realizations, we propagate the full density $\boldsymbol{\rho}(t)$
deterministically, so $\mathcal{W}(\tau)$ is the exact ensemble-averaged work for
each $\tau$---a single number with no statistical error, making the comparison
with $\Delta\mathcal{F}$ exact rather than asymptotic.  

\begin{table*}[!tbhp]
\caption{
Maximum dissipated work magnitude $\max |\mathcal{W}_{\mathrm{diss}}(\zeta_t)|$ calculated for the bare and LCD-driven protocols across the double-well and harmonic potentials at varying driving periods.}
\label{tab:work}
\begin{ruledtabular}
\begin{tabular}{l ccccc}
    & \multicolumn{2}{c}{\textbf{Double-well potential}} & \multicolumn{3}{c}{\textbf{Harmonic trap}} \\
    \cline{2-3} \cline{4-6}
    \rule{0pt}{3ex}
   Protocol duration $(\tau)$ & Bare & LCD & Bare & LCD & Analytic \\
    \hline
    \rule{0pt}{3ex}
    $10^{-3}$($\Delta t=10^{-7}$)  & 1.40 & $8.08 \times 10^{-12}$ & 0.80 & $6.80\times 10^{-12}$ & $9.43\times 10^{-13}$ \\
    $0.01$ ($\Delta t=10^{-6}$)  & 1.40 & $1.94 \times 10^{-12}$ & 0.79 & $5.70\times 10^{-12}$ & $6.22\times 10^{-14}$ \\
    $0.10$ ($\Delta t=10^{-5}$)  & 1.37 & $2.85 \times 10^{-12}$ & 0.72 & $5.15\times 10^{-12}$ & $8.02\times 10^{-14}$ \\
    $0.25$ ($\Delta t=10^{-5}$)  & 1.32 & $1.40 \times 10^{-12}$ & 0.61 & $3.97\times 10^{-12}$ & $4.14\times 10^{-14}$ \\
    $0.50$ ($\Delta t=10^{-5}$)  & 1.26 & $1.02 \times 10^{-12}$ & 0.48 & $3.31\times 10^{-12}$ & $5.21 \times 10^{-14}$ \\
    0.75 ($\Delta t=10^{-5}$)   & 1.20 & $1.88 \times 10^{-12}$ & 0.39 & $2.48\times 10^{-12}$ & $3.59 \times 10^{-14}$ \\
    1.00 ($\Delta t=10^{-5}$)   & 1.15 & $2.08\times 10^{-12}$  & 0.33 & $2.27\times 10^{-12}$ & $2.03\times10^{-14}$ \\
    2.00 ($\Delta t=10^{-5}$)   & 0.99 & $1.55\times10^{-12}$   & 0.19 & $1.28\times10^{-12}$  & $2.92\times 10^{-14}$   \\
    4.00 ($\Delta t=10^{-4}$)   & 0.76 & $1.23\times 10^{-13}$  & 0.10 & $7.07\times 10^{-13}$  & $1.20\times10^{-14}$\\
    6.00 ($\Delta t=10^{-4}$)   & 0.62 & $3.95\times10^{-13}$   & 0.07 & $4.33\times 10^{-13}$  & $8.77 \times 10^{-15}$\\
    10.0 ($\Delta t=10^{-4}$)  & 0.43 & $9.82\times 10^{-13}$   & 0.04 & $2.68\times10^{-13}$ & $6.55 \times 10^{-15}$  \\
\end{tabular}
\end{ruledtabular}
\end{table*}
\subsection{Spectral Truncatation Performance}\label{sec:spectral_truncation_main} 
Our Liouvillian CD correction term Eq.~\eqref{eq:ACD_spectral} admits a truncation, where one can restrict to a first set of spectral modes $M < N$ (or even $M \ll N$). We detail this in Section~\ref{sec:spectral_truncation}. Although, the efficacy of such a truncation is fundamentally governed by two quantities in Eq.~\eqref{eq:ACD_spectral} (i) the speed of the driving protocol $|\dot{\zeta}_t|$, which affects the relaxation to other higher modes and (ii) the instantaneous inverse relaxation gaps $\Delta_n^{-1}(t) \equiv |\lambda_0(t) - \lambda_n(t)|^{-1} = |\lambda_n(t)|^{-1}$, which dictate a closing spectral gap and the presence of nearly-degenerate modes. Although for simpler potentials such as DW or harmonic traps, omitting highly excited modes yields an incomplete cancelation of the non-adiabatic lag (imperfect escorting), as shown in Figures~\ref{fig:dw_trunc} [DW] and \ref{fig:harm_trunc} [harmonic] and Table~\ref{tab:truncation_analysis} for $\tau=0.1$. As already noted in Section~\ref{sec:rank1}, the full counterdiabatic operator is rank one,
the spectral decomposition of that rank-one correction into
individual mode contributions identifies
which spectral modes dominate the non-adiabatic lag and how many must be resolved for a given level of imperfect escorting. This information is inaccessible in the closed-form expression but essential for
understanding the physics of finite-time driving.

We show this for varied protocol speeds in Tables~(\ref{tab:dw_trunc}, \ref{tab:harmonic_trunc}), it circumvents the computationally prohibitive requirement of full matrix diagonalization while achieving modest improvements in tracking fidelity. This enables a favorable numerical trade-off in terms of computing free-energy simulations, where the spectral resolution needed to achieve a target tracking fidelity can be reduced with improved convergence compared to fully uncontrolled schemes. 
\begin{figure}[!tbhp]
  \centering
  \includegraphics[width=\linewidth]{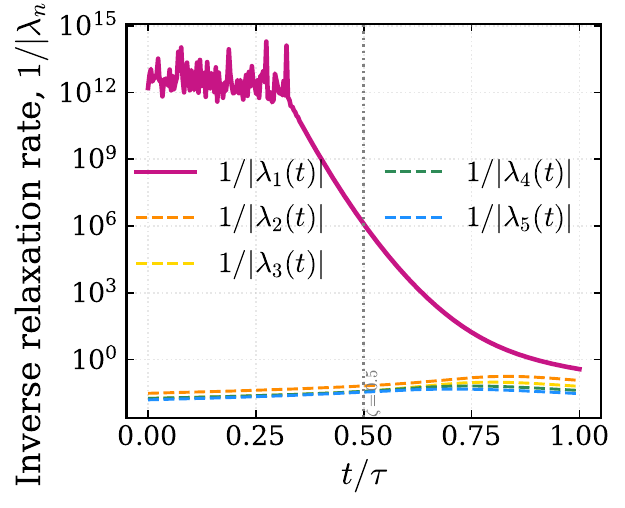}
  \caption{\textbf{Inverse relaxation rate for the quartic potential.} Inverse gap $1/|\lambda_n(t)|$ for the five
  slowest modes along the quartic coalescence protocol. The ground
  ($n=1$) mode spans \emph{twelve orders of magnitude}, from a
  round-off-dominated plateau ($t/\tau \approx 0.33–0.4$, where the true gap
  is below machine precision) down to $O(1)$ as $\zeta\to1$. Modes
  $n=2$--$5$ remain well-conditioned ($O(0.3$--$0.5)$) throughout,
  confirming that only the ground-state gap $\Delta_1(t) \equiv |\lambda_1(t) - \lambda_0(t)|$ ($\lambda_0 = 0$) is the bottleneck.}
  \label{fig:quartic-spectral-gap}
\end{figure}
\subsection{Spectral framework and vanishing spectral gap}
\label{sec:spectral_gap_limit}
As a stress test of the spectral framework, we apply the LCD
to the quartic coalescence potential~\citep{lee2026estimatingfreeenergydifferences},
$V(x,\zeta_t) = x^4 - 16(1-\zeta_t)x^2$, whose two minima at
$\pm\sqrt{8(1-\zeta_t)}$ merge at the origin as
$\zeta_t \to 1$, annihilating a barrier of height
$[8(1-\zeta_t)]^2$.
The exact free energy difference is
$\Delta\mathcal{F} = 62.9407...$ at $\beta = 1$.
Table~\ref{tab:quartic-results} reports the terminal work and
dissipation for protocol durations
$\tau \in [0.1,\,0.5]$: the LCD consistently drives $\mathcal{W}(\tau)$
closer to $\Delta\mathcal{F}$ than the bare dynamics, with
improvements ranging from $\sim 33\times$ at $\tau = 0.1$ to
$\sim 1400\times$ at $\tau = 0.5$.

Nevertheless, the residual $\mathcal{W}_{\rm diss}$ for LCD remains
non-negligible even at moderate driving speeds, in sharp
contrast to the double-well and harmonic potentials where
$|\mathcal{W}_{\rm diss}| \lesssim 10^{-12}$.
The origin of this degradation is visible in
Fig.~\ref{fig:quartic-spectral-gap}: the spectral gap
$|\lambda_1(t)|$ drops below $10^{-12}$ for
$t/\tau \approx 0.33–0.4$, reflecting the exponentially slow
inter-well relaxation in the presence of a barrier exceeding
$16\,k_BT$.
In this regime the spectral coefficients
$\sim 1/|\lambda_n|$ diverge, the eigenvector matrix
$\mathbf{D}(t)$ becomes ill-conditioned
($\mathrm{cond}(\mathbf{D}) > 10^{8}$), and the biorthogonal
decomposition can no longer resolve the nearly degenerate
slow modes at double precision.

Here, we essentially highlight and distinguish the breakdown of our spectral LCD \emph{construction}
as opposed to the counterdiabatic driving itself. As the barrier grows the ground-state gap $|\lambda_1(\zeta)|\sim e^{-\beta\Delta V_b(\zeta)}$~\citep{10.1039/TF9353100875} falls below
double precision for $t/\tau\lesssim0.4$ [Figure~\ref{fig:quartic-spectral-gap}]; the numerically
returned $\lambda_1\sim10^{-12}$--$10^{-13}$ there is round-off dominated rather
than the true (exponentially smaller) gap, and the near-degenerate eigenvectors
become ill-conditioned ($\mathrm{cond}(\bm{D})>10^{8}$). The modal coefficient $\bm{\ell}_1(\partial \Lop/\partial \zeta)\bm{r}_0/\lambda_1$ is then a quotient of two
round-off quantities and corrupts the assembled $ \Lop_{\mathrm{CD}}$ [Eq.~\eqref{eq:ACD_spectral}].
Crucially, the exact closed form counterdiabatic operator~\eqref{eq:LCD-rank1}, by contrast remains bounded throughout this protocol and tracks $\pi_{\rm eq}(t)$ directly; the divergence is therefore a breakdown of the spectral \emph{representation} of $\Lop_{\rm CD}$, not of the
counterdiabatic control itself. For the symmetric coalescence potential the closing mode is antisymmetric, so
$\bm{\ell}_1(\partial \Lop/\partial \zeta)\bm{r}_0=0$ identically by symmetry and the inverse spectral gap
$1/\lambda_1(t)$ term multiplies a vanishing numerator (see Section~\ref{sec:counterdiabatic_generator_quartic} for details). The
symmetric coalescence thus exposes a failure of the spectral \emph{representation}, not a physical limit on engineering a counterdiabatic control. 
\color{black}
\section{Discussion}
\label{sec:discussion}
We have developed an explicit spectral construction of the counterdiabatic generator for classical stochastic systems governed by the Fokker--Planck
equation. The underlying biorthogonal formalism---the eigenbasis
$\{\mathbf{r}_n(t),\bm{\ell}_n^T(t)\}$ with $\bm{\ell}_0^T=\mathbf{1}^T$ (probability conservation) and
$\mathbf{r}_0=\pieq$---follows~\citep{Iram2021}; our contribution is in the assembly $\Lop_{\rm CD}(t)$ from these eigenpairs (different from the approach of \citep{Iram2021}, its exact reduction to a rank-one closed form~[Eq.~\eqref{eq:LCD-rank1}] under detailed balance, and the numerical characterization of when that spectral assembly is reliable, especially for reducing/eliminating dissipated work.
The resulting spectral formula for the counterdiabatic
correction $\Lop_{\rm CD}(t)$ is formally identical to
the Berry--Demirplak--Rice transitionless driving prescription (also analytically shown in \citep{Iram2021}) and earns its place as an alternative approach escorting the Fokker--Planck density and a diagnostic that ties the lag to the spectral gap. 

The numerical demonstrations on the double-well and harmonic
trap confirm two key claims.
First, under LCD the figure of merits such as TVD ($\sim $ 12-orders of magnitude improvement) and KL ($\sim $ 16-orders of magnitude improvement -- machine precision in $\mathtt{FP64}$)  clearly demonstrate that we track the instantaneous equilibrium distribution at all times compared to the uncontrolled evolution. Moreover, we show that  
$\mathcal{W}^{\rm LCD}(t) = \Delta \mathcal{F}(\zeta_t)$ holds at every instant
$t \in [0,\tau]$, making the
dissipated work $\mathcal{W}_{\rm diss}(t) \approx 0$ [$\mathcal{O}(10^{-12})$] throughout the protocol at arbitrary driving speed. Secondly, we present a spectral truncation scheme, with the number of required spectral modes increasing
modestly for faster protocols and vanishing spectral gap potentials, establishing spectral truncation as a systematic way to reduce (if not completely eliminate) the non-adiabatic lag. Unlike standard flow-field protocols which obscure the underlying spectral features, our biorthogonal Liouvillian formulation explicitly exposes the physical source of dissipation: near critical points (such as the quartic coalescence), the spectral coefficients/construction become ill-conditioned as the gap closes, while the exact rank-one correction stays bounded. This establishes that the CD driving control stays bounded and only the spectral representation fails due to numerics (conditioning-related). 

\section{Conclusion}
\label{sec:conclusion}
We introduced Liouvillian counterdiabatic driving (LCD), a
spectral method, adapting Berrys' framework for eliminating non-adiabatic lag and performing escorted dynamics in classical stochastic systems driven by finite-time protocols.
The method constructs the counterdiabatic correction by promoting to a matrix-valued approach and extracts the non-adiabatic contributions during the Fokker--Planck evolution directly
from the biorthogonal eigenpairs (spectral properties) of the Fokker--Planck
generator, by moving to the instantaneous eigenbasis (adiabatic frame) using a time-dependent gauge-type transform. Thus,  one requires no generalized coordinate transforms, learned
representations, or iterative optimisation, and yields a
closed-form spectral sum that admits systematic truncation to
the slowest relaxation modes, depending on the potential and protocol speeds. 
Numerical verification on a double-well potential, a harmonic
trap, confirms tracking of the
instantaneous equilibrium to high precision, vanishing
dissipated work $\mathcal{W}_{\rm diss}(t) \approx 0$ at all times, and robustness across protocol speeds.

The spectral framework retains three roles that Eq.~\eqref{eq:dpieq-explicit} does not make manifest, and which motivate it as the organizing object of this numerical approach: (i) it exposes the AGC $\bm{\Gamma}=\bm{D}^{-1}(\partial \bm{D}/\partial t)$ as the mechanism that pumps probability out of the instantaneous equilibrium manifold (zero mode), and hence the spectral-gap origin of the lag; (ii) it yields a systematically truncable low-rank approximation; and (iii) offers an  escorting method with natural extension to systems where closed form solutions are non-analytic~\citep{Derrida_2007} and must be obtained numerically. Thus, our framework provides a complementary route to escorting the Fokker--Planck equation alongside existing flow-field and
diffeomorphism-based approaches.

\section*{Limitations and Future Directions}
The spectral LCD formula contains inverse relaxation rates
$1/\lambda_n(t)$ in its coefficients and the spectral coefficients/construction become ill-conditioned as the gap closes, while the exact rank-one correction stays bounded. This is demonstrated explicitly on the quartic coalescence
potential (Section~\ref{sec:spectral_gap_limit}), where the LCD
still reduces $\mathcal{W}_{\rm diss}$ by one to two orders of magnitude but cannot reach the machine-precision escorting achieved on the double-well and harmonic potentials, whose gaps remain well separated throughout the protocol. We note that closed-form solutions~\citep{Guéry-Odelin_2023} or diffeomorphism-based approaches such as
normalizing flows~\citep{DBLP:conf/nips/0035KN20} which operate
entirely in configuration space are therefore insensitive
to the spectral gap of $\Lop(t)$; the quartic
coalescence regime that limits the present spectral framework
may thus remain tractable for flow-based methods, suggesting
that the two approaches have complementary domains of
applicability. Moreover, the $\mathcal{O}(N^3)$ cost of full diagonalization is mitigated by three complementary strategies: exploiting the symmetric form of the rate
matrix~[Eq.~\eqref{eq:symm_liouville}], which admits
tridiagonal solvers at $12$--$15\times$ speedup
(Section~\ref{sec:symm_rep_numerics}) for our 1D use cases; spectral truncation to
$M < N$ (or even $M \ll N$) modes (Section~\ref{sec:spectral_truncation}) at the expense of varied degrees of imperfect escorting accuracies depending on the mode truncation; and
sparse Lanczos eigensolvers~\citep{Lanczos1950, golub1996matrix} that exploit the $(2d+1)N$
nearest-neighbor sparsity of the Sasa--Tasaki discretization
at a cost of $\mathcal{O}(MdN)$
(Section~\ref{sec:scalability}).
Together, these establish a viable path toward
higher-dimensional systems where full diagonalisation is
infeasible.

Looking ahead, several directions remain open: the most important direction is to extend our spectral methodology in the absence of the detail-balance condition. For such non-equilibrium steady state systems, the steady-state density $\pi_{ss}(x, t)$ is not known analytically, and requires to be computed numerically. In this work, we took the first steps by implementing the spectral framework to known baselines, subsequently to tackle complex nonequilibrium steady state (NESS) systems as a future direction, which require careful consideration of the non-conservative forces and housekeeping (Hatano-Sasa) heat terms~\citep{PhysRevLett.95.040602, PhysRevLett.86.3463}. Secondly, potentials which leads to vanishing spectral gaps, wherein time-dependent near-degeneracies cause the biorthogonal
decomposition to become ill-conditioned and demand
regularisation strategies beyond the standard eigensolver,
requiring diagonalization-free methodologies operating in Krylov spaces~\citep{wbbs-s8fs, shrestha2026shortcutsadiabaticitynonhermitiansystems}, together with a careful analysis of the thermodynamic cost and interpretation of counterdiabatic driving in this
regime.  Thirdly, establishing the precise functional relationship
between the eigenbasis gauge connection
$\bm{D}^{-1}(\partial\bm{D}/\partial t)$ and the
escorting flow field
$\mathbf{u}(\mathbf{x},t)$~\cite{PhysRevLett.100.190601}
would clarify the similarities and differences between the two approaches. Finally, extending the framework to two- and three-dimensional potentials with non-trivial barrier structure, where sparse Lanczos-based truncated LCD can be benchmarked in regimes
where full eigendecomposition is no longer available, is the
natural next step toward practical applications.
\subsection*{Reproducibility statement}
All simulations were carried out in \texttt{Python} (3.9.25) using \texttt{NumPy}
(1.26.4), \texttt{SciPy} (1.13.1), and Matplotlib (3.9.3). \texttt{NumPy's} dense
eigendecomposition (\texttt{numpy.linalg.eig}), used throughout for the
biorthogonal decomposition of the Liouville generator $\Lop(t)$, was
linked against OpenBLAS 0.3.23 (dynamic architecture dispatch,
\texttt{DYNAMIC\_ARCH}), which parallelizes the underlying LAPACK calls
across available cores; wall-clock timings reported in this work (e.g.
the dense-vs-symmetric-tridiagonal speedup benchmarks) reflect this
multi-threaded BLAS backend and will vary with core count and BLAS
implementation on other hardware, though the relative
scaling/speedup trends are expected to hold generally.

Computations were run on a dual-socket workstation with two Intel Xeon
Platinum 8468 processors (48 cores/socket, 2 threads/core; 192 logical
CPUs total) and 2\,TiB RAM, running Rocky Linux 9.8 (kernel
5.14.0-687.17.1.el9\_8.x86\_64). No GPU acceleration was used; all
reported computations are CPU-only.

Code, a curated set of example results, and a demonstration notebook
are available at{\hypersetup{urlcolor=magenta} 
    \href{https://github.com/Sancran19/LCD-FPE}{%
        \raisebox{-0.3em}{\includegraphics[width=1.15em]{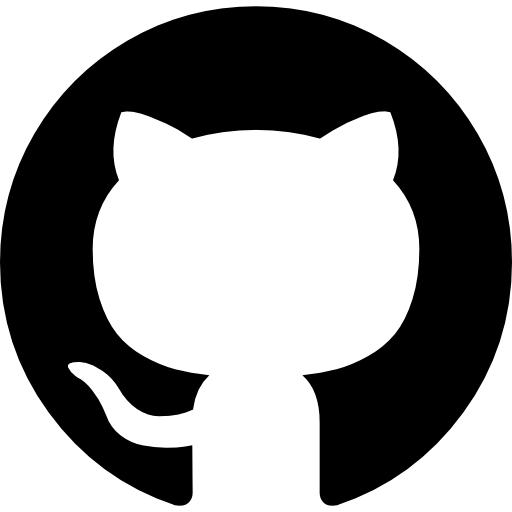}}
        \textbf{\texttt{LCD-FPE}}}}

\section*{Acknowledgements}
The ELLIS Unit Linz, the LIT AI Lab, the Institute for Machine Learning, are supported by
the Federal State Upper Austria. We thank the projects FWF AIRI FG 9-N (10.55776/FG9),
AI4GreenHeatingGrids (FFG- 899943), Stars4Waters (HORIZON-CL6-2021-CLIMATE-01-01). We
thank NXAI GmbH, Audi AG, Silicon Austria Labs (SAL), Merck Healthcare KGaA, GLS (Univ.
Waterloo), TÜV Holding GmbH, Software Competence Center Hagenberg GmbH, dSPACE GmbH,
TRUMPF SE + Co. KG.
Sandeep Suresh Cranganore was supported by the FWF Bilateral Artificial Intelligence initiative under Grant Agreement number 10.55776/COE12.

\bibliography{lcd_refs}

\onecolumngrid
\makeatletter
\long\def\frontmatter@abstractheading{} 
\setcounter{equation}{0}
\setcounter{figure}{0}
\setcounter{table}{0}
\setcounter{section}{0}
\setcounter{page}{1}

\makeatletter
\renewcommand{\theequation}{S\arabic{equation}}
\renewcommand{\thefigure}{S\arabic{figure}}
\renewcommand{\thetable}{S\arabic{table}}
\makeatother
\newpage 

\begin{center}
    \textbf{\large SUPPLEMENTAL MATERIAL} \\
\end{center}
\newcommand{\citesupp}[1]{[\ref{#1}]}
\makeatother

\section{Zero-Variance Free Energy Estimation via Perfect Bijective Mappings}
\label{app:zero_variance_mappings}

For an arbitrary set of invertible, bijective coordinate mapping functions $\{\mathcal{M}_i\}$, the statistical efficiency of utilizing escorted simulations to estimate the free energy difference $\Delta \mathcal{F}$ depends crucially on minimizing the relative entropy between the driven and target ensembles. Because the convergence of exponential averages---such as the free energy estimator in Eq.~\eqref{eq:JE}---deteriorates rapidly with increasing non-equilibrium dissipation~\citep{PhysRevLett.100.190601, 10.1063/1.3544679}, the statistical variance is fundamentally governed by the phase-space lag between the instantaneous non-equilibrium distribution and the corresponding equilibrium target. Consequently, an optimal choice of coordinate transformations must actively suppress this lag to maximize phase-space overlap and optimize estimator convergence.

To formalize this optimization framework, one analyzes the limiting case of a ``perfect'' set of mapping functions, denoted by $\{\mathcal{M}^*_i\}$, which entirely eliminates the non-equilibrium lag. Formally, for an ensemble of trajectories initiated from the canonical equilibrium state $\pi_0 \sim \pieq^{\zeta_0}(x_0)$ [since the evolution is described for discrete time-stamps $\zeta_{t_i} = \zeta_{i}$, one succinctly denotes $\pieq(x_0, \zeta_0) = \pieq^{\zeta_0}(x_0)$] and evolved under the escorted dynamics, a perfect map ensures that the subsequent microstates $x_i$ strictly trace the instantaneous equilibrium manifold $\pieq^{\zeta_i}(x_i)$ for all discrete steps $1 \le i < N$. Geometrically, this requires the target and escorted probability densities to coincide identically. 

Under a bijective coordinate transformation $\mathcal{M}^*_i: x \to x'$, conservation of probability requires the initial distribution $\pieq^{\zeta_i}(x)$ to transform under the push-forward map as~\citep{PhysRevE.65.046122, DBLP:conf/nips/0035KN20}:
\begin{equation}
    \pieq^{\zeta_{i+1}}(x') = \frac{\pieq^{\zeta_i}(x)}{J^*_{\zeta_i}(x)},
    \label{eq:jacobian_pushforward}
\end{equation}
where $J^*_{\zeta_i}(x) = \det |\partial x' / \partial x|$ denotes the Jacobian determinant of the transformation. 

Substituting the canonical Boltzmann weight $\pieq^{\zeta_t}(x) \equiv \pieq (x, \zeta_t) = e^{-\beta \big[V(x,\zeta_t) - \mathcal{F}(\zeta_t)\big]}$ into Eq.~\eqref{eq:jacobian_pushforward} and taking the natural logarithm of both sides yields the exact microscopic energy balance for a perfect mapping:
\begin{equation}
    \delta \mathcal{W}^{\text{esc}}_{i} \equiv V_{\zeta_{i+1}}(x') - V_{\zeta_i}(x) - \beta^{-1} \log J^*_{\zeta_i}(x) = \mathcal{F}_{\zeta_{i+1}} - \mathcal{F}_{\zeta_i}.
    \label{eq:microscopic_work_balance}
\end{equation}
where $\mathcal{W}^{\text{esc}}$ corresponds to the escorted work. Summing this relation over the full discrete protocol reveals that the total accumulated work with the Jacobian determinant scaled density (escorting) along any arbitrary phase-space trajectory $\gamma$ collapses identically to the total free energy difference, $\mathcal{W}^{\text{esc}}[\gamma] = \Delta \mathcal{F}$. Thermodynamically, this condition suppresses all sample-to-sample fluctuations in the work distribution, forcing the non-equilibrium work probability density to contract to a singular Dirac delta function:
\begin{equation}
    P_\mathcal{F}(W) = \delta(\mathcal{W} - \Delta \mathcal{F}).
    \label{eq:delta_work_distribution}
\end{equation}
Crucially, for a perfect set of mappings $\{\mathcal{M}^*_i\}$, the exponential estimator [Eq.~(\ref{eq:eje})] yields a deterministic, zero-variance evaluation of $\Delta \mathcal{F}$ from a single trajectory invocation. Under these conditions, i.e., a perfect sequence of mappings for the forward protocol, an identical zero-variance condition is structurally preserved in the time-reversed process.

\section{Discretization Framework and the Continuum Limit}
We consider the overdamped stochastic evolution of a classical system in a time-varying potential $V(x,t)$. In the continuum limit, the probability density $\rho(x,t)$ evolves according to the Fokker-Planck equation (FPE), which can be written compactly in terms of the continuous Liouville operator as:
\begin{equation}
    \frac{\partial \rho(x,t)}{\partial t} = \frac{\partial}{\partial x} \left[ \left(\frac{\partial V(x,t)}{\partial x}\right) \rho(x,t) \right] + \beta^{-1} \frac{\partial^2 \rho(x,t)}{\partial x^2},
    \label{eq:continuous_fpe}
\end{equation}
where $\beta = 1/(k_B T)$ represents the inverse temperature of the thermal reservoir. To treat this evolution numerically and analyze its spectral properties, we project the continuous spatial coordinate onto a uniform lattice $\{x_i\}_{i=0}^{N-1}$ with an associated grid spacing $\Delta x$. The continuous probability density is mapped onto a discrete state vector $\boldsymbol{\rho}(t) = (\rho_0, \dots, \rho_{N-1})^T \in \mathbb{R}^N$, where $\rho_i(t) \approx \rho(x_i, t)\Delta x$, subject to the strict normalization constraint $\sum_i \rho_i = 1$. Under this projection, the spatial partial differential equation maps onto a discrete Master Equation governing the probability exchange via local fluxes:
\begin{equation}
    \frac{\partial \rho_i}{\partial t} = k_{\text{fwd}}[i-1]\rho_{i-1} + k_{\text{bwd}}[i]\rho_{i+1} - \big(k_{\text{fwd}}[i] + k_{\text{bwd}}[i-1]\big)\rho_i,
    \label{eq:master_equation}
\end{equation}
where $k_{\text{fwd}}[i]$ and $k_{\text{bwd}}[i]$ represent the transition rates out of site $i$ to neighboring sites $i+1$ and $i-1$, respectively. To preserve local detailed balance with respect to the instantaneous Boltzmann distribution $\pi_{\text{eq}}(x,t) \propto \exp[-\beta V(x,t)]$, we select Sasa-Tasaki rates defined as:
\begin{align}
    k_{\text{fwd}}[i] &= \frac{1}{\beta \Delta x^2} \exp\left(-\frac{\beta}{2}\Delta V_i\right), \label{eq:rate_fwd} \\
    k_{\text{bwd}}[i] &= \frac{1}{\beta \Delta x^2} \exp\left(+\frac{\beta}{2}\Delta V_i\right), \label{eq:rate_bwd}
\end{align}
where $\Delta V_i \equiv V_{i+1} - V_i$ denotes the forward potential difference across the lattice bond.

To rigorously demonstrate that the discrete version structurally converges to the continuous FPE, we perform a multivariable Taylor expansion of the fields $\rho(x,t)$ and $V(x,t)$ around the reference grid site $x_i$ up to third order in $\Delta x$. Let $\partial f/\partial x \equiv f'$ and $\partial^2 f/\partial x^2 \equiv f''$. The non-local spatial shifts and potential gradients map to local expansions according to:
\begin{align}
    \rho_{i \pm 1} &= \rho \pm \Delta x \rho' + \frac{\Delta x^2}{2}\rho'' \pm \mathcal{O}(\Delta x^3), \\
    \Delta V_i &= \Delta x V' + \frac{\Delta x^2}{2}V'' + \mathcal{O}(\Delta x^3), \\
    \Delta V_{i-1} &= \Delta x V' - \frac{\Delta x^2}{2}V'' + \mathcal{O}(\Delta x^3).
\end{align}
Expanding the exponential terms in the Sasa-Tasaki rate definitions to second order in $\Delta x$ yields the incoming rate operators feeding probability into site $i$:
\begin{align}
    k_{\text{fwd}}[i-1] &= \frac{1}{\beta \Delta x^2}\left[ 1 - \frac{\beta\Delta x}{2}V' + \Delta x^2 \left(\frac{\beta}{4}V'' + \frac{\beta^2}{8}(V')^2\right) \right], \\
    k_{\text{bwd}}[i] &= \frac{1}{\beta \Delta x^2}\left[ 1 + \frac{\beta\Delta x}{2}V' + \Delta x^2 \left(\frac{\beta}{4}V'' + \frac{\beta^2}{8}(V')^2\right) \right].
\end{align}
Taking the respective products of these incoming rates with the shifted state densities $\rho_{i \pm 1}$, we evaluate the total incoming probability flux up to order $\mathcal{O}(\Delta x^0)$:
\begin{align}
    k_{\text{fwd}}[i-1]\rho_{i-1} &+ k_{\text{bwd}}[i]\rho_{i+1} \nonumber \\
    = \frac{1}{\beta \Delta x^2} \bigg[ &2\rho + \Delta x^2 \left( \rho'' + \beta V'\rho' \right. + \left. \left. \frac{\beta}{2}V''\rho + \frac{\beta^2}{4}(V')^2\rho \right) \right] \nonumber \\
    + \mathcal{O}(\Delta x).
    \label{eq:incoming_flux}
\end{align}

Correspondingly, we evaluate the outgoing escape rates moving away from site $i$:
\begin{align}
k_{\text{fwd}}[i] &= \frac{1}{\beta \Delta x^2}\left[1 - \frac{\beta\Delta x}{2}V' - \frac{\beta\Delta x^2}{4}V'' + \frac{\beta^2\Delta x^2}{8}(V')^2 \right], \\
k_{\text{bwd}}[i-1] &= \frac{1}{\beta \Delta x^2}\left[1 + \frac{\beta\Delta x}{2}V' - \frac{\beta\Delta x^2}{4}V'' + \frac{\beta^2\Delta x^2}{8}(V')^2 \right].
\end{align}
Summing these rate coefficients establishes the total escape rate acting on the local state density $\rho_i$:
\begin{align}
    \big(k_{\text{fwd}}[i] + k_{\text{bwd}}[i-1]\big)\rho_i \nonumber \\ 
    = \frac{1}{\beta \Delta x^2} \left[ 2\rho + \Delta x^2 \left( -\frac{\beta}{2}V''\rho + \frac{\beta^2}{4}(V')^2\rho \right) \right].
    \label{eq:outgoing_flux}
\end{align}

Finally, substituting the explicit incoming flux expansion Eq.~\eqref{eq:incoming_flux} and outgoing escape expansion Eq.~\eqref{eq:outgoing_flux} back into the master equation structure Eq.~\eqref{eq:master_equation}, the singular leading-order term $2\rho/(\beta \Delta x^2)$ cancels out identically. Grouping the surviving structural components yields:
\begin{equation}
    \frac{\partial \rho}{\partial t} = V'\rho' + V''\rho + \beta^{-1}\rho''.
\end{equation}
Recognizing that $V'\rho' + V''\rho \equiv \partial_x (V'\rho)$, we reconstruct the derivative components to obtain:
\begin{equation}
    \frac{\partial \rho(x,t)}{\partial t} = \frac{\partial}{\partial x}\left[\left(\frac{\partial V(x, t)}{\partial x} \right)\rho(x, t) \right] + \beta^{-1} \frac{\partial^2 \rho(x, t)}{\partial x^2}.
\end{equation}
This confirms that the discrete grid network updates perfectly recover the continuous Fokker-Planck dynamics in the continuum limit $\Delta x \to 0$.

\section{Spectral Representation of the Gauge Connection}\label{sec:spectral_derivation}

To establish the exact equivalence between the the AGC term $\bm{D}^{-1} (\partial \bm{D}/\partial t)$ and the spectral-gap-dependent Berry-type formulation, we begin with the instantaneous eigenvalue equation for the right eigenvectors of the Liouvillian operator,
\begin{equation}
\Lop(t) \mathbf{r}_m(t) = \lambda_m(t) \mathbf{r}_m(t).
\label{eq:eigen_eq}
\end{equation}
Differentiating Eq.~\eqref{eq:eigen_eq} with respect to time $t$ yields
\begin{equation}
\big(\partial \Lop/ \partial t\big) \mathbf{r}_m + \Lop \big(\partial \mathbf{r}_m/\partial t\big) = \big(\partial \lambda_m/\partial t\big) \mathbf{r}_m + \lambda_m \big(\partial \mathbf{r}_m/\partial t\big).
\label{eq:diff_eigen}
\end{equation}
We now project Eq.~\eqref{eq:diff_eigen} onto the dual space by left-multiplying with the left eigenvector $\boldsymbol{\ell}_n^T(t)$ corresponding to the index $n \neq m$. This yields
\begin{equation}
\boldsymbol{\ell}_n^T \big(\partial \Lop/\partial t\big) \mathbf{r}_m + \boldsymbol{\ell}_n^T \Lop \big(\partial \mathbf{r}_m/\partial t\big) = \big(\partial \lambda_m/\partial t\big) \boldsymbol{\ell}_n^T \mathbf{r}_m + \lambda_m \boldsymbol{\ell}_n^T \big(\partial \mathbf{r}_m/\partial t\big).
\end{equation}
Invoking the left-eigenvalue relation $\boldsymbol{\ell}_n^T \Lop = \lambda_n \boldsymbol{\ell}_n^T$ and the biorthogonality condition $\boldsymbol{\ell}_n^T \mathbf{r}_m = \delta_{nm}$, the first term on the right-hand side vanishes for $n \neq m$. Simplifying the left-hand side leads to
\begin{equation}
\boldsymbol{\ell}_n^T \big(\partial \Lop/\partial t\big) \mathbf{r}_m + \lambda_n \boldsymbol{\ell}_n^T \big(\partial \mathbf{r}_m/\partial t\big) = \lambda_m \boldsymbol{\ell}_n^T \big(\partial \mathbf{r}_m/\partial t\big).
\end{equation}
Rearranging terms directly isolates the projection of the eigenvector velocity onto the reciprocal basis,
\begin{equation}
\boldsymbol{\ell}_n^T \big(\partial \mathbf{r}_m/\partial t\big) = \frac{\boldsymbol{\ell}_n^T \big(\partial \Lop/\partial t\big) \mathbf{r}_m}{\lambda_m - \lambda_n} \quad (n \neq m).
\label{eq:biorthogonal_projection}
\end{equation}
The left-hand side of Eq.~\eqref{eq:biorthogonal_projection} defines the off-diagonal elements of the geometric gauge connection matrix in the adiabatic moving frame, $\Gamma_{nm} \equiv \boldsymbol{\ell}_n^T \big(\partial_t \mathbf{r}_m\big)$. 

This result highlights a key physical insight: the gauge connection $\Gamma(t) = \mathbf{D}^{-1}\partial_t \mathbf{D}$ is fundamentally a spectral-gap-dependent object. When computed numerically, the inverse relaxation gaps $(\lambda_m - \lambda_n)^{-1}$ are implicitly encoded in the gradients of the state-space trajectory $\partial \mathbf{r}_m/\partial t$, which diverge near avoided crossings or spectral degeneracies.

To express the full non-adiabatic correction in the laboratory frame, we map the off-diagonal components of the connection back via the similarity transformation $\mathbf{D} \Gamma_{\rm off} \mathbf{D}^{-1}$. Utilizing the completeness relation of the biorthogonal basis, we obtain
\begin{equation}
\mathbf{D} \Gamma_{\rm off} \mathbf{D}^{-1} = \sum_{n \neq m} \Gamma_{nm} \mathbf{r}_n \boldsymbol{\ell}_m^T = \sum_{n \neq m} \frac{\boldsymbol{\ell}_n^T \big(\partial \Lop/\partial t\big) \mathbf{r}_m}{\lambda_m - \lambda_n} \mathbf{r}_n \boldsymbol{\ell}_m^T.
\label{eq:full_lab_frame_connection}
\end{equation}
The spectral counterdiabatic driving term $\Lop_{\rm CD}(t)$ corresponds to the instantaneous equilibrium distribution $m = 0$ column of the full gauge connection. Given that $\lambda_0 = 0$, setting $m=0$ in Eq.~\eqref{eq:full_lab_frame_connection} recovers the spectral correction
\begin{equation}
\Lop_{\rm CD}(t) = -\sum_{n \neq 0} \frac{\boldsymbol{\ell}_n^T(t) \big(\partial \Lop/\partial t\big) \mathbf{r}_0(t)}{\lambda_n(t)} \mathbf{r}_n(t) \boldsymbol{\ell}_0^T,
\label{eq:LCD_reconstructed}
\end{equation}
which is mathematically identical to the localized Berry formulation, as adapted here to the Fokker--Planck evolution.

\section{Numerics Specifics}
\subsection{Eigendecomposition of the Liouville Operator}

At each instant $t$, one can perform exact diagonlization of the rate matrix:

\begin{equation}
  \Lop(t) = \bm{D}(t)\,\bm{\Lambda}(t)\,\bm{D}^{-1}(t),
\label{eq:eigendecomp}
\end{equation}

where $\bm{D}(t)\in\mathbb{R}^{N\times N}$ contains the right
eigenvectors $\bm{r}_n$ as its columns,
$\bm{D}^{-1}(t)$ has the left eigenvectors as its rows, and

\begin{equation}
  \bm{\Lambda}(t) = \mathrm{Diag}(\lambda_0,\lambda_1,\ldots,\lambda_{N-1}),
  \quad
  0 = \lambda_0 > \lambda_1 \geq \lambda_2 \geq \cdots
\label{eq:Lambda}
\end{equation}
We verify numerically that,
\begin{equation}
  \left\|\bm{D}^{-1}\Lop\bm{D} - \bm{\Lambda}\right\|_\infty
  < 10^{-6}
  \quad \forall\ \text{protocol values visited during the anneal},
\label{eq:diag_check}
\end{equation}
confirming exact diagonalisation. We report
the numerics for the exact diagonalized Liouville operator, for both
systems studied in this work, below.

\subsection{Exact diagonalization verification}\label{sec:exact_diag}

\subsubsection{Double well}

\textbf{Problem setup.} For our experimental model: the external time-varying
potential $V(x,\zeta_t) = x^4 - 2x^2 + \zeta_t x$ on a grid of $N=80$
points over $x\in[-2.5,2.5]$, $\beta=1$. The protocol is a smooth
sigmoid from $\zeta_i=-1$ to $\zeta_f=+1$:

\begin{equation}
  \zeta(t) = \zeta_i + (\zeta_f-\zeta_i)\,
  s^3(6s^2-15s+10),\quad s=t/\tau,
\label{eq:protocol}
\end{equation}

evaluated at $t/\tau=0,\,0.1,\,0.25,\,0.5,\,0.75,\,1$. Since $\zeta(t)$
depends only on $s=t/\tau$, this check is independent of the total
anneal duration $\tau$.

\begin{figure}[!tbhp]
  \centering
  \includegraphics[width=\linewidth]{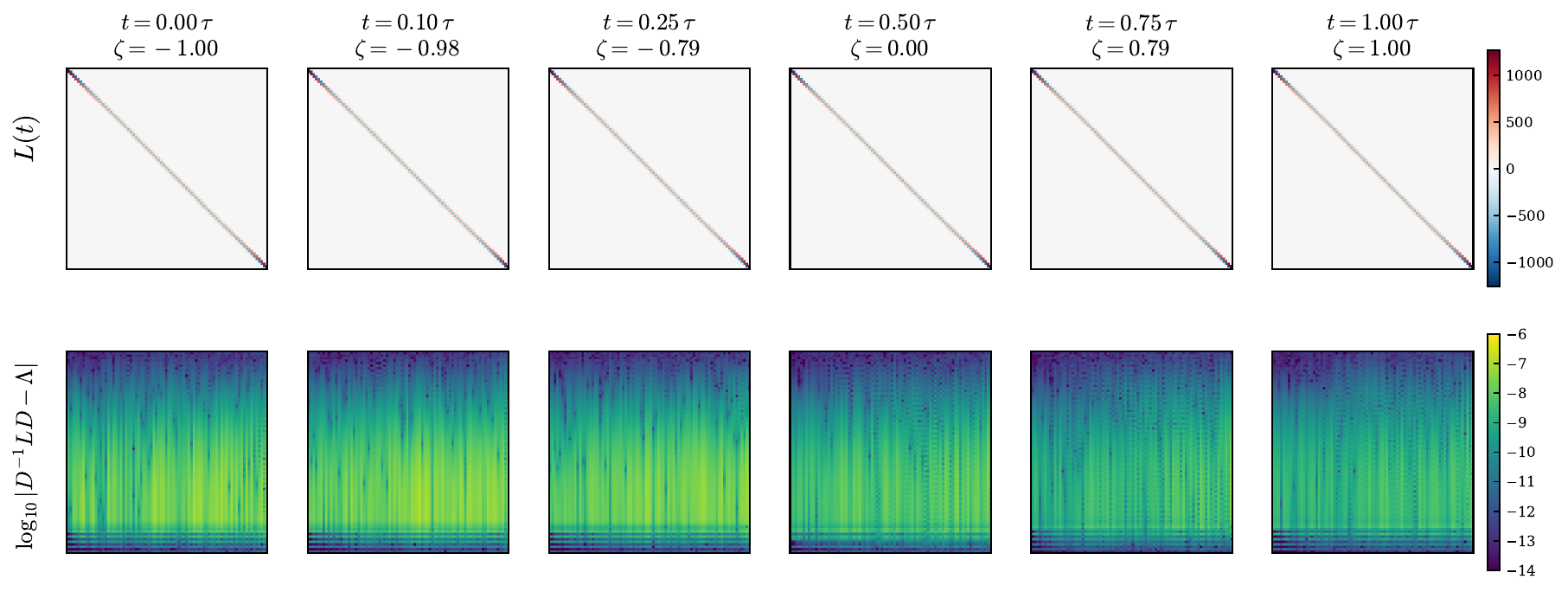}
  \caption{Double well: $\Lop(t)$ (top) and the diagonalization residual
  $\log_{10}|\bm{D}^{-1}\Lop \bm{D}-\bm{\Lambda}|$ (bottom) at six points along the anneal.
  The residual is uniformly at the $10^{-14}$--$10^{-6}$ noise floor at
  every snapshot, with no visible structure.}
  \label{fig:matrix-dw}
\end{figure}

\medskip
\textbf{Verification of $\bm{D}^{-1}\Lop \bm{D} = \bm{\Lambda}$.}
Table~\ref{tab:diag-dw} reports the off-diagonal norm of
$\bm{D}^{-1}\Lop \bm{D}$ and the biorthogonality residual
$\|\bm{D}^{-1}\bm{D} - \mathbb{I}\|_\infty$ at the six snapshots.
Both are at floating-point precision ($10^{-7}$--$10^{-11}$),
confirming Eq.~\eqref{eq:diag_check}. Figure~\ref{fig:matrix-dw} shows
$\Lop(t)$ itself (top row -- the tridiagonal birth--death structure is
visible throughout, magnitude growing toward the ends of the anneal
where the wells are most separated) alongside
$\log_{10}|\bm{D}^{-1}\Lop\bm{D}-\bm\Lambda|$ (bottom row), confirming
the residual shows no structure beyond numerical noise at every
snapshot.

\begin{table}[!tbhp]
\caption{
  \textbf{Double well:} numerical verification of the biorthogonal
  diagonalisation at six points along the anneal.
  $\varepsilon_\Lambda = \|\bm{D}^{-1}\Lop \bm{D} - \bm{\Lambda}\|_\infty$
  (reconstruction residual);
  $\varepsilon_I = \|\bm{D}^{-1}\bm{D} - \mathbb{I}\|_\infty$
  (biorthogonality residual).
}
\label{tab:diag-dw}
\begin{ruledtabular}
\begin{tabular}{cccc}
$t/\tau$ & $\zeta$ & $\varepsilon_\Lambda$ & $\varepsilon_I$ \\
\hline
0.00 & $-1.0000$            & $1.00\times10^{-7}$ & $1.60\times10^{-10}$ \\
0.10 & $-0.9829$            & $1.65\times10^{-7}$ & $1.00\times10^{-10}$ \\
0.25 & $-0.7930$            & $7.85\times10^{-8}$ & $6.35\times10^{-11}$ \\
0.50 & $\phantom{-}0.0000$  & $2.80\times10^{-8}$ & $2.55\times10^{-11}$ \\
0.75 & $\phantom{-}0.7930$  & $3.93\times10^{-8}$ & $1.51\times10^{-11}$ \\
1.00 & $\phantom{-}1.0000$  & $2.49\times10^{-8}$ & $1.17\times10^{-11}$ \\
\end{tabular}
\end{ruledtabular}
\end{table}

\newpage 
\subsubsection{Harmonic trap}
\textbf{Problem setup.} The harmonic trap,
$V(x,t)=\tfrac12\kappa_0(t)x^2$, on a grid of $N=80$ points over
$x\in[-4,4]$, $\beta=1$, annealed via the same smoothstep schedule
Eq.~\eqref{eq:protocol} with $\kappa_0:1\to4$, evaluated at the same six
$t/\tau$ points.

\begin{figure}[!tbhp]
  \centering
  \includegraphics[width=\linewidth]{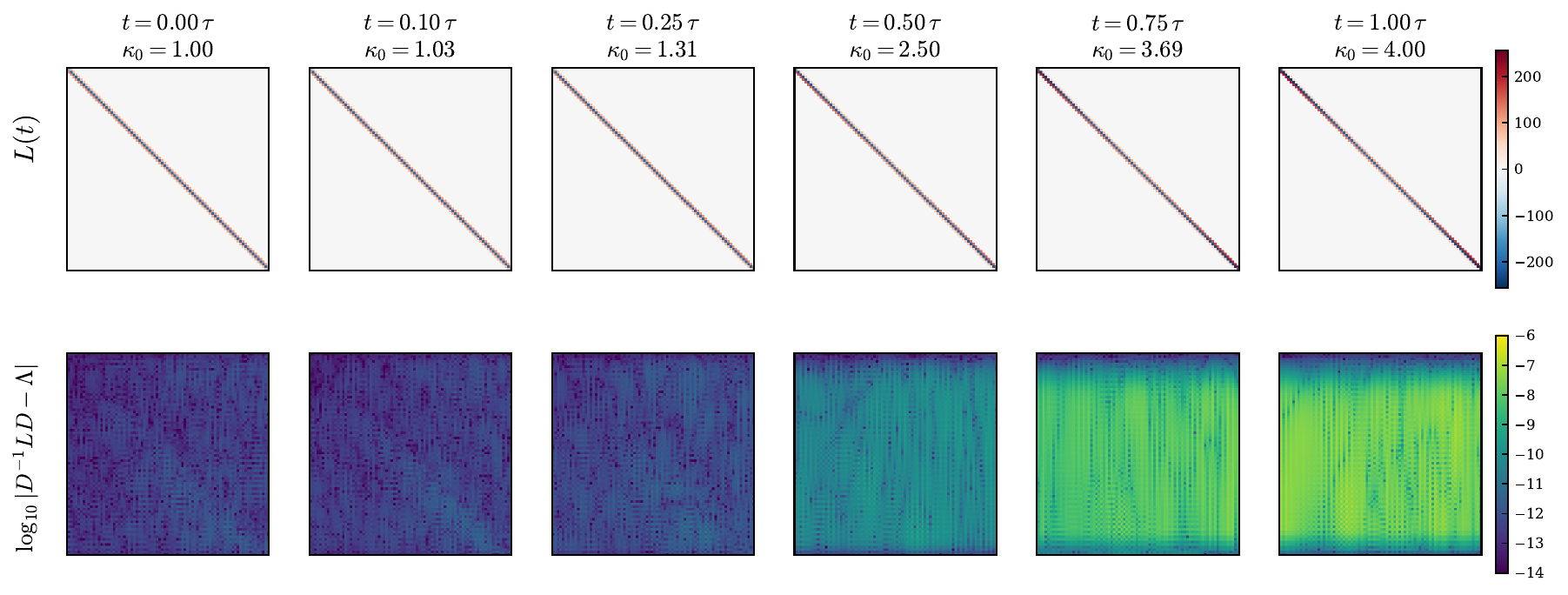}
  \caption{Harmonic trap: $\Lop(t)$ (top) and the diagonalization residual
  $\log_{10}|\bm{D}^{-1} \Lop \bm{D}-\bm{\Lambda}|$ (bottom) at six points along the anneal.
  The residual grows visibly but remains at floating-point precision
  throughout, tracking the growth of $\kappa_0$.}
  \label{fig:matrix-harm}
\end{figure}

\medskip
\textbf{Verification of $\bm{D}^{-1}\Lop \bm{D} = \bm{\Lambda}$.}
Table~\ref{tab:diag-harm} reports the same two residuals at the same
six snapshots. Both remain at floating-point precision throughout
($10^{-8}$--$10^{-15}$), though $\varepsilon_\Lambda$ and
$\varepsilon_I$ grow by roughly four orders of magnitude from
$\kappa_0=1$ to $\kappa_0=4$ -- visible directly in
Fig.~\ref{fig:matrix-harm}'s residual row, which brightens monotonically
left to right. This tracks the generator's overall matrix norm growing
with $\kappa_0$ (a stiffer trap gives larger transition rates
$k_f,k_b\propto D_c\,e^{\mp\beta\Delta V/2}$), which sets the scale of
floating-point rounding in $\bm{D}^{-1}\Lop\bm{D}$; the residual remains
many orders of magnitude below any physically relevant scale throughout.

\begin{table}[!tbhp]
\caption{
  Harmonic trap: numerical verification of the biorthogonal
  diagonalisation at six points along the protocol.
}
\label{tab:diag-harm}
\begin{ruledtabular}
\begin{tabular}{cccc}
$t/\tau$ & $\kappa_0$ & $\varepsilon_\Lambda$ & $\varepsilon_I$ \\
\hline
0.00 & 1.0000 & $4.36\times10^{-12}$ & $5.24\times10^{-15}$ \\
0.10 & 1.0257 & $5.71\times10^{-12}$ & $4.30\times10^{-15}$ \\
0.25 & 1.3105 & $1.17\times10^{-11}$ & $9.60\times10^{-15}$ \\
0.50 & 2.5000 & $6.55\times10^{-10}$ & $5.65\times10^{-13}$ \\
0.75 & 3.6895 & $3.88\times10^{-8}$  & $1.74\times10^{-10}$ \\
1.00 & 4.0000 & $9.82\times10^{-8}$  & $2.04\times10^{-10}$ \\
\end{tabular}
\end{ruledtabular}
\end{table}
\newpage 
\section{Dynamics in the Laboratory Frame}
\label{sec:normalization}

The transformation from the laboratory frame to the adiabatic frame of reference via a change of basis,
$\tilde{\vrho}(t) = \bm{D}^{-1}(t)\,\vrho(t)$,
does \emph{not} produce a probability vector.
Its components $\tilde{\rho}_n(t) = \bm{\ell}_n^T  \,\vrho(t)$
are projections of $\vrho$ onto the left eigenvectors of
$\Lop(t)$ and are not constrained to be non-negative
or to sum to unity. Thus, the normalization condition, i.e. $\sum_i \tilde{\rho}_i = 1$, need not hold in this representation. Nevertheless, the normalisation of $\vrho$ in the original
frame imposes a precise constraint on $\tilde{\vrho}$, which
we now derive.

\subsection{Dynamics under change of basis}
Separating the $n=0$ term in Eq.~\eqref{eq:rho_expansion}
and using $\tilde{\rho}_0 = 1$ and $\bm{r}_0(t) = \pieq(t)$:
\begin{equation}
  \vrho(t)
  = \underbrace{\pieq(t)}_{\tilde{\rho}_0 = 1}
  + \sum_{n=1}^{N-1} \tilde{\rho}_n(t)\, \bm{r}_n(t).
\label{eq:rho_decomp}
\end{equation}

Thus, the vector-valued density $\vrho(t)$ can be viewed as a linear combination of relaxation modes weighted by
their excitation amplitudes $\tilde{\rho}_n(t)$. The first term is the instantaneous equilibrium distribution;
the sum is over higher modes that will have larger excitation amplitudes induced by finite-time protocols $\zeta_t$ during the Fokker--Planck evolution. The leakage to relaxation modes other than $\pieq(t)$ are responsible for non-adiabatic lag. 
Since $\bbone^T \bm{r}_n = 0$ for $n \geq 1$ (biorthogonality relation), the deviation
term carries zero net probability:
\begin{equation}
  \bbone^T\!\sum_{n=1}^{N-1}\tilde{\rho}_n(t)\bm{r}_n
  = \sum_{n=1}^{N-1}\tilde{\rho}_n(t)\,\bbone^T \bm{r}_n
  = 0,
\label{eq:deviation_zero_sum}
\end{equation}
confirming that $\vrho(t)$ in the laboratory frame remains normalised for any values of $\tilde{\rho}_n$. The component $\tilde{\rho}_n(t) = \bm{\ell}^T_n \,\vrho(t)$ measures
the projection of the current distribution onto the $n$-th
relaxation mode, weighted by the corresponding left
eigenvector.
By biorthogonality Eq.~\eqref{eq:biorthog}:
\begin{equation}
  \tilde{\rho}_n(t) = 0
  \;\Longleftrightarrow\;
  \vrho(t) \perp_{\rm bio} \bm{r}_n,
\label{eq:biorthog_zero}
\end{equation}
where $\perp_{\rm bio}$ denotes orthogonality in the
biorthogonal sense.
In particular, if $\vrho(t) = \pieq(t) = \bm{r}_0(t)$, then:
\begin{equation}
  \tilde{\rho}_n(t)
  = \bm{\ell}^T_n \bm{r}_0
  = \delta_{n0}.
\label{eq:equilibrium_components}
\end{equation}
Thus, for $n \geq 1$
$\tilde{\rho}_n = \boldsymbol{\ell}_n^T\boldsymbol{\rho}$
is the projection of $\vrho$
onto relaxation mode $n$. It can be positive, negative, or zero. It is not a probability — it is a mode amplitude. So $\tilde{\vrho}(t) = (1, 0, 0, \ldots, 0)^T$ when and only
when $\vrho(t) = \pieq(t)$.
\subsection{Counterdiabatically evolved Liouville operator}
\emph{Bare dynamics} (no CD correction):
The adiabatic gauge potential term in Eq.~\eqref{eq:tilde_rho_dynamics}$   ~\Gamma_{nm}$ is non-zero for
$n \neq m$, continuously pumping excitation from the zero
mode into the relaxation modes, thus yielding an excess non-zero work in the form of dissipation, i.e. $\mathcal{W}_{\rm diss} = \mathcal{W} - \Delta \mathcal{F} > 0$.
In the long-time driven steady state,
$\tilde{\rho}_n \neq 0$ for $n \geq 1$, meaning
$\vrho(t) \neq \pieq(t)$ and the system lags behind
equilibrium:
\begin{equation}
  \vrho(t)
  = \pieq(t) + \sum_{n\geq 1}\tilde{\rho}_n(t) \bm{r}_n (t),
  \quad \tilde{\rho}_n \neq 0.
\label{eq:lag}
\end{equation}

The scalar-valued version of Eq.~\eqref{eq:lag} was introduced in \citep{Iram2021}. On adding the matrix-valued CD control terms $\mathbf{L}_{\rm CD}$, one exactly cancels $\Gamma$
in Eq.~\eqref{eq:tilde_rho_dynamics} which hinders the adiabatic tracking of the instantaneous equilibrium distribution. This yields the following evolution in the adiabatic frame:
\begin{equation}
  \frac{\partial \tilde{\rho}_n}{\partial t} = \lambda_n\,\tilde{\rho}_n,
  \quad n \geq 1.
\label{eq:cd_dynamics}
\end{equation}
For a system initialized in equilibrium,
$\vrho(0) = \pieq(0)$, Eq.~\eqref{eq:cd_dynamics} then indicates that $\tilde{\rho}_n(t) = 0$ for all $t \in [0, \tau]$, and therefore:
\begin{equation}
  \vrho(t) = \pieq(t), \quad \forall\, t \in [0, \tau].
\label{eq:perfect_tracking}
\end{equation}
Under finite-speed bare driving, $\tilde{\rho}_{n\geq1}(t) \neq 0$
develops as the protocol pumps amplitude into the relaxation modes;
the counterdiabatic correction cancels this pumping exactly,
preserving $\tilde{\rho}_{n\geq1}(t) = 0$ for all $t \in [0, \tau]$. Thus, the CD correction preserves the initial condition
$\tilde{\vrho} = (1,0,\ldots,0)^T$ exactly, ensuring
perfect tracking of the instantaneous equilibrium at
arbitrary driving speed.
\medskip 

\textbf{Probability conservation of $\Lop_{\mathrm{CD}}(t)$.} Since $\bm{\ell}^T_0 = \bbone^T$ and $\bbone^T \bm{r}_n = \bm{\ell}_0 \bm{r}_n = \delta_{0n} = 0$ for
$n\neq 0$, every column sum of $\Lop_{\mathrm{CD}}$ vanishes:
$\bbone^T \Lop_{\mathrm{CD}} = \bm{0}^T$.
The corrected generator $\Lop_{\mathrm{eff}} := \Lop+\Lop_{\mathrm{CD}}$. Thus, it is a probability  conserving quantity. We emphasize, however, that $\Lop_{\mathrm{eff}}$ is \emph{not} a stochastic rate matrix: $\Lop_{\mathrm{CD}}=(\partial \pieq/\partial t)\mathbf{1}^{T}$ is dense with sign-indefinite off-diagonal entries, so $ \Lop_{\mathrm{eff}}$ generally has
negative off-diagonals.

\subsection{Magnus fourth-order integrator}\label{sec:magnus_fourth}
Time integration of the modified Fokker--Planck equation
\begin{equation}
  \frac{d\bm{\rho}(t)}{dt}
  = \bigl[\Lop(t) + \Lop_{\rm CD}(t)\bigr]\bm{\rho}(t),
\label{eq:ode}
\end{equation}
requires a method that respects the structure of the generator at
each step.
The standard midpoint (Euler--exponential) scheme
$\boldsymbol{\rho}(t+\Delta t)
= e^{\mathcal{G}(t+\Delta t/2)\Delta t}\boldsymbol{\rho}(t)$,
where $\mathcal{G} = \Lop + \Lop_{\rm CD}$,
is accurate to $\mathcal{O}(\Delta t^2)$ but becomes insufficient
when the generator varies rapidly within a single step, as is the
case near the midpoint of a fast protocol where
$\Lop_{\rm CD}(t)$ is largest.
We therefore adopt the \emph{fourth-order Magnus integrator}~\cite{Magnus1954,Blanes2009},
which achieves $\mathcal{O}(\Delta t^4)$ accuracy while
preserving probability normalisation exactly at each step.

The Magnus integrator advances $\boldsymbol{\rho}$ from $t$ to
$t+\Delta t$ via
\begin{equation}
  \boldsymbol{\rho}(t+\Delta t) = e^{\Omega(t,\,t+\Delta t)}\,\boldsymbol{\rho}(t),
\label{eq:magnus_step}
\end{equation}
where the Magnus exponent is approximated using the two
Gauss--Legendre quadrature nodes
\begin{equation}
  t_{1,2} = t + \left(\frac{1}{2} \mp \frac{\sqrt{3}}{6}\right)\Delta t
\label{eq:gl_nodes}
\end{equation}
as
\begin{equation}
  \Omega(t,t+\Delta t)
  = \frac{\Delta t}{2}\bigl(\mathcal{G}_1 + \mathcal{G}_2\bigr)
  + \frac{\sqrt{3}\,\Delta t^2}{12}
    \bigl[\mathcal{G}_2,\,\mathcal{G}_1\bigr],
\label{eq:magnus_omega}
\end{equation}
with $\mathcal{G}_k = \Lop(t_k) + \Lop_{\rm CD}(t_k)$
evaluated at the quadrature nodes, and
$[\mathcal{G}_2,\mathcal{G}_1]
= \mathcal{G}_2\mathcal{G}_1 - \mathcal{G}_1\mathcal{G}_2$
the matrix commutator.
The matrix exponential $e^\Omega$ is computed at each step via
the Pad\'e approximation~\cite{doi:10.1137/04061101X, baker1996padé}, as implemented in \texttt{scipy.linalg.expm}. The commutator term in Eq.~\eqref{eq:magnus_omega} captures the leading non-commutativity of the generator between $t_1$ and $t_2$,
which is the dominant source of error in the midpoint scheme.
For a slowly varying protocol the commutator is small and
Eq.~\eqref{eq:magnus_omega} reduces to the midpoint scheme;
for fast protocols, where $\Lop_{\rm CD}$ changes
appreciably within a step, the commutator correction is essential.
\subsection{Analytic evolution for the harmonic trap}\label{sec:analytic_evol}
For the harmonic potential $V(x,\kappa_0(t)) = \tfrac{1}{2}\kappa_0(t)x^2$,
the Fokker--Planck equation admits an exact reduction that
bypasses the matrix-valued Liouville evolution entirely.
Since a Gaussian initial condition remains Gaussian under harmonic
confinement, we write
\begin{equation}
  \rho(x,t)
  = \sqrt{\frac{\alpha(t)}{\pi}}\,e^{-\alpha(t)x^2},
\label{eq:gaussian_ansatz}
\end{equation}
where $\alpha(t) > 0$ is the time-dependent precision parameter.
Substituting Eq.~\eqref{eq:gaussian_ansatz} directly into the
Fokker--Planck equation $\frac{\partial \rho}{\partial t} = \frac{\partial (\kappa x \rho)}{\partial x} + \frac{1}{\beta \gamma} \frac{\partial^2 \rho}{\partial x^2}$ and requiring the equality to hold for
all $x$ yields the scalar-valued Riccati equation
\begin{equation}
  \dot{\alpha}(t)
  = -\frac{2\kappa(t)}{\gamma}\,\alpha(t)
  - \frac{4\alpha^2(t)}{\beta \gamma},
\label{eq:alpha_riccati}
\end{equation}
with initial condition $\alpha(0) = \beta\kappa_i/2$,
corresponding to the equilibrium distribution for the initial
stiffness $\kappa_i$.
For the bare protocol $\kappa(t) = \kappa_0(t)$, the solution
$\alpha(t)$ lags behind the instantaneous equilibrium target
$\alpha_{\rm eq}(t) = \beta\kappa_0(t)/2$.
For the analytic counterdiabatic protocol of
Ref.~\citep{Martínez2016, Patra2017}, replacing $\kappa_0(t)$ with
$\kappa_{\rm CD}(t) = \kappa_0(t) +
\gamma\dot\kappa_0(t)/(2\kappa_0(t))$
enforces $\alpha(t) = \alpha_{\rm eq}(t)$ exactly for all
$t\in[0,\tau]$.
Equation~\eqref{eq:alpha_riccati} is integrated using a
high-order adaptive solver
(\texttt{scipy.solve\_ivp}, \texttt{DOP853},
$\mathrm{rtol}=10^{-12}$, $\mathrm{atol}=10^{-14}$),
with dense output evaluated at every reporting time.
This approach carries no spatial discretisation error: $\rho(x,t)$
is represented analytically via $\alpha(t)$ at every instant and
projected onto the spatial grid $\{x_i\}$ only when evaluating
scalar diagnostics such as the TVD, KL divergence, and dissipated
work.
The residual $|\alpha(t) - \alpha_{\rm eq}(t)|$ reaches at most
$\sim 10^{-11}$, consistent with the ODE solver tolerance, and
is entirely free of the $\mathcal{O}(\Delta x^2)$
discretisation error that affects the matrix-valued spectral LCD
on a finite grid. The harmonic trap therefore serves as a clean benchmark: the
analytic reference isolates the $\mathcal{O}(\Delta x^2)$ grid
error in the spectral CD condition and the $\mathcal{O}(\Delta t^2)$
heat-quadrature error in the dissipated work, neither of which
is present in the scalar evolution.

\subsection{Numerical evaluation of excess dissipated work}\label{sec:simpson}
At each Magnus step $t_k$ (Appendix~\ref{sec:magnus_fourth}), the
instantaneous power is recorded as
\begin{equation}
  P_k = \dot{\zeta}(t_k)\sum_{i=1}^N
        \frac{\partial V(x_i,\zeta_{t_k})}{\partial\zeta}\,
        \rho_i(t_k),
\label{eq:power_discrete}
\end{equation}
where $\rho_i(t_k)$ is the $i$-th component of the actual
Magnus4-evolved distribution at time $t_k$.
The cumulative work $\mathcal{W}(t_k)$ is obtained by integrating the
discrete power series $\{P_k\}$ using Simpson's rule,
\begin{equation}
  \mathcal{W}(t_{2m})
  = \sum_{j=0}^{m-1}
    \frac{\Delta t}{3}
    \bigl(P_{2j} + 4P_{2j+1} + P_{2j+2}\bigr),
\label{eq:simpson}
\end{equation}
which is accurate to $\mathcal{O}(\Delta t^4)$.
This is essential: a lower-order quadrature such as the
midpoint or trapezoidal rule ($\mathcal{O}(\Delta t^2)$)
introduces a spurious residual in $\mathcal{W}_{\rm diss}$
that dominates the genuine tracking error and obscures the
escorting equality. With the fourth-order Simpson integration, $|\mathcal{W}_{\rm diss}(t)|$
converges to much lower errors, consistent with the $\mathcal{O}(\Delta t^4)$ Magnus integrator accuracy and the independently measured TVD, consistently in the range of $\sim 10^{-13}-10^{-12}$, confirming the trajectory-wise escorting equality at a level set by the time-discretisation of the dynamics rather than by any deficiency of the counterdiabatic
construction.

\section{Additional experimental results}
\subsection{Free energy difference for the driven harmonic trap}
\label{sec:free_energy_estimates}

For the harmonic potential $V(x,t) = \tfrac{1}{2}\kappa(t) x^2$,
the canonical partition function at inverse temperature $\beta$
is the Gaussian integral
\begin{equation}
  \mathcal{Z}(\kappa)
  = \int_{-\infty}^{\infty} e^{-\beta\kappa x^2/2}\,\mathrm{d}x
  = \sqrt{\frac{2\pi}{\beta\kappa}}\,.
\label{eq:Z_harmonic}
\end{equation}
The Helmholtz free energy follows as
\begin{equation}
  \mathcal{F}(\kappa)
  = -\beta^{-1}\ln \mathcal{Z}(\kappa)
  = \frac{1}{2\beta}\ln\!\left(\frac{\beta\kappa}{2\pi}\right).
\label{eq:F_harmonic}
\end{equation}
The free energy difference between the final stiffness
$\kappa_f$ and the initial stiffness $\kappa_i$ is therefore
\begin{equation}
  \Delta\mathcal{F}
  = \mathcal{F}(\kappa_f) - \mathcal{F}(\kappa_i)
  = \frac{1}{2\beta}\ln\!\left(\frac{\kappa_f}{\kappa_i}\right),
\label{eq:DF_harmonic}
\end{equation}
where the $2\pi$ factors cancel exactly.
For the parameters used throughout this work,
$\kappa_i = 1$, $\kappa_f = 4$, and $\beta = 1$, this gives the free-energy difference value to be:
\begin{equation}
  \Delta\mathcal{F}
  = \tfrac{1}{2}\ln 4 = \ln 2 \approx 0.6931\,.
\label{eq:DF_value}
\end{equation}

In the numerical simulations, the free energy is computed from
the discrete partition function on the spatial grid
$\{x_i\}_{i=1}^N$,
\begin{equation}
  \mathcal{Z}_{\rm d}(\kappa)
  = \sum_{i=1}^{N} e^{-\beta\kappa x_i^2/2}\,,
\label{eq:Z_discrete}
\end{equation}
with $\mathcal{F}_{\rm d}(\kappa) = -\beta^{-1}\ln \mathcal{Z}_{\rm d}(\kappa)$.
For our grid of $N = 80$ points on $q\in[-4,\,4]$, the
discrete free energy difference
$\Delta\mathcal{F}_{\rm d}
= \mathcal{F}_{\rm d}(\kappa_f) - \mathcal{F}_{\rm d}(\kappa_i) = 0.6931$
agrees with the continuum result~\eqref{eq:DF_value} to
$\sim 5\times 10^{-5}$, confirming that the spatial
discretisation introduces negligible error in the
thermodynamic reference.

\begin{table*}[!tbhp]
\caption{\textbf{Comparison of terminal work and dissipation metrics across varying protocol durations $\tau$.} For the system driven by the harmonic potential $V_{\rm h}(x, \zeta_t)$, the metrics contrast the uncontrolled dynamics (bare) against the exact analytical counterdiabatic drive (analytic) and our numerical Liouvillian framework (LCD). All values are reported relative to the exact equilibrium free energy change $\Delta \mathcal{F} = 0.6931$.}
\begin{ruledtabular}
\begin{tabular}{cccccccc}
$\tau$ & $\mathcal{W}(\tau)$ & $\mathcal{W}^{\rm analytic}(\tau)$ & $\mathcal{W}^{\rm LCD}(\tau)$ & $\Delta \mathcal{F}$ & $\mathcal{W}_{\rm diss}(\tau)$ & $\mathcal{W}_{\rm diss}^{\rm analytic}(\tau)$ & $\mathcal{W}_{\rm diss}^{\rm LCD}(\tau)$ \\ 
\hline
\rule{0pt}{3ex}%
0.001 & 1.4977 & 0.6931 & 0.6931 & 0.6931 & 0.8046 & $4.005 \times 10^{-13}$  & $-6.867 \times 10^{-12}$ \\
0.003 & 1.4958 & 0.6931 & 0.6931 & 0.6931 & 0.8027 & $2.265 \times 10^{-14}$  & $-5.163 \times 10^{-12}$ \\
0.010 & 1.4890 & 0.6931 & 0.6931 & 0.6931 & 0.7959 & $5.818 \times 10^{-13}$  & $-5.701 \times 10^{-12}$ \\
0.030 & 1.4702 & 0.6931 & 0.6931 & 0.6931 & 0.7771 & $2.898 \times 10^{-14}$  & $-4.944 \times 10^{-12}$ \\
0.100 & 1.4097 & 0.6931 & 0.6931 & 0.6931 & 0.7166 & $5.933 \times 10^{-13}$  & $-5.149 \times 10^{-12}$ \\
0.300 & 1.2727 & 0.6931 & 0.6931 & 0.6931 & 0.5796 & $3.186 \times 10^{-13}$  & $-3.715 \times 10^{-12}$ \\
1.000 & 1.0215 & 0.6931 & 0.6931 & 0.6931 & 0.3284 & $-5.635 \times 10^{-13}$ & $-2.268 \times 10^{-12}$ \\
3.000 & 0.8274 & 0.6931 & 0.6931 & 0.6931 & 0.1343 & $5.285 \times 10^{-14}$  & $-8.511 \times 10^{-13}$ \\
10.00 & 0.7339 & 0.6931 & 0.6931 & 0.6931 & 0.0408 & $-2.345 \times 10^{-13}$ & $-2.666 \times 10^{-13}$ \\
\end{tabular}
\end{ruledtabular}
\label{tab:work_metrics_harmonic}
\end{table*}

Table~\ref{tab:work_metrics_harmonic} reports the terminal work
$\mathcal{W}(\tau)$ and dissipated work
$\mathcal{W}_{\rm diss}(\tau)
= \mathcal{W}(\tau) - \Delta\mathcal{F}$
for protocol durations spanning four decades,
$\tau \in [10^{-3},\,10]$.
Under bare dynamics, the work $\mathcal{W}(\tau)$
exceeds $\Delta\mathcal{F}$ at all finite $\tau$, with
$\mathcal{W}_{\rm diss}$ ranging from $0.80$
($\tau = 10^{-3}$, strongly irreversible) to $0.04$
($\tau = 10$, near quasi-static), consistent with the
expected $\mathcal{W}_{\rm diss} \sim 1/\tau$ scaling in
linear response.
Both the analytical CD and the spectral LCD yield
$\mathcal{W}(\tau) = \Delta\mathcal{F}$ to within
$\sim\!\mathcal{O}(10^{-12})$ across all protocol durations,
with residuals attributable to the ODE solver tolerance
(analytic) and the fourth-order Simpson work
quadrature (LCD).
This confirms that the escorting condition
$\mathcal{W}_{\rm diss}(\tau) \approx 0$ holds irrespective
of driving speed, eliminating the dissipative bias in the
free energy estimate without statistical sampling.

\section{Stress test: the quartic coalescence model}\label{sec:quartic_fehler}
The double-well and harmonic-trap systems studied above are both
\emph{well-conditioned}: the biorthogonal decomposition underlying the
spectral LCD generator remains numerically stable throughout the
protocol, and the resulting LCD driven density tracks the
instantaneous equilibrium distribution to within machine precision. To probe the
limits of this construction, we consider a deliberately adversarial
model in which the slowest relaxation mode's spectral gap closes
\emph{exponentially} over most of the protocol.

\subsection{Model and protocol}

The quartic coalescence potential is
\begin{equation}
  V(x,\zeta) \;=\; x^4 - 16(1-\zeta)\,x^2 ,
  \qquad \zeta(t) = \min(t/\tau,\,1) ,
  \label{eq:quartic-potential}
\end{equation}
annealed linearly from $\zeta=0$ to $\zeta=1$ over duration $\tau$, at
inverse temperature $\beta=1$. The discrete generator $\Lop(t)$ is
built on a grid of $N=80$ points on $x\in[-4.5,4.5]$ via the same
detailed-balance-preserving (Sasa--Tasaki-type) discretization~\citep{Sasa2006-gc} used
throughout this work, Eq.~\eqref{eq:rates} .

\begin{figure}[!tbhp]
  \centering
  \includegraphics[width=0.90\linewidth]{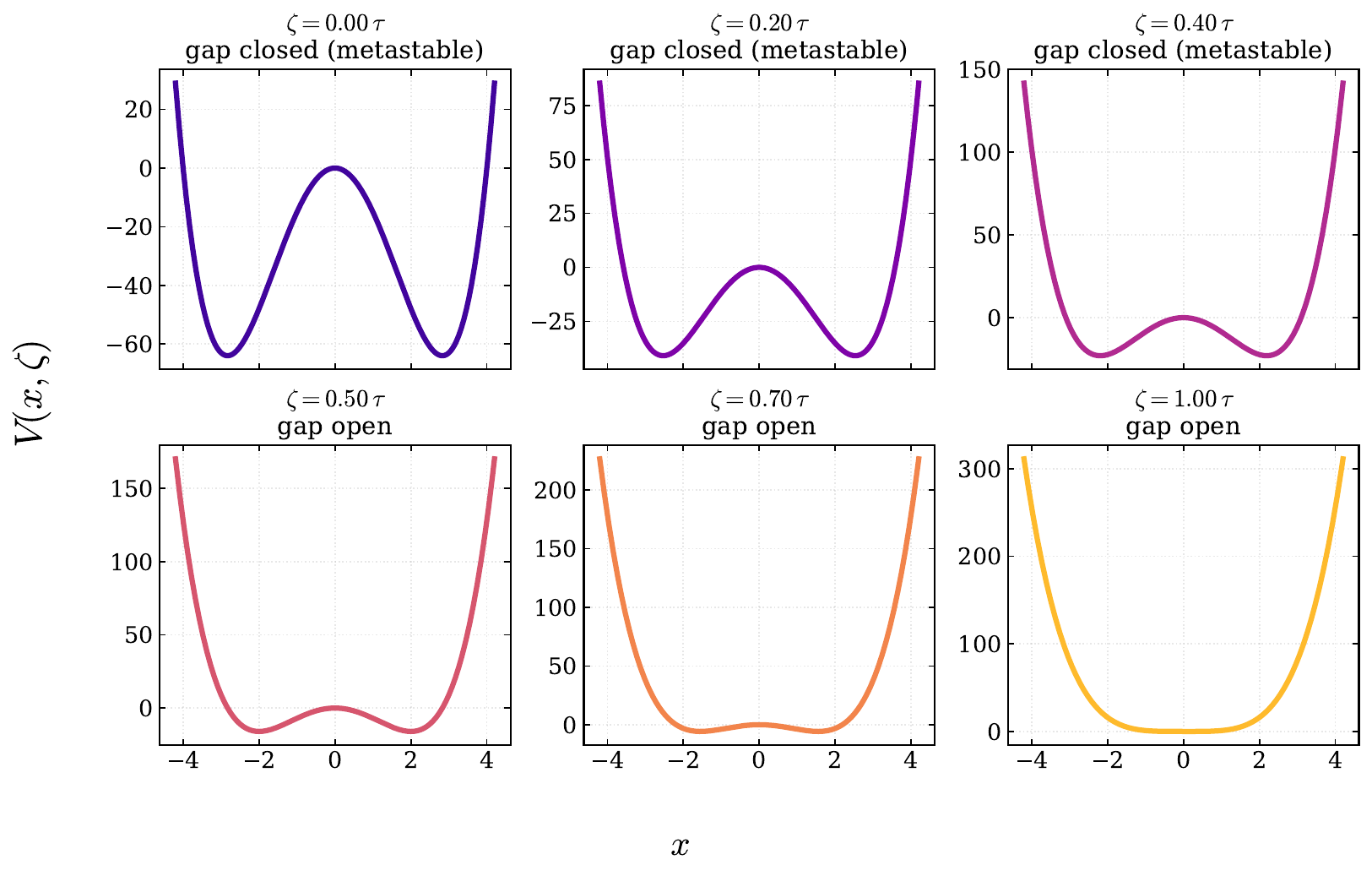}
  \caption{\textbf{The quartic coalescence potential:} The potential $V(x,\zeta)$ depicted at six points
  along the anneal. For $\zeta<0.5$ (top row) the potential is a deep,
  symmetric double well; for $\zeta\ge0.5$ (bottom row) the barrier has
  dropped below $\Delta V_b=16$ and continues to collapse smoothly into
  the single quartic well at $\zeta=1$.}
  \label{fig:quartic-potential-snapshots}
\end{figure}

At $\zeta=0$, Eq.~\eqref{eq:quartic-potential} is a symmetric double
well with minima at $x=\pm\sqrt{8}$ and a central barrier of height
\begin{equation}
  \Delta V_b(\zeta) \;=\; V(0,\zeta) - V(x_{\min},\zeta) \;=\; 64\,(1-\zeta)^2 ,
  \label{eq:barrier-height}
\end{equation}
so $\Delta V_b(0)=64$. As $\zeta\to1$ the two minima coalesce
(hence \emph{coalescence model}) into the single well $V(x,1)=x^4$,
with $\Delta V_b(1)=0$. Figure~\ref{fig:quartic-potential-snapshots}
shows the potential at six representative points spanning this
transition, split explicitly at $\zeta=0.5$ (where $\Delta V_b=16$).

\subsection{Exponentially closing spectral gap}

The slowest non-trivial eigenvalue $\lambda_1(\zeta_t)$ of $\Lop(t)$
governs the inter-well (symmetric $\leftrightarrow$ antisymmetric)
relaxation rate of the double well. For an overdamped, symmetric
double-well potential, Kramers/Eyring theory~\citep{10.1039/TF9353100875, Laidler1983} predicts this rate is
exponentially suppressed by the barrier height,
\begin{equation}
  |\lambda_1(\zeta)| \;\sim\; C(\zeta)\,\exp\!\big[-\beta\,\Delta V_b(\zeta_t)\big]
  \;=\; C(\zeta)\,\exp\!\big[-64(1-\zeta)^2\big] ,
  \label{eq:kramers}
\end{equation}
with $C(\zeta)$ a slowly-varying prefactor set by the local curvature
at the well and barrier. We verified Eq.~\eqref{eq:kramers} directly
against the numerically diagonalized $\Lop(t)$: the exponential decay
$\exp[-64(1-\zeta)^2]$ reproduces the computed $|\lambda_1(\zeta)|$ to
within a prefactor $C(\zeta)\in[2,7]$ across the entire range
$\zeta\in[0.5,1]$ (Table~\ref{tab:kramers-check}). Below
$t/\tau \approx0.33$--$0.4$, $\Delta V_b\gtrsim36$ and
$\exp(-\Delta V_b)\lesssim10^{-16}$ falls below double-precision
machine epsilon: $\lambda_1(\zeta)$ is no longer numerically resolvable
and the dense eigensolver returns a value dominated by floating-point
round-off (the noisy plateau at $1/|\lambda_1|\sim10^{12\text{--}13}$
visible in Fig.~\ref{fig:quartic-spectral-gap} for
$t/\tau\lesssim0.33$), rather than the true (exponentially smaller)
physical gap.

\begin{table*}[!tbhp]
\caption{\textbf{Verification of the Kramers-type scaling.} Numerical corroboration of Eq.~\eqref{eq:kramers}. The table monitors the first non-vanishing relaxation eigenvalue $|\lambda_1(\zeta)|$ against the classical barrier-height scaling prediction $\exp[-\Delta V_b(\zeta)]$ across varying parameter configurations $\zeta$.}
\begin{ruledtabular}
\begin{tabular}{ccccc}
$\zeta$ & $\Delta V_b(\zeta)$ & $|\lambda_1(\zeta)|$ (numeric) & $e^{-\Delta V_b}$ & ratio \\ 
\hline
\rule{0pt}{3ex}%
0.50 & 16.11 & $7.30 \times 10^{-7}$ & $1.01 \times 10^{-7}$ & 7.2 \\
0.60 & 10.31 & $1.89 \times 10^{-4}$ & $3.33 \times 10^{-5}$ & 5.7 \\
0.70 & 5.80  & $1.23 \times 10^{-2}$ & $3.03 \times 10^{-3}$ & 4.1 \\
0.80 & 2.58  & $2.00 \times 10^{-1}$ & $7.60 \times 10^{-2}$ & 2.6 \\
0.90 & 0.64  & 1.04                  & $5.25 \times 10^{-1}$ & 2.0 \\
1.00 & 0.00  & 2.72                  & 1.00                  & 2.7 \\
\end{tabular}
\end{ruledtabular}
\label{tab:kramers-check}
\end{table*}

\subsection{Consequence for the counterdiabatic generator}\label{sec:counterdiabatic_generator_quartic}

The spectral LCD generator weights each mode by
$c_n(t) = \dot{\zeta}_t \bm{\ell}_n (\partial \Lop / \partial \zeta) \bm{r}_0 / \lambda_n$. Because $\lambda_1(\zeta)$ is
exponentially small for $\zeta\lesssim0.5$, $c_1$ is exponentially
\emph{large} exactly where the counterdiabatic correction is most
needed to suppress inter-well leakage -- and the biorthogonal
decomposition itself becomes numerically ill-conditioned in the same
regime ($\mathrm{cond}(R)$ reaches $\sim10^{12}$ near $\zeta=0$). This
is a fundamentally different obstruction from the truncation-order
question studied for the double well and harmonic trap
(Section~\ref{sec:spectral_truncation}): it is not that more modes are needed for
convergence, but that the \emph{dominant} mode's own coefficient is
unreliable.

Table~\ref{tab:quartic-results} confirms the practical consequence.
Unlike the double well and harmonic trap, where the LCD generator
suppresses $\mathcal{W}_{\rm diss}$ by $9$--$12$ orders of magnitude relative to
bare Fokker--Planck dynamics, here it achieves only a modest,
$\tau$-dependent improvement ($\sim$33--1500$\times$): $\mathcal{W}_{\rm diss}$, rather than collapsing to the $\sim10^{-12}$
floor seen elsewhere in this work. The exact equilibrium free-energy
difference,
\begin{equation}
  \Delta \mathcal{F} \;=\; \mathcal{F}(\zeta{=}1) - \mathcal{F}(\zeta{=}0) \;=\; 62.9407... ,
  \label{eq:quartic-DF}
\end{equation}
computed directly from the discrete partition functions
\begin{equation}
  \mathcal{F}(\zeta) \;=\; -\beta^{-1}\ln\sum_i e^{-\beta V(x_i,\zeta)} ,
\end{equation}
is in agreement with the reported reference value
$\Delta \mathcal{F} = 62.9407...$~\cite{oberhofer2005biased}.

\begin{table*}[!tbhp]
 \caption{\textbf{Quartic coalescence model.} Work, dissipation at the
  terminal time step $t/\tau = 1.0$. Comparison of bare vs.\ spectral LCD, for $\Delta \mathcal{F} = 62.9407...$
  ($\beta=1$, $N=80$; $\Delta t=10^{-3}$ for $\tau \in [0.1, 0.5]$.}
\begin{ruledtabular}
\begin{tabular}{cccccccc}
$\tau$ & $\mathcal{W}(\tau)$ & $\mathcal{W}^{\rm LCD}(\tau)$ & $\Delta \mathcal{F}$ & $\mathcal{W}_{\rm diss}(\tau)$ & $\mathcal{W}_{\rm diss}^{\rm LCD}(\tau)$ & improvement \\ 
\hline
\rule{0pt}{3ex}%
0.10 & 86.48  & 63.66 & 62.94 & 23.54 & $0.71$ &  $33\times$ \\
0.20 & 78.87  & 63.03 & 62.94 & 15.92 & $0.09$ &  $176\times$ \\
0.50 & 71.48  & 62.95 & 62.94 & 8.54  & $0.006$ &  $1423\times$ \\
\end{tabular}
\end{ruledtabular}
\label{tab:quartic-results}
\end{table*}

We include this model precisely because it fails gracefully: it
delineates the regime -- deep, near-degenerate metastability, with a
barrier large compared to $k_BT$ over a substantial fraction of the
protocol -- in which the spectral biorthogonal construction should
\emph{not} be expected to deliver near-machine-precision counterdiabatic
driving, in contrast to the double-well and harmonic-trap results of
Sections~\ref{sec:metrics_and_fidelity}--\ref{sec:traj_dissipated}.
\medskip 

It is essential to distinguish the numerical breakdown of the spectral \emph{construction} from a genuine divergence of counterdiabatic driving. For the symmetric coalescence potential, reflection symmetry $\mathcal{P}\Lop\mathcal{P}=\Lop$ [where $(\mathcal{P}f)_i=f_{N-1-i}$] dictates that the stationary state $\mathbf{r}_0 = \pieq$ and the parametric perturbation $\partial \Lop/ \partial \zeta$ are even, whereas the slowest relaxation mode $\mathbf{r}_1$ is odd. Consequently, the transition matrix element $ \boldsymbol{\ell}_1^T(\partial \Lop/\partial \zeta)\mathbf{r}_0$ vanishes identically by parity: escorting a symmetric $\pieq$ never requires transporting probability across the central barrier. Where the eigenvalues are well-separated [e.g., at $\zeta=0.8$ with $|\lambda_1|\approx0.20$], we confirm this parity protection numerically, observing an odd-to-even overlap suppression of $|\boldsymbol{\ell}_1^T(\partial \Lop/ \partial \zeta)\mathbf{r}_0|/|\boldsymbol{\ell}_2^T(\partial \Lop/ \partial \zeta)\mathbf{r}_0|\sim10^{-11}$. However, as the barrier height grows ($\zeta\lesssim0.4$, $\Delta V_b\gtrsim36$), the Kramers gap $|\lambda_1|\sim e^{-\beta\Delta V_b}$ drops below the double-precision machine floor ($\sim10^{-12}$--$10^{-13}$), rendering the eigenvectors severely ill-conditioned [$\mathrm{cond}(\mathbf{D})>10^8$]. In this regime, numerical eigensolvers fail to resolve near-degenerate eigenvalues into definite-parity eigenvectors; the protective zero-overlap is lost to floating-point round-off, and the assembled modal coefficient $c_1=\boldsymbol{\ell}_1^T(\partial \Lop/ \partial \zeta)\mathbf{r}_0/\lambda_1$ degenerates into an unphysical $0/0$ numerical quotient that corrupts the spectral reconstruction of $\Lop_{\mathrm{CD}}$ [Eq.~\eqref{eq:ACD_spectral}].

By contrast, the numerics for the \emph{exact} counterdiabatic operator remains completely bounded throughout the protocol. Because the closed-form rank-one representation $\Lop_{\mathrm{CD}}=(\partial \pieq/\partial t)\mathbf{1}^T$ [Eq.~\eqref{eq:LCD-rank1}] bypasses eigenbasis inversion entirely, it never constructs $c_1$, maintains an $\mathcal{O}(1)$ operator norm, and tracks $\boldsymbol{\pi}_{\mathrm{eq}}(t)$ to a fidelity of $\mathrm{TVD}\sim10^{-6}$ (limited solely by spatial grid resolution and integrator tolerance). This confirms that the barrier-closure failure is an artifact of the spectral representation, not a physical limit on control. A genuine divergence of the classical counterdiabatic term—the true statistical analogue of a quantum level crossing—arises only when the target distribution itself must move rapidly across a closing gap, such as during an asymmetric barrier crossing or a first-order bifurcation of $\pieq(t)$. In those asymmetric geometries, $\|\partial \pieq/\partial t\|$ grows without bound. Thus, unlike the quantum adiabatic gauge potential, whose norm diverges universally whenever an energy gap closes, the classical Liouvillian control diverges only when probability must be actively transported through that closure.

\subsection{Spectral truncation as imperfect escorting}\label{sec:spectral_truncation}
\begin{figure*}[!tbhp]
    \centering
    \begin{subfigure}[t]{0.49\textwidth}
        \centering
        \includegraphics[width=\textwidth]{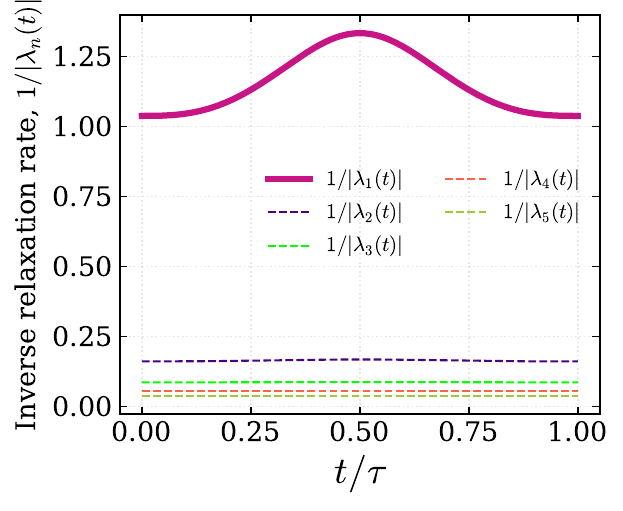}
        \caption{Inverse spectral gap magnitude of an overdamped FP equation driven by a double-well potential for $\tau = 0.1$.}
        \label{fig:inv_gap_dw}
    \end{subfigure}
    \hfill
    \begin{subfigure}[t]{0.49\textwidth}
        \centering
        \includegraphics[width=\textwidth]{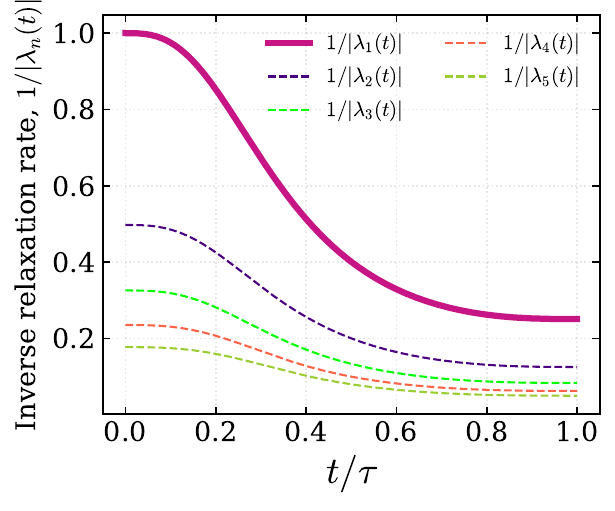}
        \caption{Inverse spectral gap magnitude of an overdamped FP equation driven by a harmonic-trap potential for $\tau = 0.1$.}
        \label{fig:inv_gap_harm}
    \end{subfigure}

    \caption{\textbf{Temporal evolution of the inverse spectral gap.} 
    The inverse spectral gap $\Delta_n^{-1}(t) \equiv |\lambda_0(t) - \lambda_n(t)|^{-1}$ computed between the ground state (instantaneous equilibrium distribution) zero-mode ($\lambda_0 = 0$) and the first few dominant biorthogonal eigenvalues $\lambda_n(t)$ as a function of time $t$ for (a) the double-well potential and (b) the harmonic trap landscape. The trajectories map the instantaneous relaxation timescales governing the non-symmetric transition rate matrix throughout the driving protocol.}
    \label{fig:spectral_gap_comparison}
\end{figure*}

The counterdiabatic correction $\Lop_{\rm CD}(t)$ [Eq.~\eqref{eq:ACD_spectral}] is a sum over all $N-1$
non-zero relaxation modes.
In practice, the sum can be truncated to the $M$ slowest
modes, $\Lop_{\rm CD}^{(M)}(t)
  = -\sum_{n=1}^{M}
    \frac{\boldsymbol{\ell}_n^T(t)\,
          (\partial\Lop/\partial t)\,
          \mathbf{r}_0(t)}
         {\lambda_n(t)}\;
    \mathbf{r}_n(t)\,\boldsymbol{\ell}_0^T,$
yielding an \emph{imperfect} but systematically improvable
escorting: the driven distribution no longer tracks
$\boldsymbol{\pi}_{\rm eq}(t)$ exactly, but the tracking
error decreases monotonically as $M$ increases, interpolating
between the uncontrolled bare dynamics ($M=0$) and the full
LCD ($M=N-1$). The effectiveness of this truncation rests on the spectral
structure of $\Lop(t)$.
Since the relaxation rates $|\lambda_n|$ grow with mode
index (Figure~\ref{fig:spectral_gap_comparison}), the
coefficients $c_n \sim |\lambda_n|^{-1}$ decay, and the
dominant contribution comes from the modes whose relaxation
timescales $|\lambda_n|^{-1}$ are comparable to or longer
than the driving timescale $\tau$.
The number of modes required for a given accuracy therefore
depends on the protocol speed: for slow protocols fewer modes suffice to suppress
the TVD by several orders of magnitude; for faster protocols modes with shorter relaxation timescales
contribute more significantly, and $M \sim (N-1)/2$ is
needed to achieve comparable improvement.
\begin{table*}[!tbhp]
\caption{\textbf{Truncated Liouvillian counterdiabatic driving.}
  Maximum $\text{TVD}$, maximum $\mathcal{D}_{\text{KL}}$, and maximum absolute intermediate dissipated work $\max|\mathcal{W}_{\text{diss}}(t)|$ evaluated for several truncation thresholds $M$ across the double-well and harmonic trap potentials at a fixed driving period $\tau = 0.1$ ($\Delta t = 10^{-6}$).}
\label{tab:truncation_analysis}
\begin{ruledtabular}
\begin{tabular}{l ccc c ccc}
 & \multicolumn{3}{c}{\textbf{Double Well}} & & \multicolumn{3}{c}{\textbf{Harmonic Trap}} \\
\cline{2-4} \cline{6-8}
\rule{0pt}{3ex}
Truncation ($M$) & $\max$ TVD & $\max \mathcal{D}_{\text{KL}}$ & $\max|\mathcal{W}_{\text{diss}}(t)|$ & & Max TVD & Max $D_{\text{KL}}$ & $\max|\mathcal{W}_{\text{diss}}(t)|$ \\
\hline
\rule{0pt}{3ex}
 5 & $2.66\times10^{-3}$ & $3.33\times10^{-5}$ & $1.34\times10^{-4}$ & & $2.07\times10^{-4}$ & $1.17\times10^{-4}$ & $5.69\times10^{-5}$ \\
10 & $7.44\times10^{-5}$ & $2.71\times10^{-8}$ & $1.97\times10^{-7}$ & & $6.56\times10^{-5}$ & $9.01\times10^{-6}$ & $2.34\times10^{-5}$ \\
15 & $1.35\times10^{-6}$ & $1.10\times10^{-11}$ & $1.96\times10^{-10}$ & & $3.21\times10^{-5}$ & $2.50\times10^{-6}$ & $6.10\times10^{-6}$ \\
20 & $1.39\times10^{-7}$ & $1.16\times10^{-13}$ & $5.42\times10^{-12}$ & & $1.32\times10^{-5}$ & $2.64\times10^{-7}$ & $7.28\times10^{-7}$ \\
25 & $7.77\times10^{-9}$ & $8.54\times10^{-16}$ & $2.86\times10^{-12}$ & & $7.86\times10^{-6}$ & $5.04\times10^{-8}$ & $1.61\times10^{-7}$ \\
30 & $1.35\times10^{-9}$ & $4.53\times10^{-16}$ & $2.85\times10^{-12}$ & & $3.97\times10^{-6}$ & $3.74\times10^{-9}$ & $1.41\times10^{-8}$ \\
35 & $1.53\times10^{-10}$ & $4.54\times10^{-16}$ & $2.84\times10^{-12}$ & & $2.64\times10^{-6}$ & $5.94\times10^{-10}$ & $3.71\times10^{-9}$ \\
Full ($M=79$) & $2.14\times10^{-12}$ & $3.88\times10^{-16}$ & $2.85\times10^{-12}$ & & $1.95\times10^{-12}$ & $4.26\times10^{-16}$ & $5.15\times10^{-12}$ \\
\end{tabular}
\end{ruledtabular}
\end{table*}

\begin{table*}[!tbhp]
\caption{\textbf{Spectral truncation convergence metrics for the time-varying DW potential.} performance for fast ($\tau = 0.01$) and slow ($\tau = 0.50$) protocol durations. Errors drop systematically as the number of retained modes $n_{\rm trunc}$ approaches the full discrete grid size ($N = 79$).}
\begin{ruledtabular}
\begin{tabular}{ccccc}
$\tau$ & Truncation $(M)$ & Max TVD & Max $\mathcal{D}_{\rm KL}$ & Max $\vert\mathcal{W}_{\rm diss}(t)\vert$ \\ 
\hline
\rule{0pt}{3ex}\textbf{0.01} & 5  & $7.139 \times 10^{-3}$ & $4.686 \times 10^{-4}$ & $2.697 \times 10^{-4}$ \\
($dt=10^{-6}$)             & 10 & $2.788 \times 10^{-4}$ & $1.033 \times 10^{-6}$ & $6.129 \times 10^{-7}$ \\
                            & 15 & $8.263 \times 10^{-6}$ & $1.083 \times 10^{-9}$ & $9.541 \times 10^{-10}$ \\
                            & 20 & $8.460 \times 10^{-7}$ & $1.291 \times 10^{-11}$ & $1.678 \times 10^{-11}$ \\
                            & 25 & $6.220 \times 10^{-8}$ & $6.315 \times 10^{-14}$ & $2.056 \times 10^{-12}$ \\
                            & 30 & $1.051 \times 10^{-8}$ & $2.268 \times 10^{-15}$ & $1.948 \times 10^{-12}$ \\
                            & 35 & $1.383 \times 10^{-9}$ & $3.891 \times 10^{-16}$ & $1.943 \times 10^{-12}$ \\
                            & Full ($n=79$) & $1.709 \times 10^{-12}$ & $3.631 \times 10^{-16}$ & $1.942 \times 10^{-12}$ \\[1ex]
\hline
\rule{0pt}{3ex}\textbf{0.50} & 5  & $6.605 \times 10^{-4}$ & $1.594 \times 10^{-6}$ & $3.490 \times 10^{-5}$ \\
\textit{(Slow)}             & 10 & $1.594 \times 10^{-5}$ & $1.069 \times 10^{-9}$ & $4.347 \times 10^{-8}$ \\
                            & 15 & $2.784 \times 10^{-7}$ & $4.111 \times 10^{-13}$ & $4.035 \times 10^{-11}$ \\
                            & 20 & $2.809 \times 10^{-8}$ & $4.860 \times 10^{-15}$ & $1.344 \times 10^{-12}$ \\
                            & 25 & $1.566 \times 10^{-9}$ & $4.409 \times 10^{-16}$ & $1.021 \times 10^{-12}$ \\
                            & 30 & $2.709 \times 10^{-10}$ & $3.959 \times 10^{-16}$ & $1.018 \times 10^{-12}$ \\
                            & 35 & $3.059 \times 10^{-11}$ & $4.004 \times 10^{-16}$ & $1.019 \times 10^{-12}$ \\
                            & Full ($n=79$) & $1.077 \times 10^{-12}$ & $4.125 \times 10^{-16}$ & $1.020 \times 10^{-12}$ \\
\end{tabular}
\end{ruledtabular}
\label{tab:dw_trunc} 
\end{table*}

\begin{table*}[!tbhp]
\caption{\textbf{Spectral truncation convergence metrics for a harmonically confined Brownian particle.} Trajectory-wise quantities converge rapidly with the truncation scale $M$, demonstrating highly predictable low-rank approximations.}
\begin{ruledtabular}
\begin{tabular}{ccccc}
$\tau$ & Truncation $(M)$ & Max TVD & Max $\mathcal{D}_{\rm KL}$ & Max $\vert\mathcal{W}_{\rm diss}(t)\vert$ \\ 
\hline
\rule{0pt}{3ex}\textbf{0.01} & 5  & $2.293 \times 10^{-4}$ & $2.597 \times 10^{-4}$ & $1.532 \times 10^{-4}$ \\
\textit{(Fast)}             & 10 & $8.761 \times 10^{-5}$ & $6.485 \times 10^{-5}$ & $6.675 \times 10^{-5}$ \\
                            & 15 & $4.981 \times 10^{-5}$ & $3.564 \times 10^{-5}$ & $3.457 \times 10^{-5}$ \\
                            & 20 & $2.558 \times 10^{-5}$ & $1.432 \times 10^{-5}$ & $1.650 \times 10^{-5}$ \\
                            & 25 & $1.753 \times 10^{-5}$ & $9.819 \times 10^{-6}$ & $1.060 \times 10^{-5}$ \\
                            & 30 & $1.062 \times 10^{-5}$ & $5.084 \times 10^{-6}$ & $5.961 \times 10^{-6}$ \\
                            & 35 & $7.863 \times 10^{-6}$ & $3.352 \times 10^{-6}$ & $3.982 \times 10^{-6}$ \\
                            & Full ($n=79$) & $2.202 \times 10^{-12}$ & $4.930 \times 10^{-16}$ & $5.703 \times 10^{-12}$ \\[1ex]
\hline
\rule{0pt}{3ex}\textbf{0.50} & 5  & $1.516 \times 10^{-4}$ & $6.234 \times 10^{-6}$ & $5.381 \times 10^{-6}$ \\
\textit{(Slow)}             & 10 & $3.211 \times 10^{-5}$ & $4.117 \times 10^{-8}$ & $2.684 \times 10^{-7}$ \\
                            & 15 & $1.255 \times 10^{-5}$ & $4.537 \times 10^{-9}$ & $5.349 \times 10^{-8}$ \\
                            & 20 & $4.063 \times 10^{-6}$ & $4.555 \times 10^{-10}$ & $8.275 \times 10^{-9}$ \\
                            & 25 & $2.165 \times 10^{-6}$ & $1.372 \times 10^{-10}$ & $3.123 \times 10^{-9}$ \\
                            & 30 & $9.791 \times 10^{-7}$ & $3.107 \times 10^{-11}$ & $9.346 \times 10^{-10}$ \\
                            & 35 & $6.216 \times 10^{-7}$ & $1.335 \times 10^{-11}$ & $4.667 \times 10^{-10}$ \\
                            & Full ($n=79$) & $1.299 \times 10^{-12}$ & $4.558 \times 10^{-16}$ & $3.308 \times 10^{-12}$ \\
\end{tabular}
\end{ruledtabular}
\label{tab:harmonic_trunc} 
\end{table*}
 
\newpage 
\subsection{Efficient simulations in symmetric representation}\label{sec:symm_rep_numerics}
Because the Fokker-Planck operator $\Lop(t)$ is non-symmetric due to the spatial discretization construction, computing its spectral decomposition directly requires dense, non-symmetric eigensolvers that scale poorly [$\mathcal{O}(N^3)$] and are prone to numerical instabilities, such as spurious complex eigenvalue pairs. Under conditions of detailed balance, this bottleneck can be systematically bypassed. By defining a time-dependent similarity transformation to the symmetric gauge,
\begin{equation*}
\Lop_{\rm sym}(t) = \boldsymbol{\pi}_{\rm eq}^{-1/2}(t) \Lop(t) \boldsymbol{\pi}_{\rm eq}^{1/2}(t),
\label{eq:symmetric_gauge_transform}
\end{equation*}
the generator is mapped to a real symmetric tridiagonal matrix. This structure allows the use of highly optimized tridiagonal algorithms (e.g., standard bisection and inverse iteration), which drastically reduce the computational complexity of eigenbasis reconstruction.

To rigorously validate this acceleration scheme, we assess both its numerical fidelity and performance across a wide parameter space. We evaluate the framework at a spatial grid resolution of $N=80$ across uniform temporal snapshots $t/\tau \in \{0, 0.25, 0.50, 0.75, 1.00\}$ for driving protocols varying by three orders of magnitude ($\tau \in \{0.01, 0.1, 1.0\}$). Benchmarked execution times are reported as mean values over $100$ iterations after a $10$-cycle warm-up phase.

The results for the double-well and driven harmonic landscapes are detailed in Table~\ref{tab:sym_double_well} and Table~\ref{tab:sym_harmonic}, respectively. Across all protocols, the symmetry of the gauge-transformed operator is maintained near the double-precision limit, with the non-symmetry metric strictly bounded by $\max \Vert \Lop_{\rm sym} - \Lop_{\rm sym}^T \Vert \sim \mathcal{O}(10^{-14} - 10^{-13})$. Spectral preservation under Eq.~\eqref{eq:symmetric_gauge_transform} is corroborated by checking the maximum absolute discrepancy between the eigenvalues of the original non-symmetric operator and its symmetric-gauge counterpart. The numerical difference remains exceptionally small, varying from $\mathcal{O}(10^{-12})$ in the smoothly behaving harmonic potential to $\mathcal{O}(10^{-8})$ in the highly driven double-well potential, where sharp boundary layers can transiently enhance numerical gradients.

Crucially, this symmetric-gauge formulation yields a dramatic performance improvement. For the double-well landscape, transitioning to the tridiagonal solver cuts the average diagonalisation time from $2.01$~ms to $0.17$~ms, representing a consistent $\sim 12\times$ speedup. For the harmonic potential, the speedup is even more pronounced, consistently exceeding $15\times$. Since counterdiabatic driving requires evaluating the Liouvillian eigenbasis at every temporal step along the driving path, this order-of-magnitude reduction in execution time is critical for making continuous non-equilibrium tracking computationally feasible in complex systems.

\begin{figure*}[!tbhp]
\centering
\includegraphics[width=0.55\textwidth]{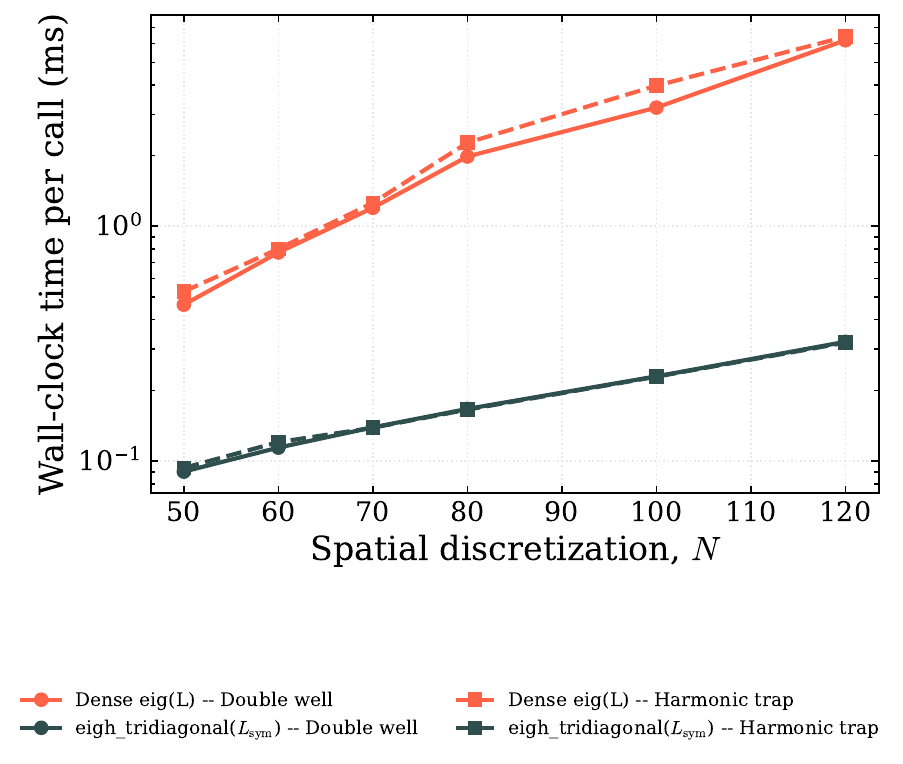}
\label{fig:speedup_diagonalization}
\caption{
  \textbf{Diagonalisation wall-clock time versus spatial
  discretisation $N$.}
  Dense non-symmetric eigensolver (applied to
  $\Lop(t)$) compared with the tridiagonal symmetric
  eigensolver applied to
  $\Lop_{\rm symm} =
  \pieq^{-1/2}\,\Lop\,
  \pieq^{1/2}$.
  The symmetrised form yields a consistent
  $12$--$15\times$ speedup, with eigenvalues agreeing to
  $10^{-8}$--$10^{-13}$.
}

\end{figure*}

\begin{table}[!tbhp]
\caption{\label{tab:sym_double_well}
  \textbf{Symmetric-gauge verification for double-well potential.}
  Symmetry residual, eigenvalue agreement, and
  diagonalisation wall-clock time (averaged over 100 runs,
  $N=80$) at five checkpoints along the protocol for three
  durations $\tau$.
  The tridiagonal solver achieves a consistent
  $11$--$12\times$ speedup.
}
\begin{ruledtabular}
\begin{tabular}{cccccccc}
$\tau$ & $t/\tau$ & $\max\|\Lop_{\rm sym}-\Lop_{\rm sym}^T\|$ & $\max\|\Delta\lambda\|$ & $t_{\rm dense}$ (ms) & $t_{\rm tri}$ (ms) & Speedup \\ 
\hline
\rule{0pt}{2.5ex}%
0.01 & 0.00 & $1.42 \times 10^{-13}$ & $5.60 \times 10^{-8}$ & 2.016 & 0.166 & $12.1\times$ \\
     & 0.25 & $1.42 \times 10^{-13}$ & $4.25 \times 10^{-8}$ & 1.906 & 0.167 & $11.4\times$ \\
     & 0.50 & $1.42 \times 10^{-13}$ & $1.55 \times 10^{-8}$ & 2.014 & 0.165 & $12.2\times$ \\
     & 0.75 & $1.42 \times 10^{-13}$ & $7.02 \times 10^{-9}$ & 2.073 & 0.167 & $12.4\times$ \\
     & 1.00 & $1.14 \times 10^{-13}$ & $5.74 \times 10^{-9}$ & 1.953 & 0.167 & $11.7\times$ \\
\hline
\rule{0pt}{2.5ex}%
0.10 & 0.00 & $1.42 \times 10^{-13}$ & $5.60 \times 10^{-8}$ & 2.005 & 0.166 & $12.1\times$ \\
     & 0.25 & $1.42 \times 10^{-13}$ & $4.25 \times 10^{-8}$ & 1.904 & 0.166 & $11.5\times$ \\
     & 0.50 & $1.42 \times 10^{-13}$ & $1.55 \times 10^{-8}$ & 1.991 & 0.166 & $12.0\times$ \\
     & 0.75 & $1.14 \times 10^{-13}$ & $1.20 \times 10^{-8}$ & 2.056 & 0.167 & $12.3\times$ \\
     & 1.00 & $1.14 \times 10^{-13}$ & $5.74 \times 10^{-9}$ & 1.941 & 0.168 & $11.6\times$ \\
\hline
\rule{0pt}{2.5ex}%
1.00 & 0.00 & $1.42 \times 10^{-13}$ & $5.60 \times 10^{-8}$ & 2.005 & 0.166 & $12.1\times$ \\
     & 0.25 & $1.42 \times 10^{-13}$ & $4.25 \times 10^{-8}$ & 1.894 & 0.166 & $11.4\times$ \\
     & 0.50 & $1.42 \times 10^{-13}$ & $1.55 \times 10^{-8}$ & 1.991 & 0.165 & $12.0\times$ \\
     & 0.75 & $1.42 \times 10^{-13}$ & $7.02 \times 10^{-9}$ & 2.059 & 0.167 & $12.3\times$ \\
     & 1.00 & $1.14 \times 10^{-13}$ & $5.74 \times 10^{-9}$ & 1.951 & 0.167 & $11.7\times$ \\
\end{tabular}
\end{ruledtabular}
\end{table}

\begin{table}[!tbhp]
\caption{\label{tab:sym_harmonic}
  \textbf{Symmetric-gauge verification: harmonic trap.}
  Same layout as Table~\ref{tab:sym_double_well}.
  The tridiagonal solver achieves a $14$--$16\times$ speedup,
  with eigenvalue agreement ranging from $10^{-12}$ to
  $10^{-8}$ along the protocol.
}
\begin{ruledtabular}
\begin{tabular}{cccccccc}
$\tau$ & $t/\tau$ & $\max\|\Lop_{\rm sym}-\Lop_{\rm sym}^T\|$ & $\max\|\Delta\lambda\|$ & $t_{\rm dense}$ (ms) & $t_{\rm tri}$ (ms) & Speedup \\ 
\hline
\rule{0pt}{2.5ex}%
0.01 & 0.00 & $4.26 \times 10^{-14}$ & $2.56 \times 10^{-12}$ & 2.444 & 0.163 & $15.0\times$ \\
     & 0.25 & $4.26 \times 10^{-14}$ & $2.27 \times 10^{-12}$ & 2.392 & 0.165 & $14.5\times$ \\
     & 0.50 & $4.26 \times 10^{-14}$ & $1.81 \times 10^{-10}$ & 2.275 & 0.166 & $13.7\times$ \\
     & 0.75 & $4.26 \times 10^{-14}$ & $2.56 \times 10^{-8}$  & 2.682 & 0.169 & $15.8\times$ \\
     & 1.00 & $5.68 \times 10^{-14}$ & $6.56 \times 10^{-8}$  & 2.568 & 0.170 & $15.1\times$ \\
\hline
\rule{0pt}{2.5ex}%
0.10 & 0.00 & $4.26 \times 10^{-14}$ & $2.56 \times 10^{-12}$ & 2.450 & 0.163 & $15.1\times$ \\
     & 0.25 & $4.26 \times 10^{-14}$ & $2.27 \times 10^{-12}$ & 2.395 & 0.165 & $14.5\times$ \\
     & 0.50 & $4.26 \times 10^{-14}$ & $1.81 \times 10^{-10}$ & 2.279 & 0.166 & $13.7\times$ \\
     & 0.75 & $2.84 \times 10^{-14}$ & $1.71 \times 10^{-8}$  & 2.649 & 0.169 & $15.7\times$ \\
     & 1.00 & $5.68 \times 10^{-14}$ & $6.56 \times 10^{-8}$  & 2.571 & 0.170 & $15.1\times$ \\
\hline
\rule{0pt}{2.5ex}%
1.00 & 0.00 & $4.26 \times 10^{-14}$ & $2.56 \times 10^{-12}$ & 2.451 & 0.163 & $15.0\times$ \\
     & 0.25 & $4.26 \times 10^{-14}$ & $2.27 \times 10^{-12}$ & 2.390 & 0.165 & $14.5\times$ \\
     & 0.50 & $4.26 \times 10^{-14}$ & $1.81 \times 10^{-10}$ & 2.267 & 0.166 & $13.7\times$ \\
     & 0.75 & $4.26 \times 10^{-14}$ & $2.56 \times 10^{-8}$  & 2.670 & 0.169 & $15.8\times$ \\
     & 1.00 & $5.68 \times 10^{-14}$ & $6.56 \times 10^{-8}$  & 2.563 & 0.170 & $15.1\times$ \\
\end{tabular}
\end{ruledtabular}
\end{table}

\subsection{Scalability: Sparsity of the Liouville Operator and Extension to Higher Dimensions}
\label{sec:scalability}

A natural question is whether our biorthogonalization framework to compute the $\Lop_{\rm CD}$ extends
beyond one spatial dimension.
We address this by analysing the structure of the
discretised Liouville operator $\Lop(t)$ and showing
that its sparsity, combined with the spectral truncation
result of Section~\ref{sec:spectral_truncation}, makes the construction
tractable in two and three dimensions. In $d$ spatial dimensions, the state space is discretized
on a regular grid of $N_{x}$ points per dimension, giving
a total of $N = N_{x}^d$ grid points.
The probability vector $\vrho(t) \in \mathbb{R}^N$ is
obtained by flattening the $d$-dimensional grid into a
single vector, and the Liouville operator becomes an
$N \times N$ rate matrix.
Table~\ref{tab:scaling} summarises the matrix size,
memory requirement, and full-diagonalisation cost for a spatial discretization of say
$N_x = 50$ across dimensions.

\begin{table}[h]
\caption{
  Scaling of the Liouville operator $\Lop(t)$
  with spatial dimension $d$ for $N_x = 50$ grid points
  per dimension.
  Memory assumes dense storage (8 bytes [$\mathtt{FLOAT64}$] per entry).
  Diagonalisation cost assumes $O(N^3)$ complexity.
}
\label{tab:scaling}
\begin{ruledtabular}
\begin{tabular}{crrrl}
$d$ & $N = N_x^d$ & Matrix size & Memory & Diag.\ cost \\
\hline
1 & $50$ & $50 \times 50$ & $<1\,\mathrm{MB}$ &
  $\sim 10^5$ ops \\
2 & $2{,}500$ & $2500 \times 2500$ & $50\,\mathrm{MB}$ &
  $\sim 10^{10}$ ops \\
3 & $125{,}000$ & $1.25\!\times\!10^5 \times 1.25\!\times\!10^5$ &
  $125\,\mathrm{GB}$ & $\sim 10^{15}$ ops \\
\end{tabular}
\end{ruledtabular}
\end{table}

Full diagonalisation is therefore already intractable in 3D for spatial discretization of say, $N_x = 50$.

The key structural property that rescues the situation
is that $\Lop(t)$ is \emph{extremely sparse}, which can be seen from Figures~(\ref{fig:matrix-dw}, \ref{fig:matrix-harm}).
In $d$ dimensions with nearest-neighbour coupling
(the Sasa--Tasaki discretisation~\citep{sasa2005steadystatethermodynamics}),
each grid point $\mathbf{i} = (i_1,\ldots,i_d)$ is
connected only to its $2d$ nearest neighbours
$\mathbf{i} \pm \mathbf{e}_k$ ($k = 1,\ldots,d$),
where $\mathbf{e}_k$ is the unit vector in direction $k$.
The column of $\Lop$ corresponding to site
$\mathbf{i}$ therefore has exactly $2d$ off-diagonal
non-zero entries (the transition rates to/from neighbours)
plus one diagonal entry (the negative total outgoing rate).
The total number of non-zero entries is thus:
\begin{equation}
  \mathrm{nnz}(\Lop) = (2d + 1)\,N
  = \mathcal{O}(d\,N),
\label{eq:nnz}
\end{equation}
growing \emph{linearly} in the number of states, not
quadratically.
The sparsity — fraction of non-zero entries — is
$(2d+1)/N$, which decreases as $N^{-1}$ for fixed $d$:
\begin{equation}
  \text{sparsity} =
  \frac{(2d+1)\,N}{N^2} = \frac{2d+1}{N}
  \xrightarrow{N\to\infty} 0.
\label{eq:sparsity}
\end{equation}
\begin{table}[!h]
\caption{
  Non-zeros per column and sparsity of $\Lop(t)$
  for $N_x = 50$ per dimension.
}
\label{tab:sparsity}
\begin{ruledtabular}
\begin{tabular}{crrl}
$d$ & nnz/column & Total nnz & Sparsity \\
\hline
1 & 3   & $150$         & $6.0\%$    \\
2 & 5   & $12{,}500$    & $0.2\%$    \\
3 & 7   & $875{,}000$   & $0.006\%$  \\
\end{tabular}
\end{ruledtabular}
\end{table}

\noindent
The sparsity arises directly from the local structure
of the Fokker--Planck equation: probability can only
flow between adjacent grid points, so the generator
$\Lop(t)$ is a \emph{graph Laplacian} on the
$d$-dimensional lattice, with non-zeros only at
positions corresponding to neighboring pairs.
In memory, $\Lop(t)$ is therefore never stored
as a dense $N\times N$ array but as a sparse matrix
with $O(dN)$ entries — reducing the 3D memory
requirement from $125\,\mathrm{GB}$ (dense) to
$\sim 7\,\mathrm{MB}$ (sparse, $N=125{,}000$,
$d=3$). As already noted in Section~\ref{sec:rank1} for the equilibrium problem the rank-one closed form is $\mathcal O(N)$, and needs none of this; the sparse/truncated construction is relevant to the diagnostic analysis and to the NESS setting where $r_0$ must be computed numerically.
\end{document}